\documentclass[11pt, a4paper]{article}
\pdfoutput=1
\usepackage{jheppub}
\usepackage{latexsym, graphicx}
\usepackage{amsmath, amssymb}
\usepackage{color}

\title{Quantum radiation by an Unruh-DeWitt detector in oscillatory motion}
\author[a]{Shih-Yuin Lin}
\affiliation[a]{Department of Physics, National Changhua University of Education, Changhua 50007, Taiwan}
\emailAdd{sylin@cc.ncue.edu.tw}
\date{11 September 2017}

\abstract{Quantum radiated power emitted by an Unruh-DeWitt (UD) detector in linear oscillatory motion in (3+1)D Minkowski space, 
with the internal harmonic oscillator minimally coupled to a massless scalar field, is obtained non-perturbatively by numerical method. 
The signal of the Unruh-like effect experienced by the detector is found to be pronounced in quantum radiation in the highly non-equilibrium 
regime with high averaged acceleration and short oscillatory cycle, and the signal would be greatly suppressed by quantum interference 
when the averaged proper acceleration is sufficiently low.
An observer at a fixed angle would see periods of negative radiated power in each cycle of motion, while the averaged radiated power over a cycle is always positive as guaranteed by the quantum inequalities. Coherent high harmonic generation and down conversion 
are identified in the detector's quantum radiation. Due to the overwhelming largeness of the vacuum correlators of the free field, the 
asymptotic reduced state of the harmonics of the radiation field is approximately a direct product of the squeezed thermal states.}


\begin{document}

\maketitle

\section{Introduction}

A uniformly, linearly accelerated point-like detector moving in the Minkowski vacuum will experience thermal fluctuations at a temperature 
proportional to the proper acceleration of the detector~\cite{Un76}. This is called the Unruh effect and the temperature is called the 
Unruh temperature. While the derivation in time-dependent perturbation theory is well established, direct experimental evidence for the 
Unruh effect is still lacking. One closely related observation in laboratories is the electron depolarization in storage rings, namely, 
the Sokorov-Ternov effect~\cite{ST63}, which can be connected to the ``circular Unruh effect"~\cite{BL83, BL87, Un98, AS07}. 
Nevertheless, the centripetal acceleration in the circular Unruh effect is quite 
different in nature from the original linear, uniform acceleration in the Unruh effect~\cite{HJ00}. To get closer to the original conditions 
in Unruh's derivation, there have been existing proposals to look at the quantum correction by the Unruh effect to the radiation emitted by a 
{\it linearly} accelerated charge or atom~\cite{CT99, SSH06, SSH08}, which is called the ``Unruh radiation".

Seeking the evidence of the Unruh temperature in quantum radiation is, however, not as straightforward as it appears.
To well define a finite temperature in an atom-field state, the atom should be in equilibrium with the field. Unfortunately a 
uniformly, linearly accelerated Unruh-DeWitt (UD) detector (analogous to an atom)~\cite{Un76, DeW79} derivatively coupled to a massless 
scalar field in the Minkowski vacuum emits {\it no} radiation in equilibrium conditions in (1+1)D Minkowski space 
\cite{Gr86, RSG91, Un92, MPB93, HR00}. 
In (3+1)D Minkowski space there will be radiation by a uniformly accelerated UD detector in steady state at late times (at a constant 
radiation rate with respect to the proper time of the detector), but the radiated energy is not converted from the one experienced by 
the detector in the Unruh effect~\cite{LH06}. The physical reason for these results is that quantum interference between the vacuum 
fluctuations driving the detector and the radiation emitted by the driven detector is perfectly destructive in equilibrium conditions. 

In laboratories, producing an eternal, constant linear acceleration for a charge or an atom is impossible, anyway. 
In Ref.~\cite{CT99} and similar proposals the charge motion would be driven by an intense laser field, which can make the acceleration 
linear, but not uniform, thus
the radiation corresponding to the Unruh effect may not be totally suppressed by interference. The only concern is that the Unruh 
temperature is not well defined in these non-equilibrium setups. Fortunately, a time-varying effective temperature whose value is close 
to the Unruh temperature of the averaged acceleration can be defined for the detectors in oscillatory motion \cite{DLMH13}. 
As we will demonstrate later, in the regime of high acceleration and short oscillating cycle of the motion, 
the signal of the effective Unruh temperature can be pronounced in the Unruh radiation \footnote{This paper is partly based on \cite{Lin16}}.

This paper is organized as follows. 
To get non-perturbative, time-dependent results of the radiation field emitted by a point-like detector in oscillatory motion, 
in Section \ref{SecTechIs} we introduce the UD harmonic-oscillator (HO) detector model considered in Refs.~\cite{LH06, DLMH13} 
and then address some technical issues.
We determine the radiation in the radiation zone defined in the Minkowski coordinates for a laboratory observer \cite{Hi02, Ja98, LH06}. 
Then we present our numerical results of the radiated power in Section \ref{SecResult}.
We show that when observed at a fixed angle, there will be negative radiated power in some periods during each cycle of the
oscillatory motion \cite{OYZ15, Lin16}. This indicates that the Unruh radiation observed at the null infinity may correspond to a 
multi-mode squeezed state of the field~\cite{KF93, UW84, SSH06, SSH08} or something similar.
So we further study the correlations in the radiation field in Section \ref{SecCorr}, where we identify the nonlinear optical 
effects such as the coherent high-harmonic generation and down-conversion \cite{SSH08}. 
Then in Section \ref{SecARS} the asymptotic reduced state of the field harmonics is constructed using the correlators of the field in the 
radiation zone. Our analysis shows that this asymptotic reduced state of the radiation field looks like a direct product of the squeezed 
thermal states. Finally, a summary of our results is given in Section \ref{SecConcl}. A few analytic results for the two-point correlators 
are given in Appendix \ref{Sec2pt}, which can help to control the singularities in our numerical calculation.

\section{Renormalized expectation values of stress-energy tensor}
\label{SecTechIs}

Consider an Unruh-DeWitt detector with its internal degree of freedom acting as a harmonic oscillator and minimally coupled to a 
massless scalar field $\Phi$ in (3+1)D Minkowski space, described by the action  
\begin{eqnarray}
  S &=& -\int d^4 x \sqrt{-g} \frac{1}{2}\partial_\mu\Phi(x) \partial^\mu\Phi(x) +\nonumber\\ & &
  \int d\tau 
   \left\{ \frac{m_0}{2}\left[\left(\partial_\tau Q\right)^2
    -\Omega_{0}^2 Q^2\right]
    + \lambda_0\int d^4 x 
    Q(\tau)\Phi (x)\delta^4\left(x^{\mu}-z^{\mu}(\tau)\right)\right\},
  \label{Stot1}
\end{eqnarray}
where $g_{\mu\nu} = {\rm diag}(-1,1,1,1)$, $z^{\mu}$ is the worldline of the detector parametrized by its proper time $\tau$, $\Omega_0$
is the bare natural frequency of the internal HO, and $\lambda_0$ is the coupling constant of the detector and the field. 
Here we take $c=\hbar=G=1$. From (\ref{Stot1}) one can derive the conjugate momenta $\hat{P}= m_0 d\hat{Q}/d\tau$ and $\hat{\Pi} = 
\partial_t\hat{\Phi}$ of the detector and the field, respectively. Below we set $m_0=1$ for simplicity.
 
As we discussed in Ref.~\cite{LH06}, by virtue of the linearity of this UD detector theory, the field operator $\hat{\Phi}^{}_x$ after the 
detector-field coupling is switched on will become a linear combination of the mode functions each associated with a creation 
($\hat{b}^\dagger_{\bf k}$) or annihilation operator ($\hat{b}^{}_{\bf k}$) of the free field mode with wave vector ${\bf k}$, or a raising 
($\hat{a}^\dagger$) or lowering operator ($\hat{a}^\dagger$) of the free internal HO of the detector $Q$. 
Also due to the linearity, a mode function of the field has the form $\phi^ \kappa_x = \phi^{{}^{[0]} \kappa}_{\; x}+
\phi^{{}^{[1]} \kappa}_{\;x}$ ($ \kappa = a, \{{\bf k}\}$; $\phi^a_x$ and $\phi^{\bf k}_x$ are associated with $\hat{a}$ and 
$\hat{b}^{}_{\bf k}$, respectively), which is the superposition of the homogeneous solution $\phi^{{}^{[0]} \kappa}_x$ corresponding to 
vacuum fluctuations of the free field and the inhomogeneous solution $\phi^{{}^{[1]} \kappa}_{\;x}$ sourced from the point-like detector. 
One can group the homogeneous solutions of the mode functions with the associated operators into $\hat{\Phi}^{{}^{[0]}}_x$ and the 
inhomogeneous solutions into 
$\hat{\Phi}^{{}^{[1]}}_x$ such that the field operator is in the form $\hat{\Phi}^{}_x=\hat{\Phi}^{{}^{[0]}}_x+\hat{\Phi}^{{}^{[1]}}_x$. 

Suppose the detector-field coupling is switched on at $t=t^{}_I$, when the combined system is initially in the factorized state 
\begin{equation}
  |\psi(0)\rangle = |g^{}_A\rangle \otimes | 0^{}_M\rangle,  
\label{initstat}
\end{equation}
which is a product of the ground state of the free UD detector $|g^{}_A\rangle$ and the Minkowski vacuum of the field $|0^{}_M\rangle$.
The wavefunction (or density matrix) is thus in a Gaussian form.
In the Heisenberg picture, the correlators of the field amplitude at different spacetime points $x$ and $x'$ after $t=t^{}_I$ are given by
\begin{equation}
   G(x,x')\equiv \langle \psi(0)| \hat{\Phi}^{}_x \hat{\Phi}^{}_{x'} |\psi(0)\rangle = \sum_{i,j=0,1} G^{[ij]}(x,x'),
\label{Gij}
\end{equation}
where $G^{[ij]}(x,x') \equiv \langle \hat{\Phi}^{{}^{[i]}}_x \hat{\Phi}^{{}^{[j]}}_{x'} \rangle$
with respect to the initial state $|\psi(0)\rangle$ in (\ref{initstat}). 
Then the expectation value of the stress-energy tensor (minimal, $\xi=0$) can be written as
\begin{equation}
  \langle T_{\mu\nu}[\Phi(x)]\rangle = 
	\lim_{x'\to x} {\rm Re}\left[ {\partial\over\partial x^\mu}{\partial\over\partial x'^\nu}
	- {1\over 2} g^{}_{\mu\nu}g_{}^{\rho\sigma}{\partial\over\partial x^\rho}{\partial\over\partial x'^\sigma}\right]G(x,x')
	\equiv \sum_{i,j=0,1} \langle T^{[ij]}_{\mu\nu}(x)\rangle
\end{equation}
where $\langle T^{[ij]}_{\mu\nu}\rangle$ is contributed by $G^{[ij]}$.

$G^{[00]}(x, x')$ is the Green's function of the free field, it diverges as $x'\to x$, so does $T^{[00]}_{\mu\nu}$.
Nevertheless, there is no physical effect from this part of the stress-energy tensor in Minkowski space, and so it can be subtracted 
in the spirit of the normal ordering in obtaining the vacuum energy in the conventional quantum field theory. 
We thus define the renormalized stress-energy tensor as
$\langle T_{\mu\nu}(x) \rangle_{\rm ren} \equiv \langle T_{\mu\nu}(x) \rangle -\langle T_{\mu\nu}^{[00]}(x)\rangle$.
Doing this is nothing but setting the zero point of vacuum stress-energy.

Suppose a UD detector is oscillating about the spatial origin of the Minkowski coordinates, which is chosen as the laboratory frame.
Suppose a set of the radiation-detecting apparatus are located at a large constant radius $r$ at different angles from the spatial origin,
namely, at $x^\mu=(x^0, x^1, x^2, x^3)= (t, r \sin\theta \cos \varphi, \,r \sin\theta \sin \varphi, \, r \cos \theta)$ in the 
Minkowski coordinates. Then 
the differential radiated power per unit solid angle measured in laboratories can be written as 
\begin{equation}
  {d{\cal P} \,\over d\Omega^{}_{\rm II}}(t,\theta,\varphi) = -\lim_{r\to\infty} r^2 \langle T_{tr}(x) \rangle_{\rm ren}   
	=  {d{\cal P}^{[11]}\over d\Omega_{\rm II}} + {d{\cal P}^{[01]}\over d\Omega_{\rm II}} + {d{\cal P}^{[10]}\over d\Omega_{\rm II}},  
	\label{dPdWIIren}
\end{equation}
where $d\Omega_{\rm II}$ is the element of the solid angle and the $[ij]$ component is defined by
\begin{equation}
	{d{\cal P}^{[ij]}\over d\Omega_{\rm II}} \equiv -\lim_{r\to\infty,x'\to x}{r^2\over 2} {\rm Re} 
	\left(\partial_t \partial_{r'} + \partial_r \partial_{t'}\right) G^{[ij]}(x,x'). \label{dP11dW}
\end{equation}
In this paper we are considering the cases with the UD detector in linear oscillatory motion in $x^3$-direction, namely, 
$z^\mu(\tau) = (z^0(\tau), 0,0, z^3(\tau))$, and the radiation will be independent of the azimuth angle $\varphi$ by symmetry. 

The whole calculation will be started with the subtracted two-point correlators of the field. However, as many quantities for quantum fields 
with infinite degrees of freedom, each of $G^{[11]}$, $G^{[10]}$ and $G^{[01]}$ is still singular in the coincidence limit.
One has to control the singularities with the hope that some of them could cancel in the measurable quantities while others could be
tamed by introducing physical cutoffs. To identify the problem, let us look into more details of the correlators.

\subsection{Two-point correlators at late times}
\label{2PCorrLT}

At late times, the retarded field has carried the initial information in the detector away to the null infinity, so the behavior of the combined system around the detector is dominated by vacuum fluctuations of the field as well as the detector's response to them.
From Ref.~\cite{LH06} and \cite{OLMH12}, one has
\begin{eqnarray}
  & & \langle Q(\tau)Q(\tau')\rangle \approx \langle Q(\tau)Q(\tau')\rangle_{\rm v} \nonumber\\ &=& 
	\frac{8\gamma\pi}{\Omega^2}\int_{\tau^{}_I\to-\infty}^\tau d\tilde{\tau} \int_{\tau^{}_I\to-\infty}^{\tau'} d\tilde{\tau}' 
	K(\tau-\tilde{\tau})K(\tau-\tilde{\tau}') D^+(z(\tilde{\tau}), z(\tilde{\tau}')), \label{QQvLT}
\end{eqnarray}
where the v-part of the correlator $\langle  Q(\tau)Q(\tau')\rangle_{\rm v}$ is defined by the mode functions associated with the field 
operators and the initial data in the field state only. Here $\gamma\equiv\lambda_0^2/(8\pi)$ is the coupling strength, $\Omega\equiv\sqrt{
\Omega_r^2-\gamma^2}$ is the natural frequency with $\Omega_r$ renormalized from the bare natural frequency of the detector $\Omega_0$, 
$K(X)\equiv e^{-\gamma X}\sin\Omega X$ is the propagator, $\tau^{}_I$ is the proper time of the detector at the initial moment, and 
\begin{equation}
   D^+(x, x') \equiv \frac{\hbar}{(2\pi)^2 (x_\mu - x'_\mu)(x^\mu - x'^\mu)} \label{WightmanD}
\end{equation}
is the positive-frequency Wightman function of the free massless scalar field in the Minkowski vacuum state, with a proper choice of the 
integration contour understood \cite{BD82}. While one could get rid of the divergence of the integrand as $\tilde{\tau}'\to \tilde{\tau}$ 
in (\ref{QQvLT}) by choosing the integration contour, the divergence of $\langle Q(\tau)Q(\tau')\rangle$ in the coincidence limit 
$\tau'\to\tau$ is unavoidable.

From (A1) in Ref.~\cite{LH06}, denoting $x^\mu=(t,{\bf x})$ and $\Phi_x=\Phi_{\bf x}(t)$, one has 
\begin{eqnarray}
G^{[11]}(x,x') &=& \langle \hat{\Phi}^{{}^{[1]}}_{\bf x}(t)\hat{\Phi}^{{}^{[1]}}_{\bf x'}(t') \rangle
  = {\lambda_0^2\over (2\pi)^2 4 {\cal R R'}}\theta(\eta_-)\theta(\eta_-') \langle\hat{Q}(\eta_-)\hat{Q}(\eta_-') \rangle, \label{G11} \\
\partial_\mu \partial_{\nu'} G^{[11]}(x,x')
	&=& {\lambda_0^2\over (2\pi)^2 4 {\cal R R'}}\theta(\eta_-)\theta(\eta_-') \times \nonumber\\ & &
  \left[ {{\cal R}_{,\mu}{\cal R}'_{,\nu'}\over {\cal R R'}}\langle \hat{Q}(\eta_-)\hat{Q}(\eta_-') \rangle +
  \eta_{-,\mu}\eta'_{-,\nu'}\langle \hat{P}(\eta_-)\hat{P}(\eta_-')\rangle \right. \nonumber\\ & & \left. 
	- {{\cal R}_{,\mu}\over {\cal R}}\eta'_{-,\nu'} \langle \hat{Q}(\eta_-)\hat{P}(\eta_-')\rangle -
	\eta_{-,\mu}{{\cal R}'_{,\nu}\over {\cal R'}}\langle \hat{P}(\eta_-)\hat{Q}(\eta_-')\rangle \right], \label{DDG11}
\end{eqnarray}
with the singular behaviors of (\ref{QQvLT}) and other correlators of the detector. Also at late times,
\begin{eqnarray}
& & G^{[10]}(x,x') = \langle \hat{\Phi}^{{}^{[1]}}_{\bf x}(t) \hat{\Phi}^{{}^{[0]}}_{\bf x'}(t') \rangle \approx \nonumber\\ & & 
  {2\gamma\theta(\eta_-(x))\over \Omega {\cal R}(x)} 
	\int_{\tau^{}_I\to-\infty}^{\tau^{}_-(x)} d\tilde{\tau} K(\tau^{}_-(x) -\tilde{\tau}) D^+(x', z(\tilde{\tau}-i\epsilon)), \label{G10} \\
& & \partial_\mu \partial_{\nu'} G^{[10]}(x, x') \approx 
   {2\hbar\gamma\theta(\eta_-(x))\over \Omega {\cal R}(x)} \times \nonumber\\ & &
	  \int_{\tau^{}_I\to-\infty}^{\tau^{}_-(x)} d\tilde{\tau} 
		\left[ -\frac{{\cal R}_{,\mu}}{\cal R}  K(\tau^{}_-(x) -\tilde{\tau})  + \eta_{-,\mu}K'(\tau^{}_-(x) -\tilde{\tau})\right]
		D^+_{,\nu'}(x', z(\tilde{\tau}-i\epsilon)),
\label{DDG10}
\end{eqnarray}
whose integrands and integrals diverge as $\tilde{\tau}'\to \tau^{}_-(x')$ and $x'\to x$, respectively.
Here we denote $K'(X) \equiv \partial_X K(X)$, $f_{,\mu}\equiv\partial f(x)/\partial x^\mu$, $g_{,\mu'}\equiv\partial g(x')/\partial x'^\mu$, 
the retarded time $\tau^{}_-$ is defined by $\sigma(x, z(\tau^{}_-(x)))=0$ subject to
$x^0 > z^0(\tau^{}_-(x))$ with Synge's world function $\sigma(x,x') \equiv  -(x_\mu - x'_\mu)(x^\mu-x'^\mu)/2$, and
${\cal R}(x)$ is the retarded distance determined by the local frame of the detector as
\begin{equation}
    {\cal R} = \left| \frac{d\sigma(x,z(\tau))}{d\tau} \right|_{\tau=\tau^{}_-},
\end{equation}
and $\eta_- \equiv \tau^{}_-(x) -\tau^{}_I$, while ${\cal R}' \equiv {\cal R}(x')$ and $\eta_-' \equiv \eta_-^{}(x')$.
Note that the retarded distance $aX/2$ and the retarded proper time for a uniformly accelerated detector in Ref.~\cite{LH06}
has been generalized to ${\cal R}$ and $\eta_-$ for a detector in oscillatory motion here
\footnote{In Eq. (A1) in Ref.~\cite{LH06}, 
$\partial_t \partial_{r'} G^{[11]}_{\rm v}(x,x')$ only counts the contribution by the 
v-parts of the correlators of the detector $\langle .. \rangle_{\rm v}$. The expression for the a-part, $\partial_t \partial_{r'} 
G^{[11]}_{\rm a}(x,x')$, has the same form as Eq. (A1) in Ref.~\cite{LH06} except the v-parts of the detector-detector correlators
$\langle .. \rangle_{\rm v}$ are replaced by the a-parts $\langle .. \rangle_{\rm a}$. Since $\langle \Phi_x \rangle =0$ in the cases
we are considering, the expression for $\partial_t \partial_{r'} G^{[11]}(x,x') =\partial_t \partial_{r'} [G^{[11]}_{\rm a}(x,x') + 
G^{[11]}_{\rm v}(x,x') ]$  is simply the same expression as Eq. (A1) in Ref.~\cite{LH06} except all the v-part of the detector-detector
correlators there are replaced by the complete one, namely, 
$\langle .. \rangle_{\rm v} \to \langle .. \rangle=\langle .. \rangle_{\rm a}+ \langle .. \rangle_{\rm v}$.}.
For an observer at the null infinity 
the more the 4-velocity of the detector is pointing towards the observer,  the smaller ${\cal R}/r$ is.

\subsection{Controlling the singularities}
\label{SecSubSing}

From the correlators of the detector $\langle QQ\rangle$, $\langle QP\rangle$, $\langle PQ\rangle$, and $\langle PP\rangle$ in (\ref{DDG11}) and thus (\ref{dP11dW}), one could extract the Unruh or the effective temperature experienced by the detector \cite{LH07}. 
So we call the all-retarded-field part of the differential radiated power $d{\cal P}^{[11]}/d\Omega_{\rm II}$ as the naive Unruh radiation.
It diverges when one takes the coincidence limit on the two-point correlators of the detector, as one can see from (\ref{dP11dW}), 
(\ref{QQvLT}) and (\ref{DDG11}).
When the trajectory of the detector is not as simple as those in uniform motion or uniform acceleration, setting consistent cutoffs in the 
double integral for the correlators such as (\ref{QQvLT}) is not easy. In Refs.~\cite{DLMH13} and \cite{OLMH12} we have dealt with these 
singularities carefully. We subtract the integral for the two-point correlators of the detector in oscillatory motion by those for a 
uniformly accelerated detector. The subtracted integral gives a finite result. Then we add the analytic results for the uniformly 
accelerated detector back, whose singular behavior are well understood and under control once the UV cutoff is introduced.

For the interference terms of the differential radiated power, $d{\cal P}^{[01]}/d\Omega_{\rm II}+ d{\cal P}^{[10]}/d\Omega_{\rm II}$, 
the situation is similar.
As we mentioned, in the integrand of (\ref{DDG10}), 
\begin{eqnarray} 
  {\partial\over\partial x^\nu} D^+(x - z(\tilde{\tau}-i\epsilon)) = {\hbar\over 2\pi^2} {z_\nu(\tilde{\tau}-i\epsilon)-x_\nu \over
	  \left[(x_\mu - z_\mu(\tilde{\tau}-i\epsilon))(x^\mu - z^\mu(\tilde{\tau}-i\epsilon))\right]^2}.
\end{eqnarray}
diverges as $\tilde{\tau}\to \tau^{}_-(x)$ and $\epsilon\to 0+$.
When $\epsilon$ is positive and non-zero, expanding $z^\mu(\tau^{}_- -i\epsilon) \approx z^\mu(\tau^{}_-)-i\epsilon 
\dot{z}^\mu(\tau^{}_-) + (-i\epsilon)^2 \ddot{z}^\mu(\tau^{}_-)/2 + \cdots$, one finds 
\begin{eqnarray}
  & &\partial_\nu D^+(x - z(\tau^{}_-(x) -i\epsilon)) = {\hbar\over 2\pi^2}\left\{ 
	    {1\over\epsilon^2} {x_\nu-z_\nu^-\over 4\left[ \dot{z}_\mu^-(x^\mu -z^\mu_-)\right]^2} + \right. \nonumber\\
  & & \left.{i\over\epsilon}\left[ {\dot{z}_\nu^-\over 4 \left[ \dot{z}_\mu^-(x^\mu -z^\mu_-)\right]^2}	
	    - {(x_\nu-z_\nu^-)(\ddot{z}_\rho^-(x^\rho-z^\rho_-) - \dot{z}_\rho^- \dot{z}^\rho_-) \over
			   4 \left[ \dot{z}_\mu^-(x^\mu -z^\mu_-)\right]^3} \right] + O(\epsilon^0)\right\}
\end{eqnarray}
with $z^\mu_- \equiv z^\mu(\tau^{}_-(x))$. To subtract out the divergent $\epsilon^{-2}$ and $\epsilon^{-1}$ terms,
one needs to introduce a reference worldline, $\tilde{z}^\mu(\tau)$ with $\tilde{z}^\mu(\tau^{}_-) = z^\mu(\tau^{}_-)$, 
$\dot{\tilde{z}}^\mu(\tau^{}_-)=\dot{z}^\mu(\tau^{}_-)$, and $\ddot{\tilde{z}}^\mu(\tau^{}_-)=\ddot{z}^\mu(\tau^{}_-)$.
For a general worldline $z^\mu$ at a specific moment $\tau$, 
the simplest reference worldline for subtraction is again the one for a uniformly accelerated detector, and
luckily, we have also obtained the analytic results of the interference terms for the uniformly accelerated detector in closed form in 
Ref.~\cite{LH06}. Similar to what we did for 
$d{\cal P}^{[11]}/d\Omega_{\rm II}$, after we get the finite result for the subtracted interference terms,
we add the analytic result back in the final step to get the complete result with the divergences well controlled. 

In the cases with $\ddot{z}^\mu(\tau^{}_-)=0$, the reference worldline and the analytic result to be added reduce to those for the detector 
in uniform motion with $\tilde{z}^\mu(\tau^{}_-) = z^\mu(\tau^{}_-)$ and $\dot{\tilde{z}}^\mu(\tau^{}_-)=\dot{z}^\mu(\tau^{}_-)$. 
Some analytic expressions of the correlators for an UD detector in uniform acceleration and uniform 
motion are given in Appendix \ref{Sec2pt}, for adding back to the subtracted numerical results.

For the reference worldlines either in uniform acceleration or in uniform motion, the UV divergence ($\Lambda_1$ in Section \ref{Sec2pt}) 
in $T^{[11]}_{\mu\nu}$ will be exactly canceled by the ones in the interference terms $T^{[10]}_{\mu\nu}+T^{[01]}_{\mu\nu}$~\cite{LH06}.  
Thus, after combining the numerical result of the subtracted power and the exact 
analytic result from the reference worldlines, the final result will be regular and independent of the UV cutoff for the detector.

\subsection{On-resonance case}

When the period of a cycle of oscillatory motion in the proper time of the UD detector is integer times of the natural period of the 
internal HO ($\tau_p = n \times (2\pi/\Omega)$, $n$ integer), it is possible to get the late-time result with a finite domain of 
integration to make the numerical calculation more economic. Our experience in calculating the effective temperature in a UD detector 
in oscillatory motion shows that such kind of the resonance condition is not catastrophic \cite{DLMH13}.
In these cases, since $z^0(\tau-\tau_p) = z^0(\tau)-t_p$ and $z^3(\tau-\tau_p)=z^3(\tau)$ for a detector in oscillatory 
motion of period $\tau_p$ in proper time and $t_p$ in the coordinate time, and $K(\tau-\tilde{\tau}+m \tau_p)= 
e^{-\gamma(\tau-\tilde{\tau}+m \tau_p)}\sin\Omega(\tau-\tilde{\tau}+m \tau_p)=e^{-m \gamma \tau_p}K(\tau-\tilde{\tau})$ with integer $m$, 
when $\tau-\tau^{}_I \to \infty$ and $\tau'-\tau^{}_I \to \infty$ at late times, (\ref{QQvLT}) becomes
\begin{eqnarray}
  \langle \hat{Q}(\tau)\hat{Q}(\tau')\rangle
  &=& \frac{2\gamma\hbar}{\pi \Omega^2}\sum_{n,n'=0}^\infty e^{-\gamma(n+n')\tau_p} 
	\int_{\tau-\tau_p}^\tau d\tilde{\tau} \int_{\tau'-\tau_p}^{\tau'} d\tilde{\tau}' \times \nonumber\\ & &
	\frac{K(\tau-\tilde{\tau})K(\tau-\tilde{\tau}')}{\left[z^3(\tilde{\tau})-z^3(\tilde{\tau}')\right]^2 -
	  \left[z^0(\tilde{\tau})-z^0(\tilde{\tau}')-(n-n')t_p\right]^2}\nonumber\\
  &=&	\frac{2\gamma\hbar}{\pi \Omega^2} \left(\frac{1}{1-e^{2\gamma\tau_p}}\right)
	  \int_{\tau-\tau_p}^{\tau} d\tilde{\tau}\int_{\tau'-\tau_p}^{\tau'} d\tilde{\tau}' \sum_{n=-\infty}^\infty e^{-|n| \gamma \tau_p}  
		\times\nonumber\\ & & \frac{ K(\tau-\tilde{\tau})K(\tau-\tilde{\tau}')}
		{\left[z^3(\tilde{\tau})-z^3(\tilde{\tau}')\right]^2 -\left[z^0(\tilde{\tau})-z^0(\tilde{\tau}')+n t_p\right]^2}.
\end{eqnarray}
The integrand can be written in closed form by noting that $\sum_{n=0}^\infty z^n(a+n)^{-s} = {\bf \Phi}(z,s,a)$, which is the 
Hurwitz-Lerch transcendent \cite{math10}.
This reduces the domain of the integral from ${\bf R}^2$ to a finite square, though the integrand diverges at $\tilde{\tau}= 
\tilde{\tau}'+n\tau_p$ for some integer $n$ and have to be treated in the way given in Section \ref{SecSubSing}.

Similarly, the late-time interference terms (\ref{G10}) and (\ref{DDG10}) in the on-resonance cases can be written as
\begin{eqnarray}
	G^{[10]}(x,x')&=& \frac{\gamma\hbar}{2\pi^2\Omega {\cal R}(x)}
	\int_{\tau^{}_-(x)-\tau_p}^{\tau^{}_-(x)}d\tilde{\tau} \sum_{n=0}^\infty e^{-n \gamma \tau_p} \nonumber\\ & &\frac{K(\tau^{}_-(x)-\tilde{\tau})}
	{\left|{\bf x}'-{\bf z}(\tilde{\tau})\right|^2 -\left[x'^0-z^0(\tilde{\tau})+n t_p\right]^2},\\
	\partial_\mu \partial_{\nu'} G^{[10]}(x,x') 
	&=& \frac{\gamma\hbar}{2\pi^2\Omega {\cal R}}
	\int_{\tau^{}_-(x)-\tau_p}^{\tau^{}_-(x)}d\tilde{\tau} \sum_{n=0}^\infty e^{-n \gamma \tau_p} (-2) \times\nonumber\\
	& & \frac{\left[ -\frac{{\cal R}_{,\mu}}{\cal R} K(\tau^{}_- -\tilde{\tau}) + \tau_{-,\mu}K'(\tau^{}_- -\tilde{\tau})\right]
	  \left[x'_\nu -z_\nu(\tilde{\tau}-n\tau_p)\right]}{\left\{ \left|{\bf x}'-{\bf z}(\tilde{\tau})\right|^2 -
	  \left[x'^0-z^0(\tilde{\tau})+n t_p\right]^2\right\}^2},
\end{eqnarray}
whose domains are reduced to finite intervals, though the integrands also diverge as 
$\tilde{\tau}-n \tau_p \to \tau^{}_-(x')$ for some positive integer $n$ and should be properly treated.

When $n \tau_p = 2\pi/\Omega$ with integer $n$, the situation is similar to the above case with $\tau_p = n \times (2\pi/\Omega)$. 
Only minor modifications on the above integrals are needed.

\section{Radiated power}
\label{SecResult}

As an example, let us consider a detector moving along the worldline given by Chen and Tajima in Ref. \cite{CT99},
\begin{equation}
  z^\mu (t) = \left( t, 0, 0, -{1\over\omega_0}\sin^{-1}{2a_0\cos\omega_0 t \over \sqrt{1+4 a_0^2}}\right) ,
	\label{CTtraj}
\end{equation}
which is the trajectory of a charge at a nodal point of magnetic field in a cavity. 
The effective temperature of the detector in this worldline has been studied in \cite{DLMH13}.
Let the coupling is switched on at $\tau^{}_I=0$ (when $t=0$). Then $\eta_- = \tau^{}_-$.
Since $\tau^{}_I$ does not go to $-\infty$, we are not really at late times here, and in the results of this section we have actually 
included the a-parts of the correlators \cite{LH06} in addition to the v-parts discussed in Section \ref{2PCorrLT}, though the contributions by the a-parts are small in the figures we are going to present. Suppose the observer is located at $x^\mu = (t_0+r, r{\bf \hat{r}})$. 
For $r\gg 2\pi\omega_0^{-1}$, one can compute $t_- = z^0 (\tau^{}_-(x))$ of the retarded time $\tau^{}_-(x)$ by solving
\begin{equation}
  t_0 - t_- \approx {\cos\theta\over \omega_0}\sin^{-1}\left( {2 a_0 \cos\omega_0 t_-\over\sqrt{1+4 a_0^2}}\right),
\end{equation}
from $\sigma(x, z(t_-(x))) 
=0$. Then ${\tau^{}_-} =F(\omega_0 t_-, -4a_0^2)/\omega_0$ and 
in the radiation zone,
\begin{eqnarray}
	{\cal R} &=& \left|v_\mu(\tau^{}_-)(x^\mu-z ^\mu(\tau^{}_-)\right|_{r\to \infty} \approx 
	    r \left[v^0 (\tau^{}_-) - v^3 (\tau^{}_-)\cos\theta \right],\\
  \partial_t \tau^{}_- 
	    &\approx&  -\partial_r \tau^{}_-\approx {r\over {\cal R}}, \label{TtCT} \\
  {\partial_t {\cal R}\over {\cal R}}  &\approx& -{\partial_r {\cal R}\over {\cal R}} \approx 
	  {a^0 (\tau^{}_-)-a^3 (\tau^{}_-)\cos\theta\over \left[ v^0 (\tau^{}_-)-v^3 (\tau^{}_-)\cos\theta\right]^2}, \label{Rt1R}
\end{eqnarray}
where $v^\mu (\tau)\equiv \dot{z}^\mu (\tau)=(\sqrt{1+4a_0^2 \sin^2\omega_0 t}, 0,0,2a_0\sin\omega_0 t)$ and $a^\mu (\tau)\equiv 
\ddot{z}^\mu (\tau)$ are the four-velocity and four-acceleration of the detector moving along the worldline $z^\mu $, respectively,
while $F(\phi, m)$ is the elliptic integral of the first kind \cite{DLMH13}.
(\ref{TtCT}) can be quickly derived by partially differentiate the equation $\sigma(x, z(\tau^{}_-(x)))=0$.
Let the period of the oscillatory motion be $t_p = 2\pi/\omega_0$ in the coordinate time and $\tau_p = \omega_0^{-1}F(2\pi, -4a_0^2)$ in 
the proper time of the detector. We define the directional proper acceleration $\alpha(t_-) =  |a(t_-)|\,{\rm sign}\,a^3(t_-) = 2a_0 
\omega_0 \cos \omega_0 t_-$ where $|a|=\sqrt{a_\mu a^\mu }$ is the proper acceleration \cite{DLMH13}. 
Then one has $a^\mu =(\alpha v^3 , 0, 0, \alpha v^0 )$ for $z^\mu $ in (\ref{CTtraj}), so that 
$|\partial_\mu {\cal R}/{\cal R}|:|\partial_\mu \tau^{}_-| \approx |\alpha| = |a|$, $\mu=0, 3$ around $\theta=0$ and $\pi$
in the radiation zone. 
In the cases with large $a_0 \omega_0$, $|\partial_\mu {\cal R}/{\cal R}|$ can dominate over 
$|\partial_\mu \tau^{}_-|$ at most of the observing angles. 
Also the reference worldline to control the singularities for an observer at $x$ would be, for ($t_-$ mod $t_p) = t_p/4$ or $3t_p/4$ 
(when $\alpha(t_-)=0$), 
$\tilde{z}^\mu_x(\tau) = z^\mu(t_-(x)) + v^\mu(t_-(x))(\tau - \tau^{}_-(x))$, otherwise 
\begin{equation}
  \tilde{z}^\mu_x(\tau) =
	\left( \frac{\sinh \left[\alpha(t_-) (\tau - \tau^{}_- +\bar{\tau})\right] }{\alpha(t_-)}+ {\cal O}^0, 0, 0, 
	\frac{\cosh \left[\alpha(t_-) (\tau-\tau^{}_-+\bar{\tau})\right] }{\alpha(t_-)}+ {\cal O}^3\right)
\end{equation}
where ${\cal O}^0 \equiv t_- - \alpha^{-1}(t_-) v^3(t_-)$, ${\cal O}^3 \equiv z^3(t_-)- \alpha^{-1}(t_-)v^0(t_-)$, and
$\bar{\tau}\equiv\alpha^{-1}(t_-) \sinh^{-1} v^3(t_-)$. The advanced time for this reference worldline or its image reads \cite{Lin03, LH06}
\begin{equation}
  \tau^{}_+(x) =\frac{1}{\alpha(t_-)}\log \left| \frac{r(1+\cos\theta)+t_0 -{\cal O}^0-{\cal O}^3}
	  {r(1-\cos\theta)+t_0 -{\cal O}^0+{\cal O}^3} \right| - \tau^{}_-.
\end{equation}

We show our numerical results for the renormalized differential radiated power (\ref{dPdWIIren}) emitted by a UD detector moving along 
(\ref{CTtraj}) in Figures \ref{CTradt}, \ref{CTradtheta} and \ref{CTvarw}.

\subsection{Negative radiated power}
\label{SecNegaE}

\begin{figure} 
\includegraphics[width=4.8cm]{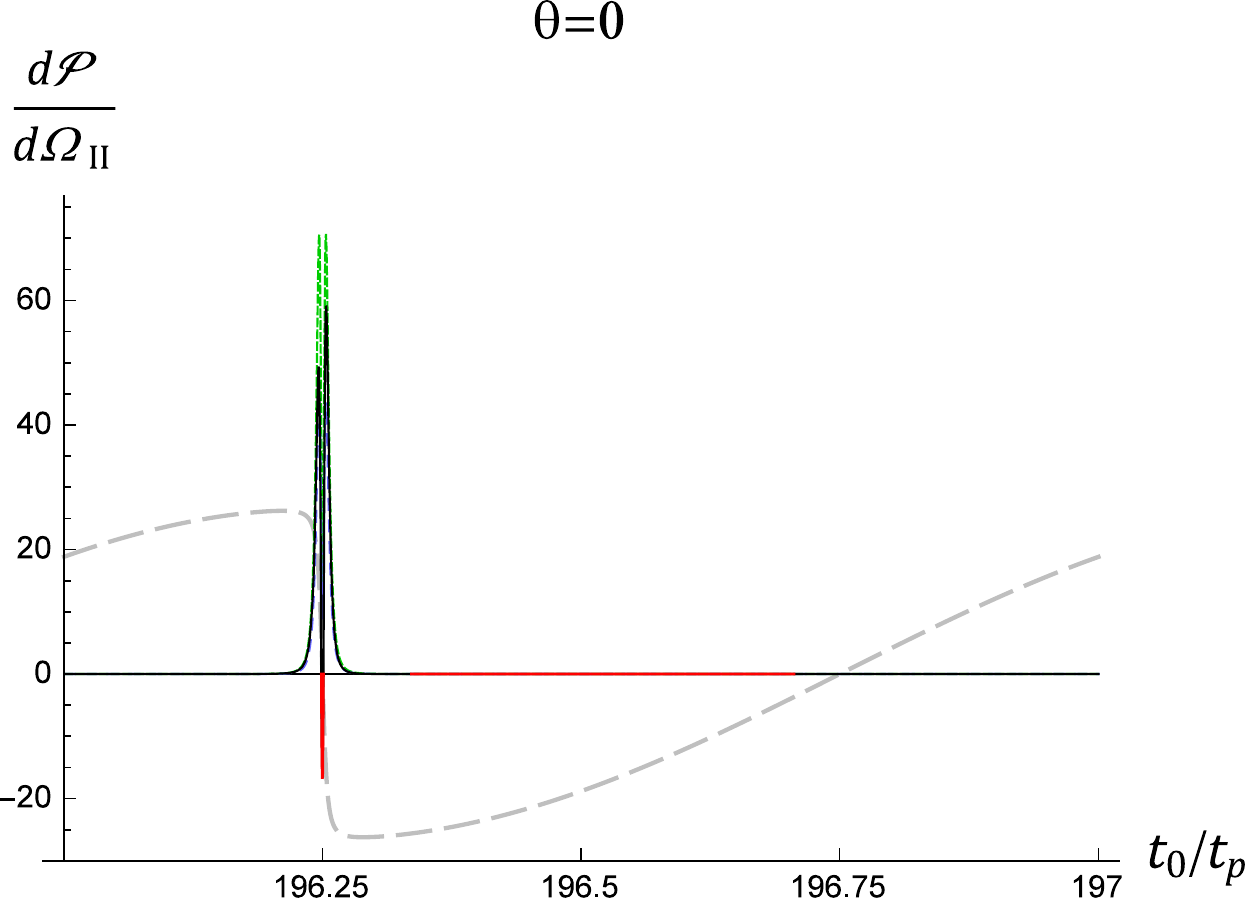}
\includegraphics[width=4.8cm]{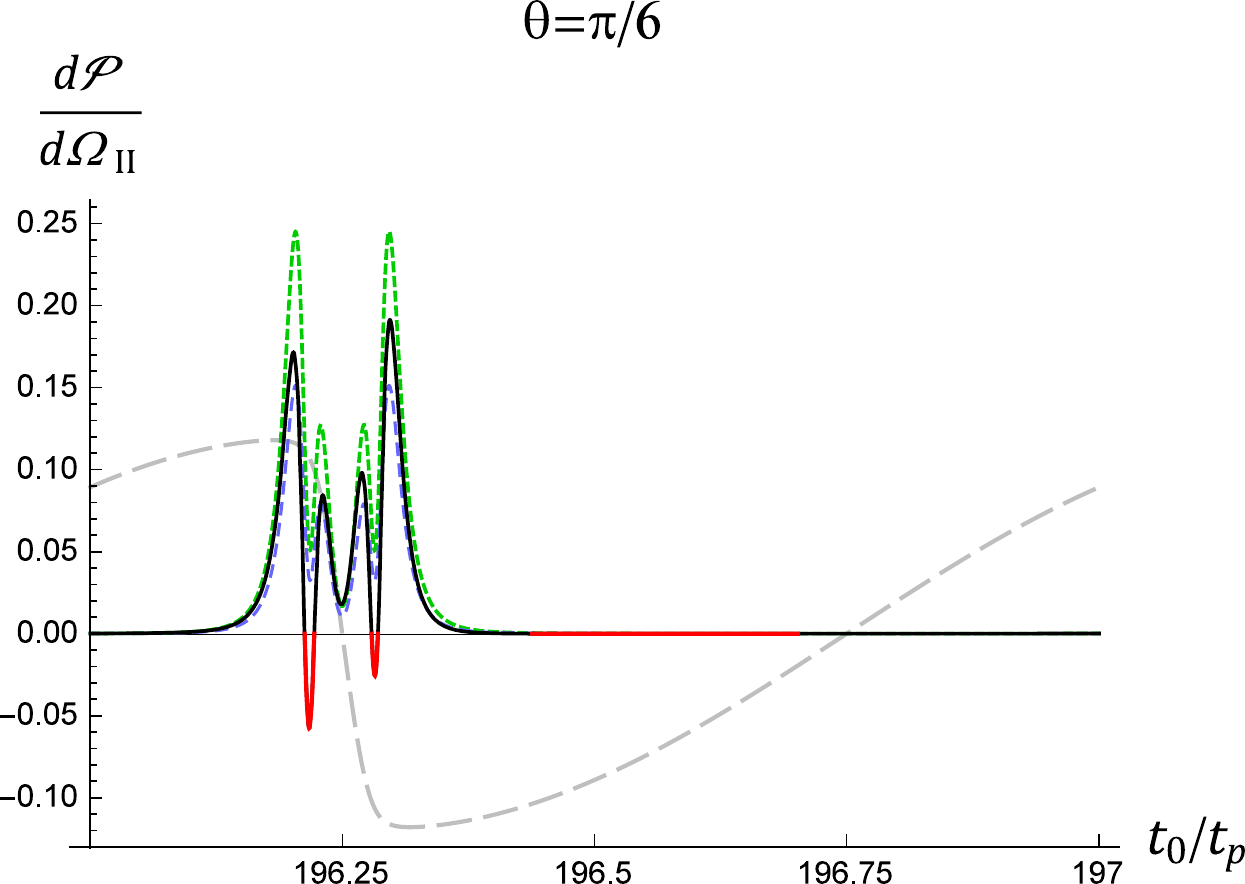}
\includegraphics[width=4.8cm]{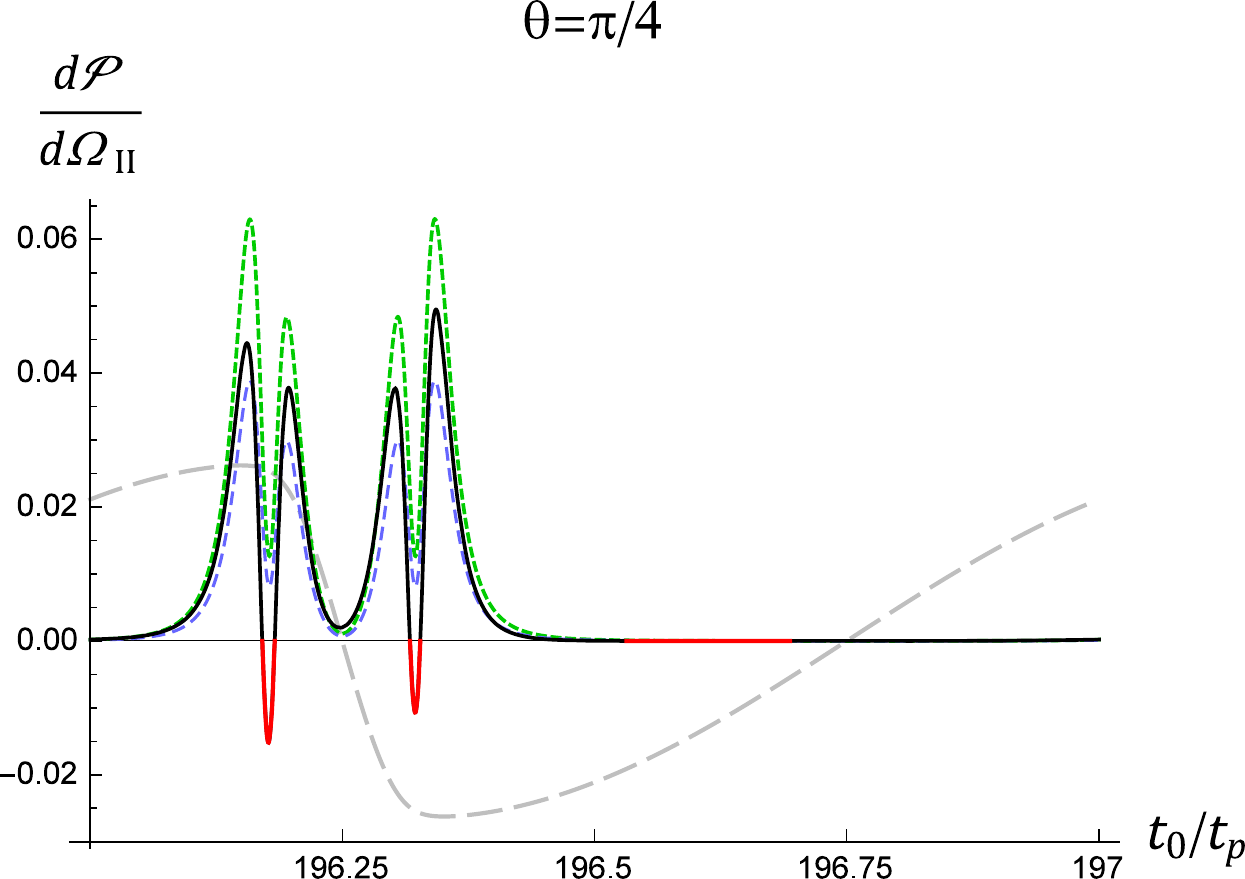}\\
\includegraphics[width=4.8cm]{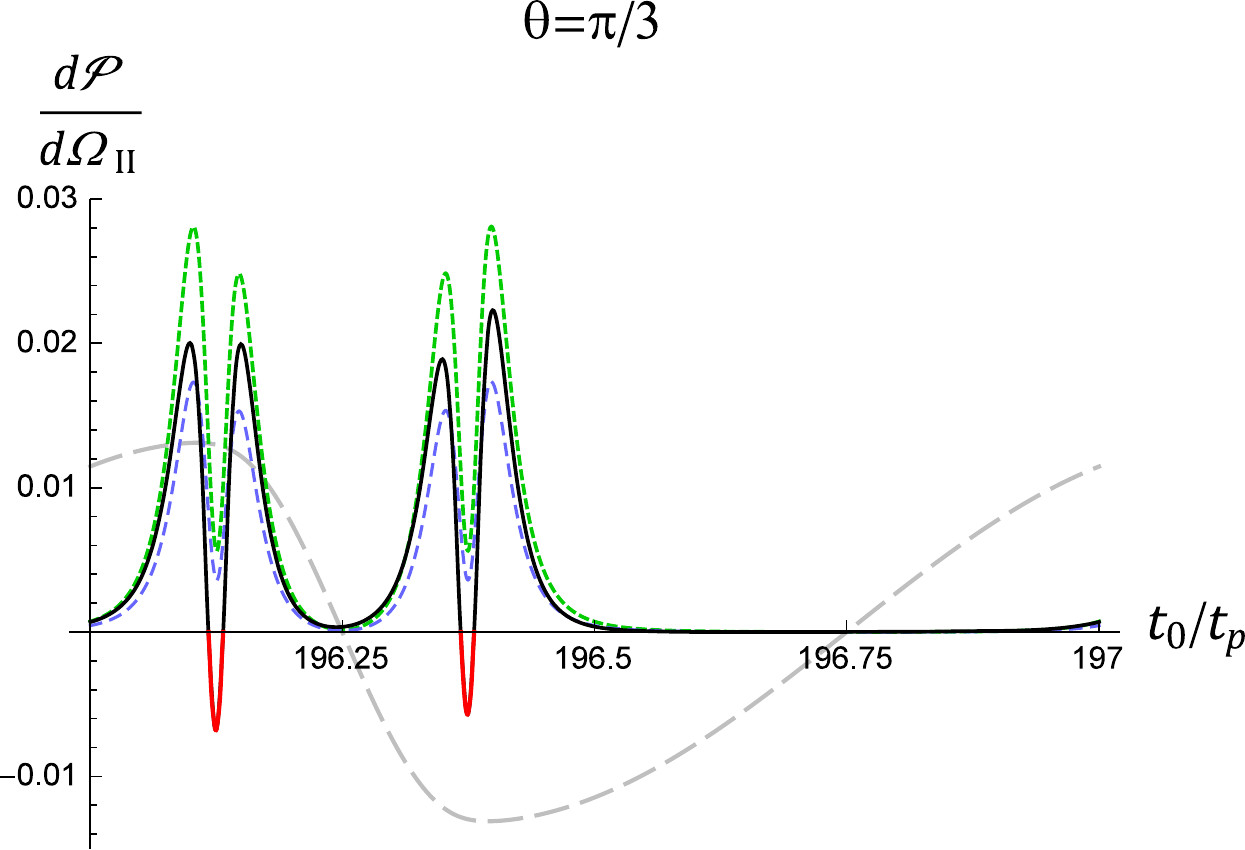}
\includegraphics[width=4.8cm]{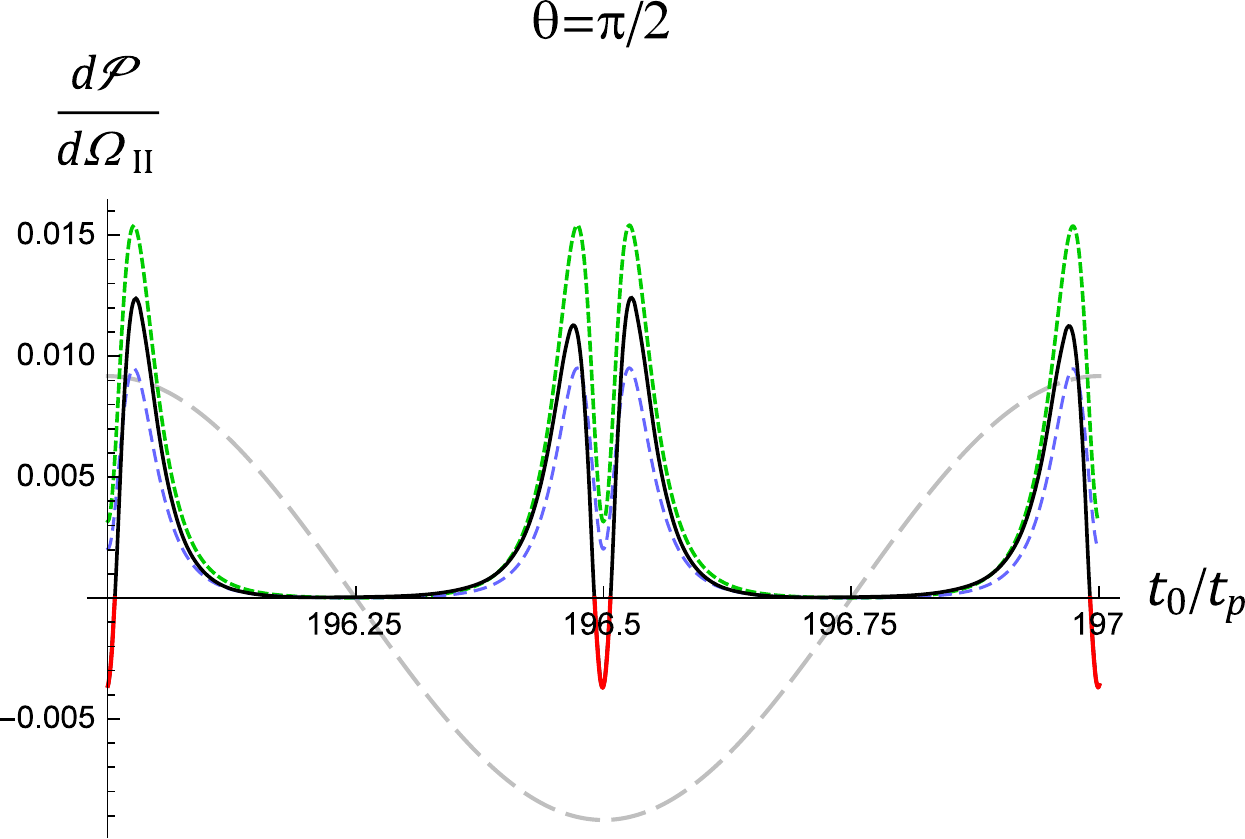}
\includegraphics[width=4.8cm]{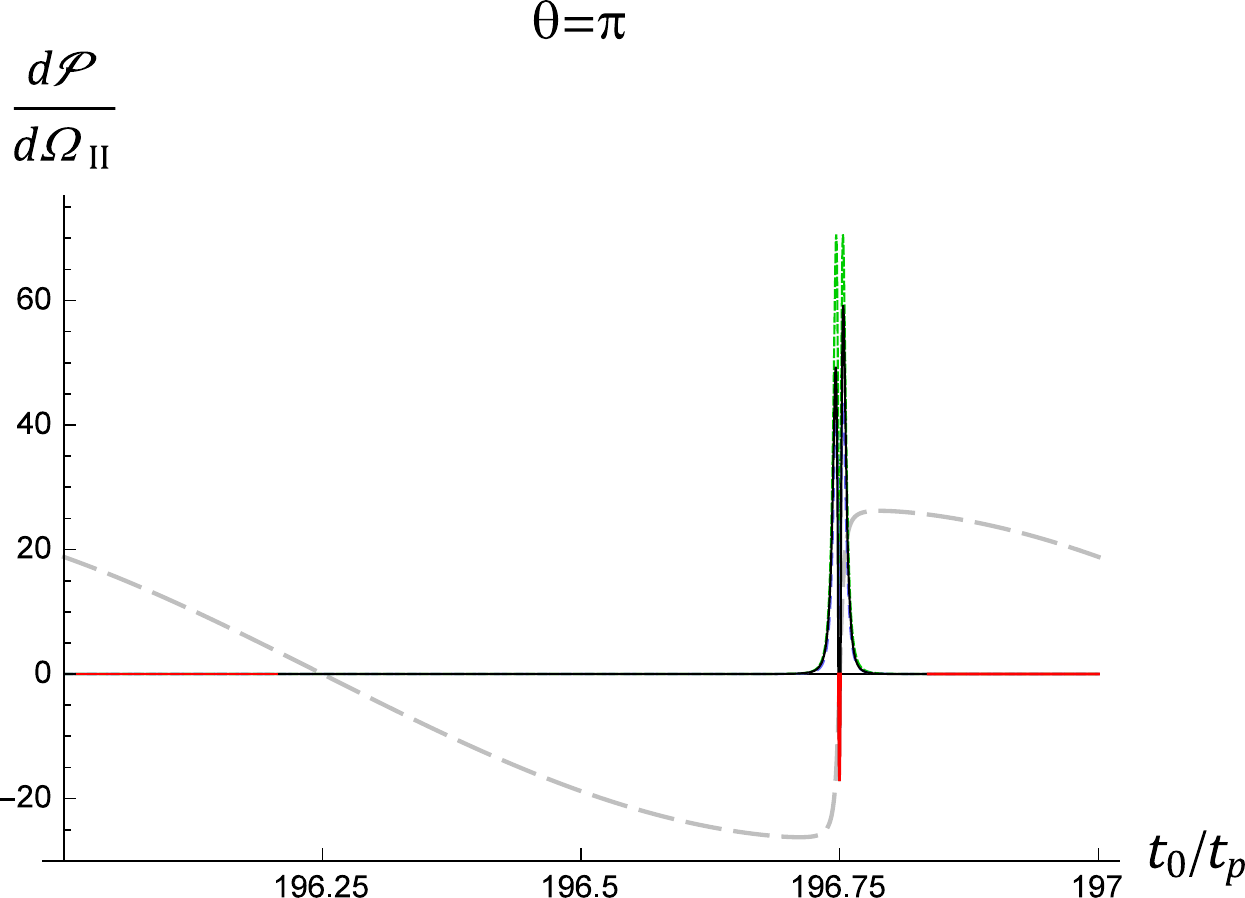} 
\caption{The differential radiated power per solid angle $d{\cal P}/d\Omega_{\rm II}$ against $t_0/t_p$ emitted by a detector in the 
worldline (\ref{CTtraj}) and observed at angle $\theta$ and time $t=t_0+r$ in the radiation zone $r\gg 2\pi/\omega_0$.
Here $\gamma=0.01$, $\Omega=2.3$, $\Lambda_1=\Lambda_0=20$, $r=10^{40}$, $\omega_0=3.277$, $a_0=2$, and so the averaged proper acceleration 
$\bar{a}=10$, the period of each cycle $t_p\equiv 2\pi/\omega_0=1.917$ in the rest frame, and $\tau_p = 0.838$ in the comoving frame. 
The solid and green dashed curves represent the differential radiated power with and without the interference terms, respectively, while the 
blue dashed curves represent the naive differential radiated power $d{\cal P}^{[11]}/d\Omega_{\rm II}$ without the Unruh effect, namely, the 
two-point correlators for the accelerated detector have been replaced by the ones for an inertial detector. The black and red sections of 
the solid curve represent the positive and negative radiated power. The gray dashed curves represent the scaled directional acceleration of the detector $\alpha(t_-(x))$ at the moment it emitted the observed radiation.}
\label{CTradt}
\end{figure}

\begin{figure} 
\includegraphics[width=4.8cm]{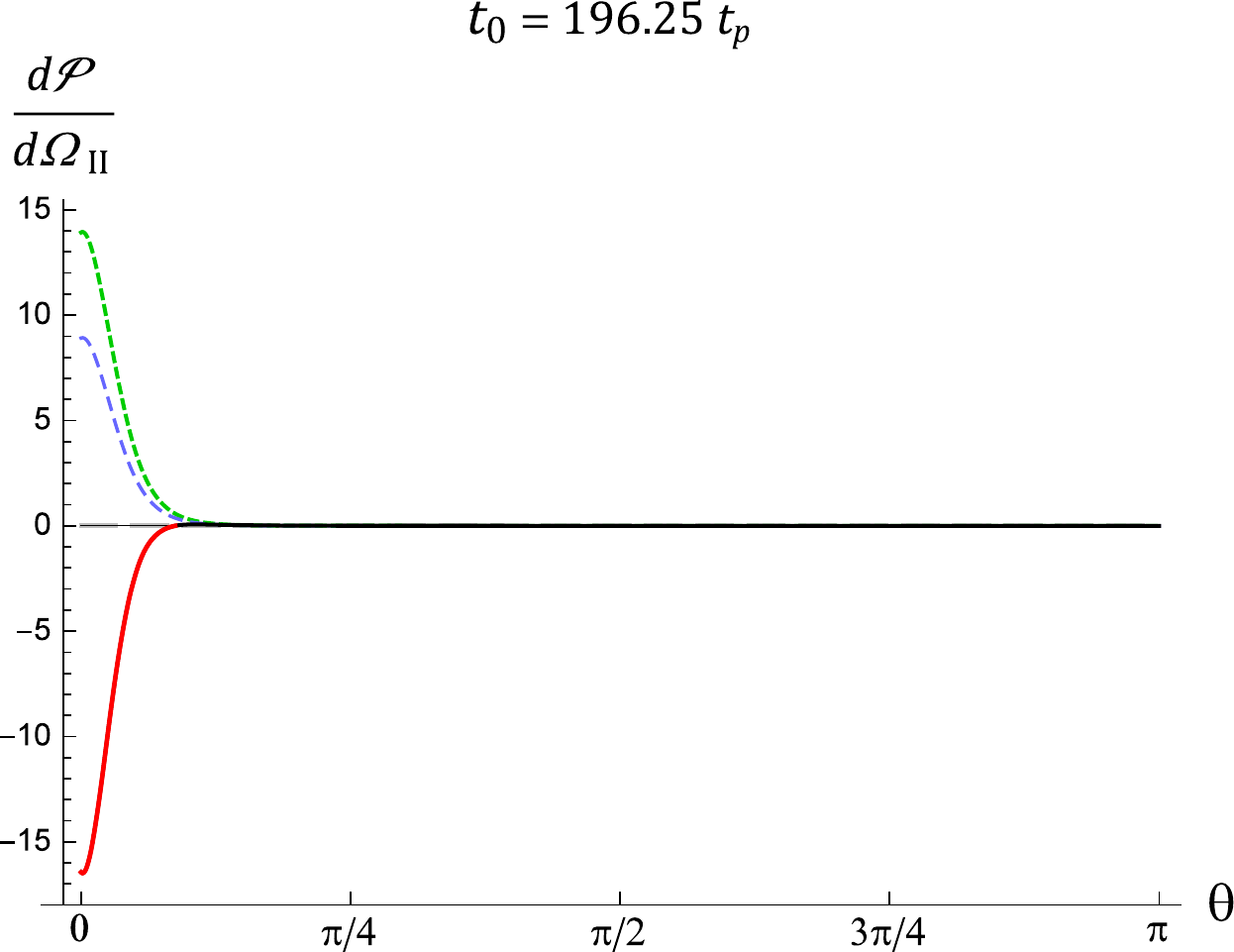}
\includegraphics[width=4.8cm]{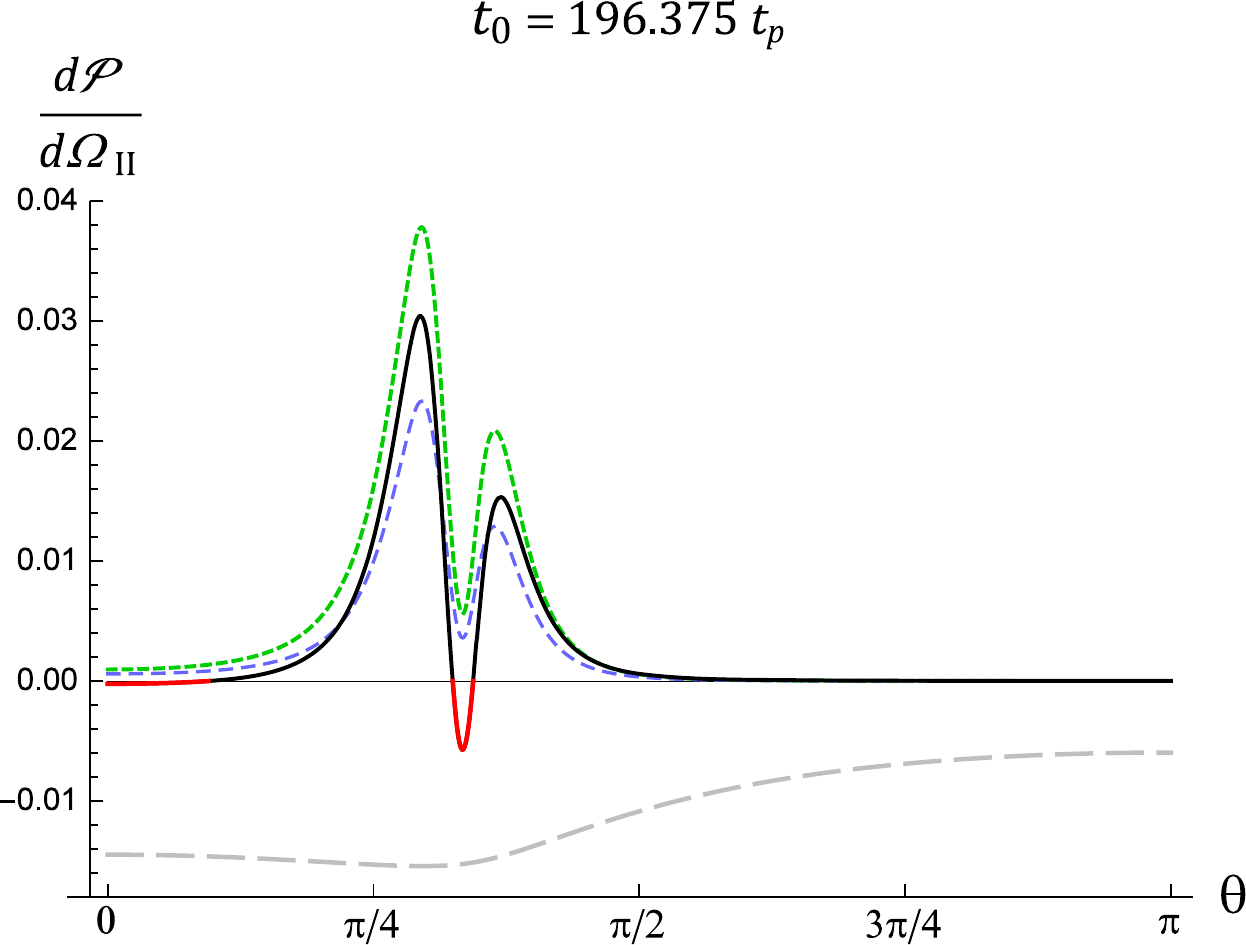}
\includegraphics[width=4.8cm]{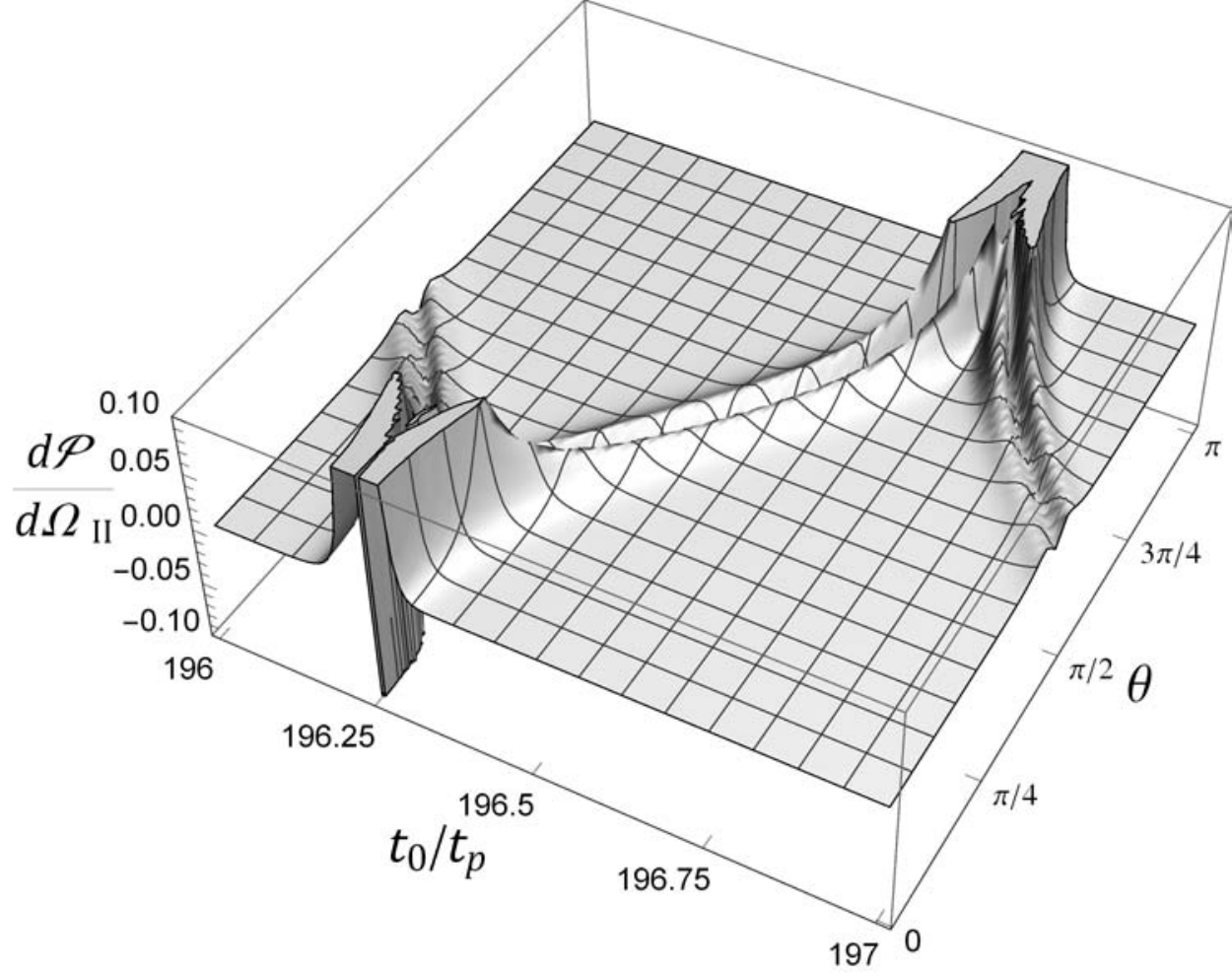}\\
\includegraphics[width=4.8cm]{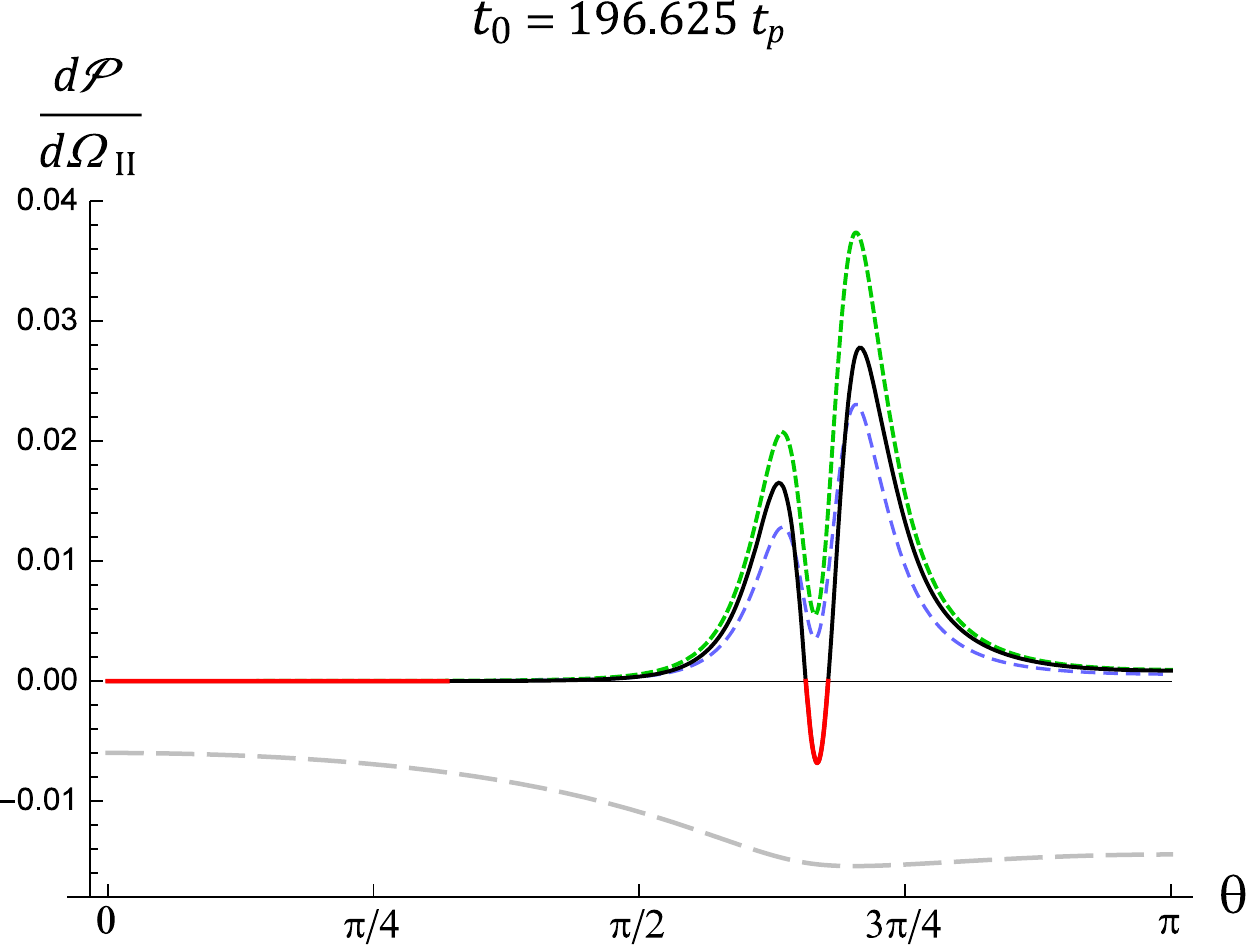}
\includegraphics[width=4.8cm]{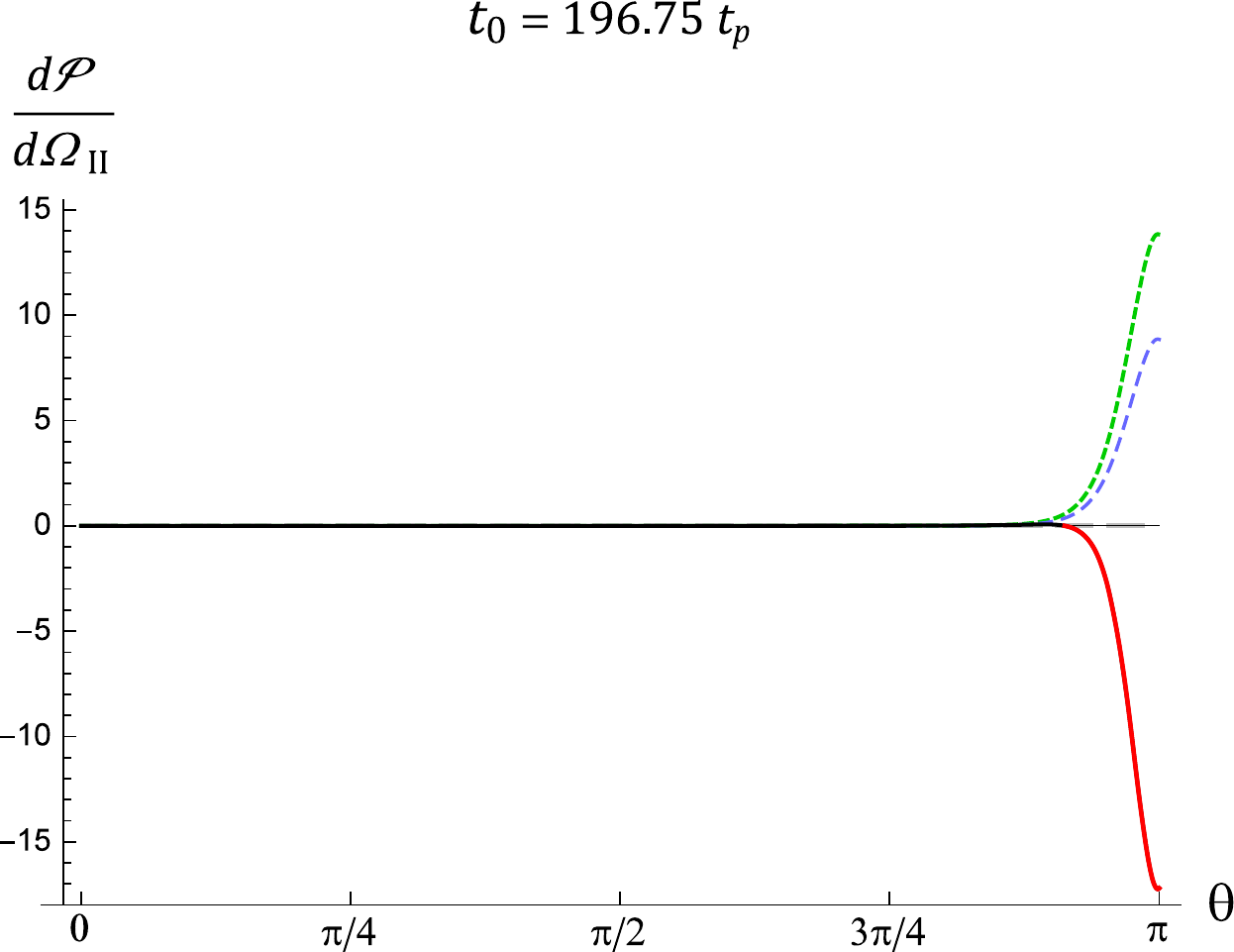}
\includegraphics[width=4.8cm]{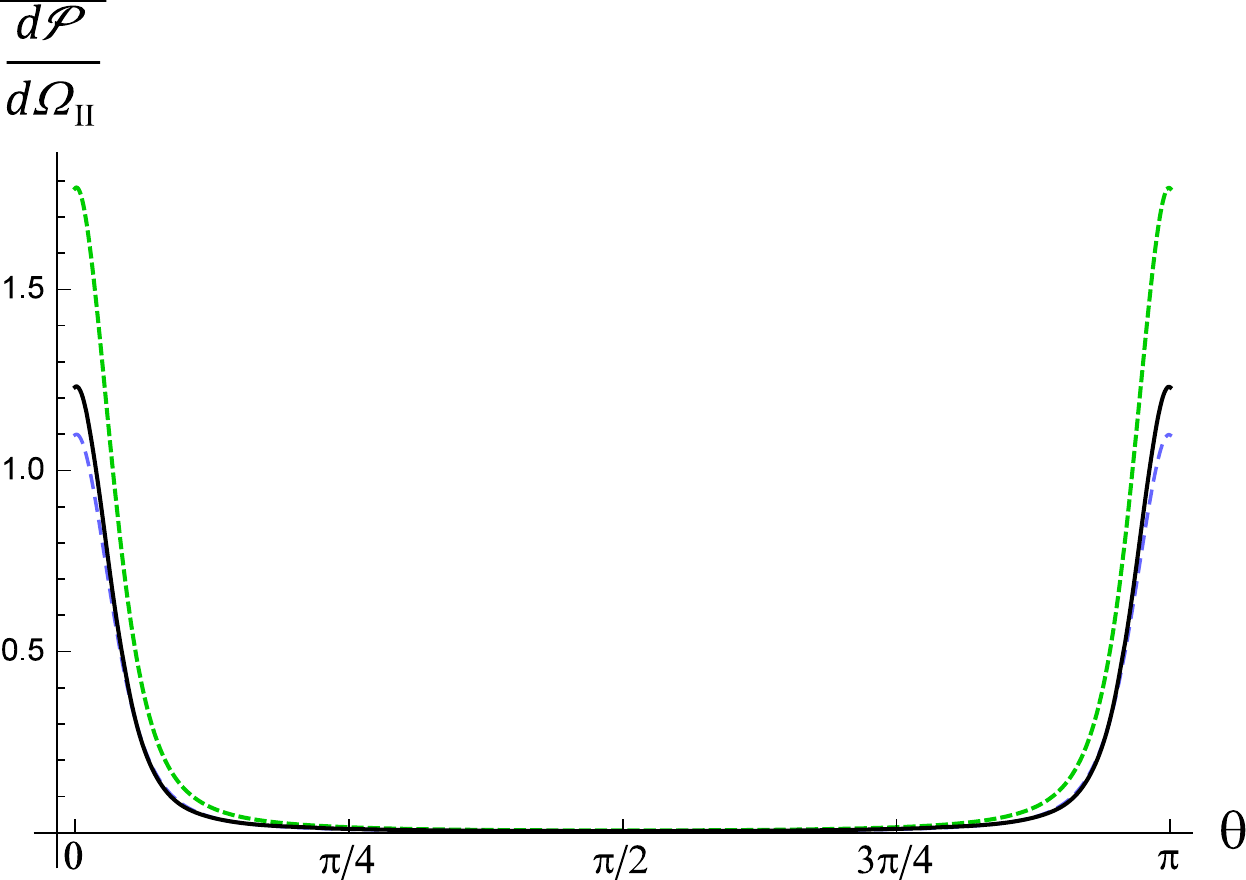}
\caption{The differential radiated power $d{\cal P}/d\Omega_{\rm II}$ against $\theta$ at fixed $t_0$ for the same detector in Figure 
\ref{CTradt}. The bottom-right plot is the differential radiated power averaged over a cycle of oscillatory motion. The radiated energy is 
positive for all $\theta$ while concentrated around $\theta=0$ and $\pi$, namely, the directions of the linear oscillatory motion.}
\label{CTradtheta}
\end{figure}

In Figure \ref{CTradt}, we demonstrate the time evolution of the full differential radiated power in one cycle at large times at fixed angles.
One can see that the time evolution of the full differential radiated power at each fixed angle with the interference terms included has a 
double-peak structure. Each pulse consists of two main peaks of positive flux, and valleys of negative flux in-between.

In the moving-mirror model in (1+1)D, Fulling and Davies found that a negative radiated power will be observed when the acceleration of the mirror is time varying while the mirror is moving towards the observer \cite{FD76}. 
Our results at $\theta=0$ and $\pi$ (on the plane of the oscillatory motion of the detector) are consistent with this observation.
However, when observed far off the oscillation plane, the negative radiated power in our examples does not occur around the moment when 
the observed proper acceleration has the most significant change. This is clear in the plots with $\theta\not
=0$ or $\pi$ in Figure \ref{CTradt}, where the double-peak pulse occurs around the maximum of the observed proper acceleration 
$|\alpha(t_-(x))|$, while the value of $|\partial_t \alpha|$ can be small in the period of negative flux between the peaks. 
Actually, the double-peak structure in Figure \ref{CTradt} has been obvious in the naive term $d{\cal P}^{[11]}/
d\Omega_{\rm II}$ (green dashed curve). We find this behavior in our example is dominated by the ${\cal R}_{,t}{\cal R}_{,r}/{\cal R}^4$ 
factor in (\ref{DDG11}) (here $a_0\omega_0=6.554$ in $\alpha$, see the statement below (\ref{Rt1R}).) When the scaled retarded distance 
${\cal R}/r$ for the observer at some fixed angle become very small, the observed radiated power will be amplified and a pulse emerges.
However, around the moment that the retarded distance reaches the minimum, one has ${\cal R}_{,t}=0$ (i.e. $a^0 (\tau^{}_-(x)) = 
a^3 (\tau^{}_-(x))\cos\theta$ by (\ref{Rt1R})) and so a valley between two positive main peaks is formed in a pulse. 
On the other hand, the negative correction from the interference terms become the most negative at some moment a little bit ahead of the 
valley of the naive term, so the total differential radiated power around the valley become negative (also see the inset of Figure \ref{HHG2} (left)).

At each fixed time, there will always exist negative radiated power around some observing angle, as shown in Figure \ref{CTradtheta}. 
Nevertheless, the averaged radiated power over a cycle of oscillation at each fixed angle 
must be positive (Figure \ref{CTradtheta} (lower-right)), as guaranteed by the quantum inequalities~\cite{Fo91}.

Looking more closely (also see Figure \ref{CTvarw}) one can see that, when the averaged proper acceleration \cite{DLMH13}
\begin{equation} 
   \bar{a}\equiv \frac{\int_0^{\tau_p} |a(\tau+\tilde{\tau})| d\tilde{\tau}}{\int_0^{\tau_p} d\tilde{\tau}} = 
	 \frac{\omega_0 \sinh^{-1}2a_0}{F(\pi/2, -4a_0^2)} 
\end{equation} 
is sufficiently small ($\bar{a}=10$ in Figure \ref{CTradt}), and the observing angle is in the vicinity of $\theta=0$ or $\pi$, there may be 
a longer period of negative radiated power between the second positive peak of one pulse and the first positive peak of the successive pulse, 
while the value of the negative radiated power is very close to zero.
In some cases 
the period of this negative radiated power can be longer than a half of the period of a cycle (e.g. Figure \ref{HHG2}).

In our examples the period of the oscillatory motion $t_p=2\pi/\omega_0$ are small compared to the time scale of the relaxation of the 
detector $1/\gamma$. Under this condition, if we increase the coupling strength $\gamma$ with other parameters fixed, the time evolution 
of the angular distribution of the differential radiated power will be similar to Figures \ref{CTradt} and \ref{CTradtheta}, except that the 
peaks will be roughly amplified as $\gamma^1$.
The ratio of the maximum amplitudes of the negative radiated flux to the positive one does not change significantly as we increase 
$a_0$ or $1/\omega_0$ but keep $\bar{a}$ fixed. 

The negative radiated power in the above result does not imply absorption, or radiation in the opposite direction. 
It can excite an UD detector at a rate lower than the case in zero energy density \cite{Gr88}, and produce 
no decrease of entropy~\cite{DOS82}. Our result reveals another resemblance between the detector theory and the 
moving-mirror models in quantum field theory in curved spacetime. 
Actually, some Unruh-DeWitt detector theories in (1+1)D have been used to describe mirrors in a more realistic way than those 
simply introducing boundary conditions for the fields at the mirror's position \cite{GBH13, WU14, WU15, SLH15}. 

\subsection{Evidence of Unruh effect in radiation}

\begin{figure} 
\includegraphics[width=4.8cm]{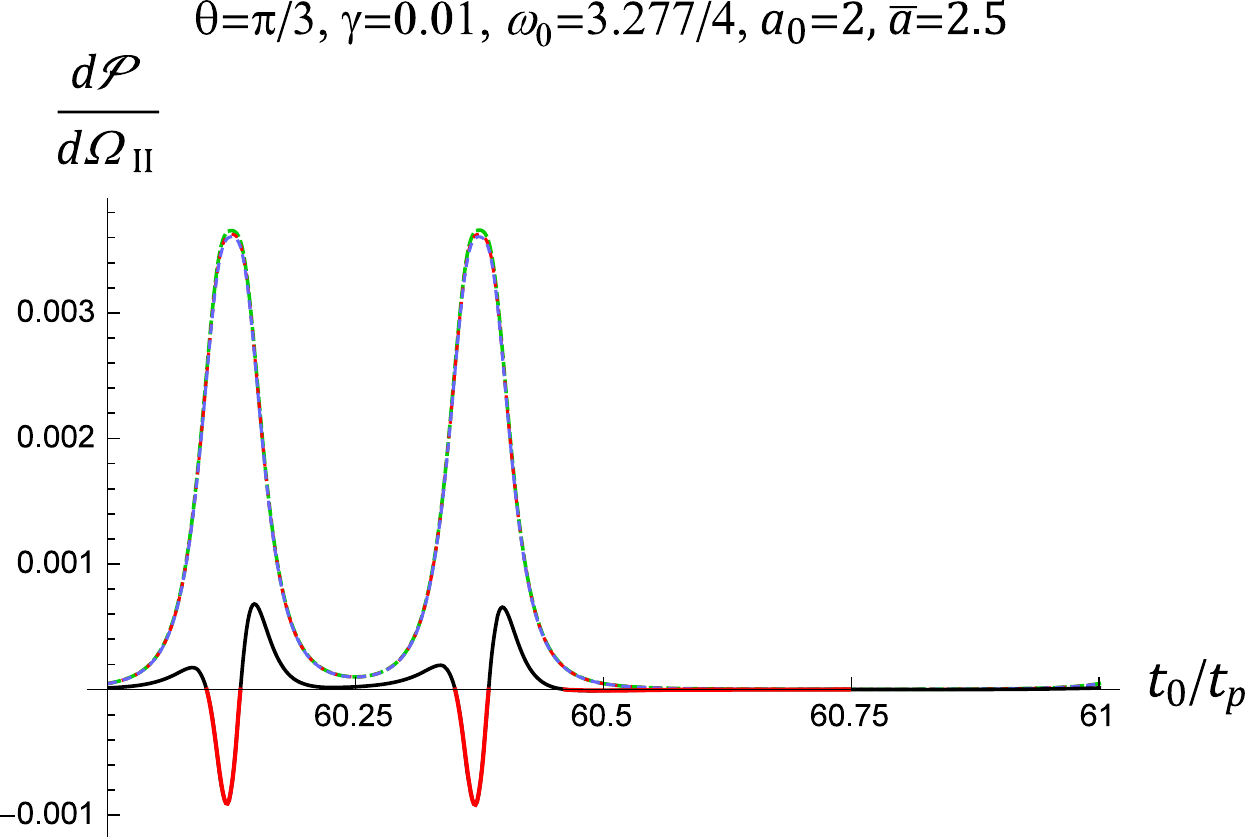}
\includegraphics[width=4.8cm]{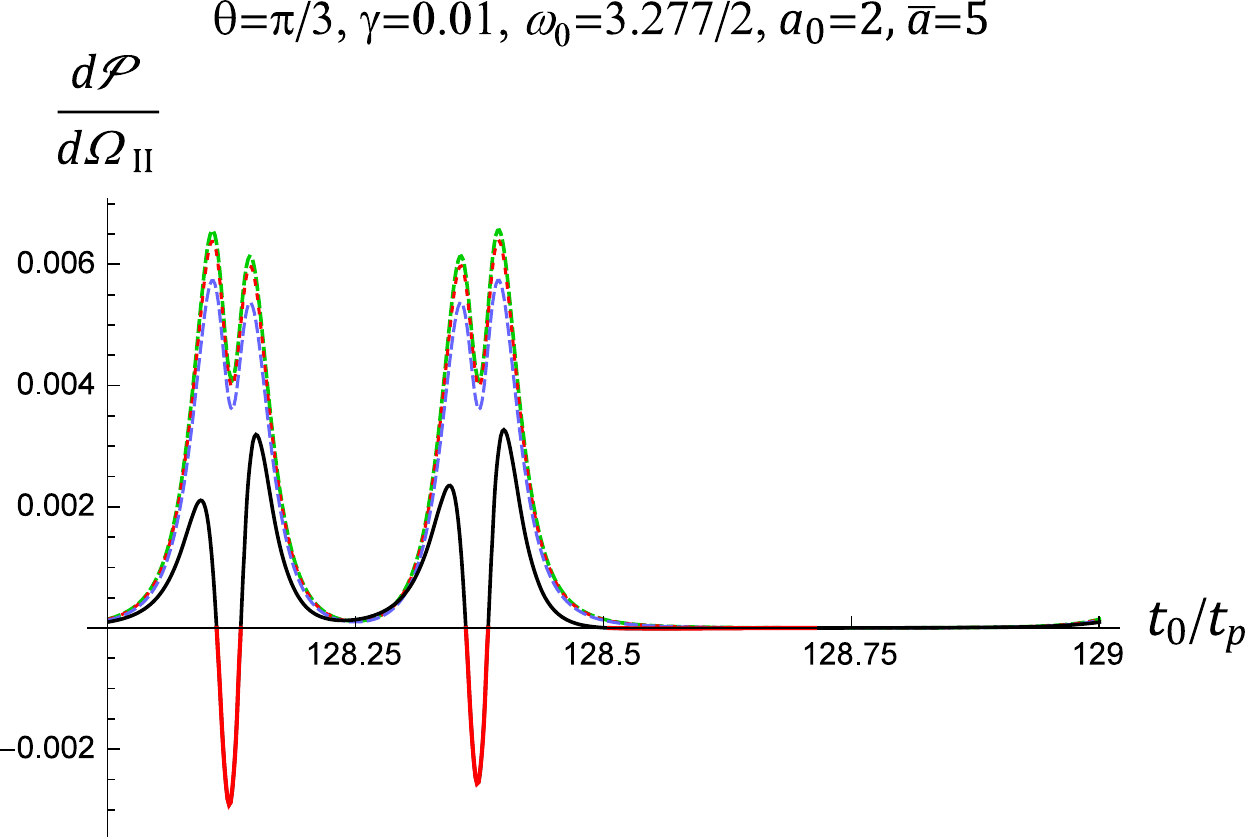}
\includegraphics[width=4.8cm]{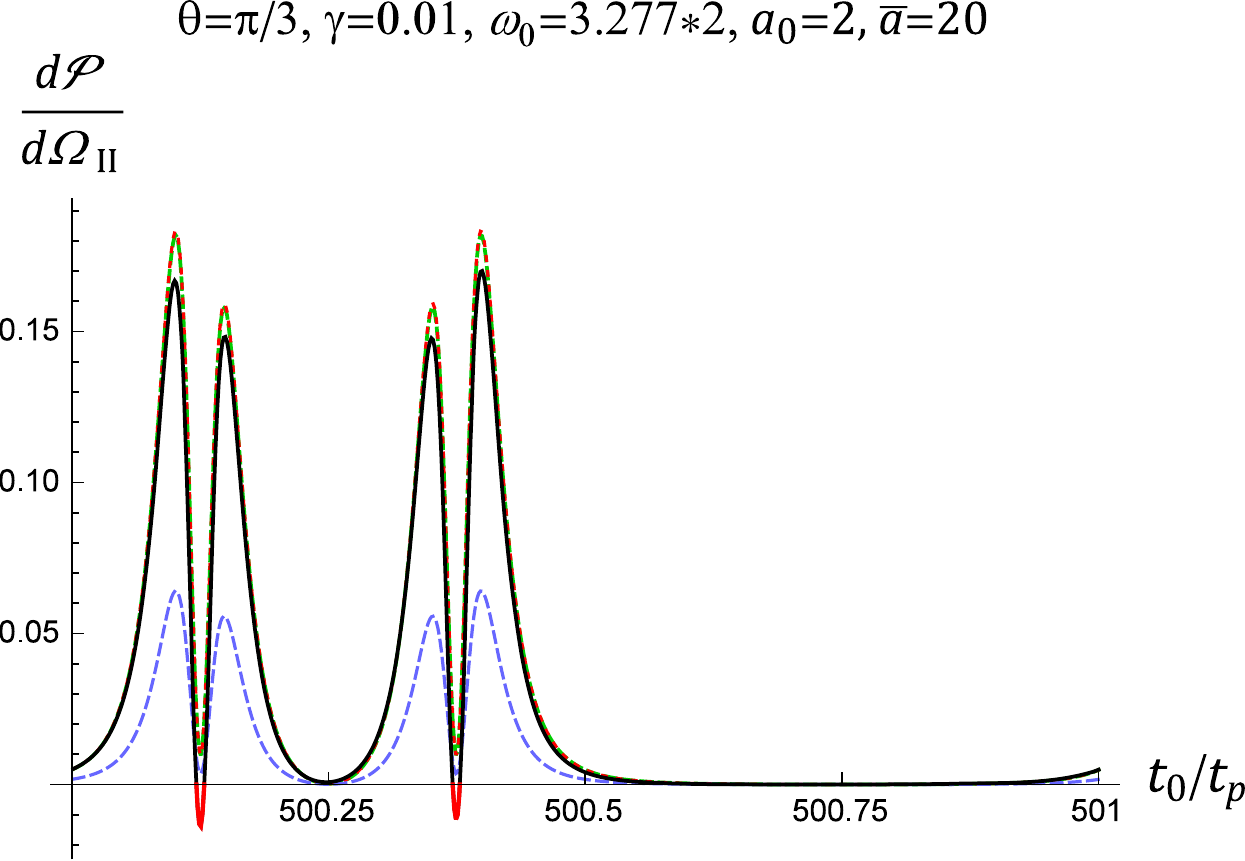}  
\caption{From left to right are the time evolutions of the differential radiated power at $\theta=\pi/3$ with the averaged accelerations from 
low to high values ($\bar{a}=2.5$, $5$, and $20$, respectively; The result with $\bar{a}=10$ has been shown in the lower-left plot,
Figure \ref{CTradt}). We compare the full result $d{\cal P}/d\Omega_{\rm II}$ (solid) with the (modified) naive terms 
$d{\cal P}^{[11]}/d\Omega_{\rm II}$:
the green dashed, red dotted, and blue dashed curves are contributed by the correlators of the detector in oscillatory motion,
in uniform acceleration at the proper acceleration $\bar{a}$, and at rest, respectively.
One can see that $d{\cal P}/d\Omega_{\rm II}$ gets closer to the naive result $d{\cal P}^{[11]}/d\Omega_{\rm II}$ at the effective or 
Unruh temperature as the averaged acceleration $\bar{a}$ increases,
while the red ``tail" of the small negative radiated power after the second pulse of the full result gradually disappears.
Also the green dashed curve is much closer to the red dotted curve than to the solid curve, indicating that the deviation of the detector's 
effective temperature from the Unruh temperature is not important for the radiated power, compared with the correction from the 
interference terms.}
\label{CTvarw}
\end{figure}

In Ref. \cite{DLMH13} we observed that, at a lower (higher) value of the averaged proper acceleration $\bar{a}$, the effective temperature 
of an UD detector in oscillatory motion tends to be higher (lower) than the naive Unruh temperature $\bar{a}/(2\pi)$ experienced by a 
uniformly accelerated detector at the proper acceleration $\bar{a}$. 
Indeed, the effective temperature in the example shown in Figures \ref{CTradt} and \ref{CTradtheta} is about $T_{\rm eff}\approx 1.6754$ to 
$1.6767$, which is higher than the naive Unruh temperature $\bar{a}/(2\pi) \approx 1.5916$  ($\bar{a}=10$),
while in Figure \ref{CTvarw} (right), the effective temperature is about $3.156$, which is lower than $\bar{a}/(2\pi) \approx 3.183$
($\bar{a}=20$). Anyway, in Figure \ref{CTvarw} one can see that the deviation of the effective temperature from the 
naive Unruh temperature $\bar{a}/(2\pi)$ due to non-uniform acceleration is negligible in the radiated power, compared with the 
correction from the interference terms.

A detector at rest still has non-zero correlators $\langle \hat{Q}^2\rangle$ and $\langle \hat{P}^2\rangle$ contributed by 
vacuum fluctuations at zero temperature, such that the naive Unruh radiation $d{\cal P}^{[11]}/d\Omega_{\rm II}$ from (\ref{DDG11}) 
is positive even at zero averaged acceleration. Since we expect that the radiation by the detector should cease as its averaged proper 
acceleration $\bar{a}\to 0$, the negative interference terms should be able to cancel the naive differential radiated power in this case.
Indeed, we find the radiated power tends to be suppressed larger by the interference terms when $\omega_0$ or $\bar{a}$ gets smaller 
(Figure \ref{CTvarw} (left)). Here, for a fixed $a_0$, a smaller $\omega_0$ on the one hand gives a smaller averaged proper acceleration 
$\bar{a}$, on the other hand it implies a longer period of oscillatory motion, so that the detector has more time to approach to the 
equilibrium conditions studied in Ref.~\cite{LH06}. Both suppress the radiated power.

In contrast, as $\omega_0$ or $\bar{a}$ increases, the importance of the interference terms decreases, and the full result of the 
differential radiated power get much closer to the naive result $d{\cal P}^{[11]}/d\Omega_{\rm II}$ at the effective or Unruh temperature 
than the naive result at zero temperature (Figure \ref{CTvarw} (middle) to (right)).
This suggests that the Unruh-like effect experienced by a UD detector could be observed in the Unruh radiation in highly non-equilibrium 
conditions, with a very short period of oscillatory motion and a very high averaged proper acceleration.

\section{Correlations in radiation}
\label{SecCorr}

To obtain the late-time two-point correlators more efficiently, we consider the on-resonance case below.

\subsection{High harmonic generation}
\label{cHHG}

\begin{figure}
\includegraphics[width=7.2cm]{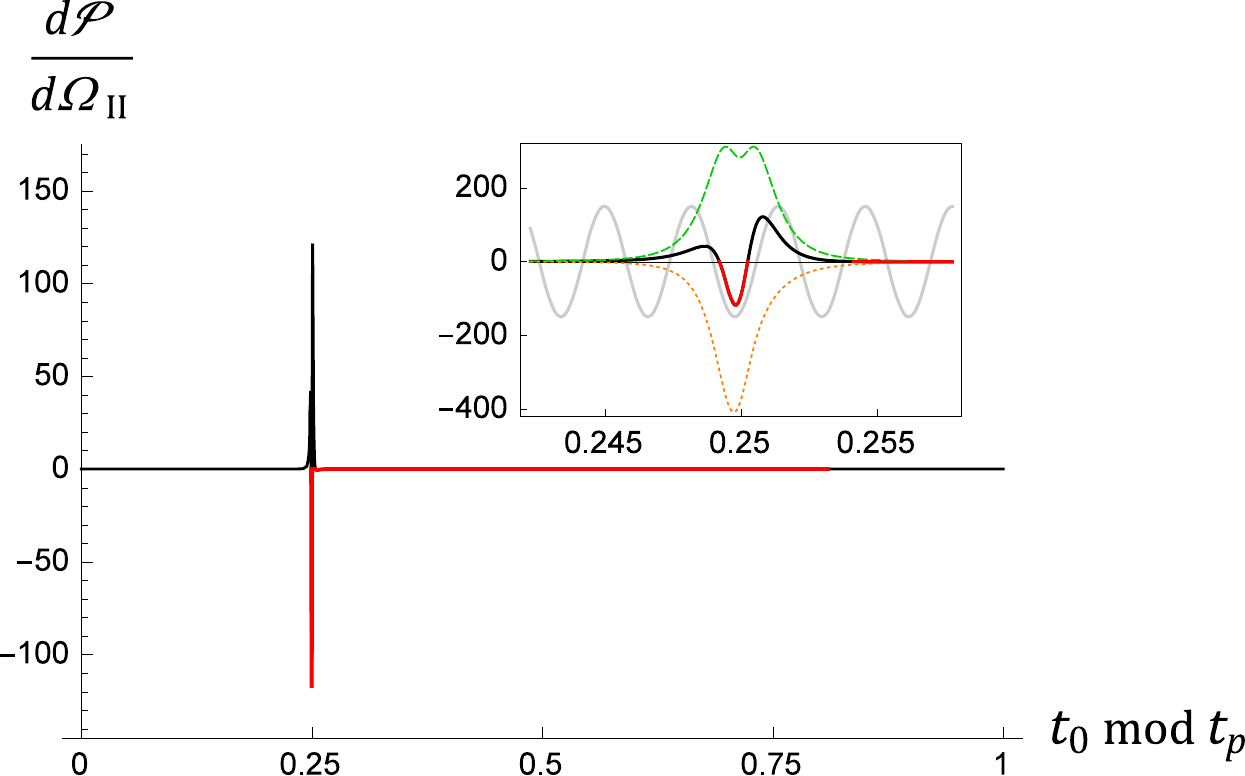} 
\includegraphics[width=7.2cm]{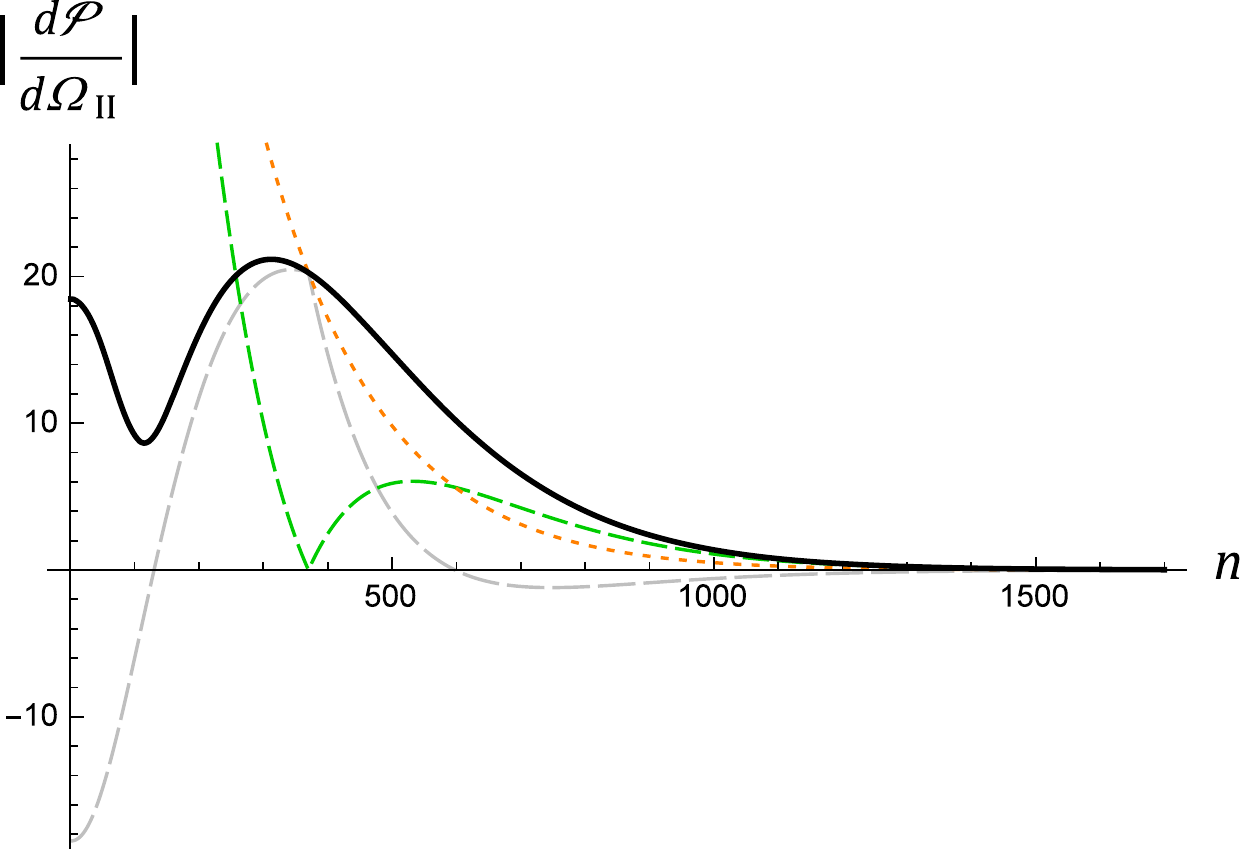} 
\caption{(Left) Time evolution of the late-time differential radiated power $d{\cal P}/d\Omega_{\rm II}$ observed at $\theta=0$ (black/red).
Here $\gamma=0.01$, $\Omega=4.3$, $\Lambda_1=\Lambda_0=20$, $a_0=4$, $\bar{a}\approx 7.6$, 
and the period of each cycle $\tau_p =2\pi/\Omega$ (on-resonance case) in the comoving frame. The black and red parts of the solid curve 
represent the differential radiated power with positive and negative values, respectively. In the inset, the green-dashed and orange-dotted curves 
represent the naive term $d{\cal P}^{[11]}/d\Omega_{\rm II}$ and the interference terms $d({\cal P}^{[01]}+{\cal P}^{[10]})/
d\Omega_{\rm II}$, respectively.
(Right) $|d{\cal P}/d\Omega_{\rm II}|$ in the frequency domain for the harmonics with $\omega=n\omega_0$, $n=1,2,3,\cdots$, after a
Fourier transform (black). The maximum of the right hump is located at $\omega = 313 
\omega_0$, at which frequency the sine wave is plotted as the gray solid curve in the inset of the left plot. 
Between the two humps the local minimum is located at $\omega=115\omega_0$.  
The gray dashed curve represents $|d({\cal P}^{[01]}+{\cal P}^{[10]})/d\Omega_{\rm II}|$ (orange dotted) subtracted by
$|d{\cal P}^{[11]}/d\Omega_{\rm II}|$ (green dashed), showing that the interference terms contribute the major part for $131\le n\le 597$.
Note that only the discrete data for $n\in {\bf Z}$ are shown in this plot.}
\label{HHG2}
\end{figure}

The frequency spectrum of the differential radiated power with respect to the observer's clock shows that there are quanta at high 
harmonics generated in a broad range of frequencies. Indeed, in Figure \ref{HHG2} (right) one can see two main humps in the frequency 
spectrum of the differential radiated power at $\theta=0$, the left one has the maximum around $\omega_0$, and the right one around 
$313 \omega_0$. 
The former is the frequency of the oscillatory motion of the detector, while the latter corresponds to the time-scale of 
the double-peak pulse around $t_- \approx t_p/4$, when the detector reaches its maximum speed $v^{}_{\rm max}= 2 a^{}_0/\sqrt{1+4a_0^2}$ 
toward the observer at $\theta=0$. The major part of the right hump around the maximum is contributed by the interference terms 
$d({\cal P}^{[01]}+{\cal P}^{[10]})/d\Omega_{\rm II}$. 
The two humps are quite broad, which is similar to the high harmonic generation (HHG) in atomic and plasma physics.
The HHG here, however, is caused by the relativistic motion \cite{MK05, LKK05, TW07}, which is not simple-harmonic in spacetime, 
rather than the bremsstrahlung during the recombination in the three-stage model \cite{KSK93, Co93, LBIHC}. 
The pulses are compressed in time and concentrated in a narrow angular distribution by the Lorentz boost of the radiated field, 
similar to the Larmor radiation of a classical charge in relativistic motion in electrodynamics \cite{Ja98}.

While the origin of the HHG sounds classical in our linear system, the harmonics generated here are quantum coherent. Later in the scaled
two-point correlators of the field $r^2\langle \hat{\Phi}_{\bf x}(t), \hat{\Phi}_{\bf x'}(t')\rangle|_{r\to\infty}$ in the 
radiation zone, we will see that the harmonics are correlated in the frequency domain.
Thus the two short pulses of the radiated energy around $\theta=0$ and $\pi$ in the upper-left and lower-right plots of Figure 
\ref{CTradt} in each cycle of the motion is associated with the coherent HHG. 
In this aspect the highly compressed pulses we found are similar to the attosecond laser.

\begin{figure}
\includegraphics[width=7.2cm]{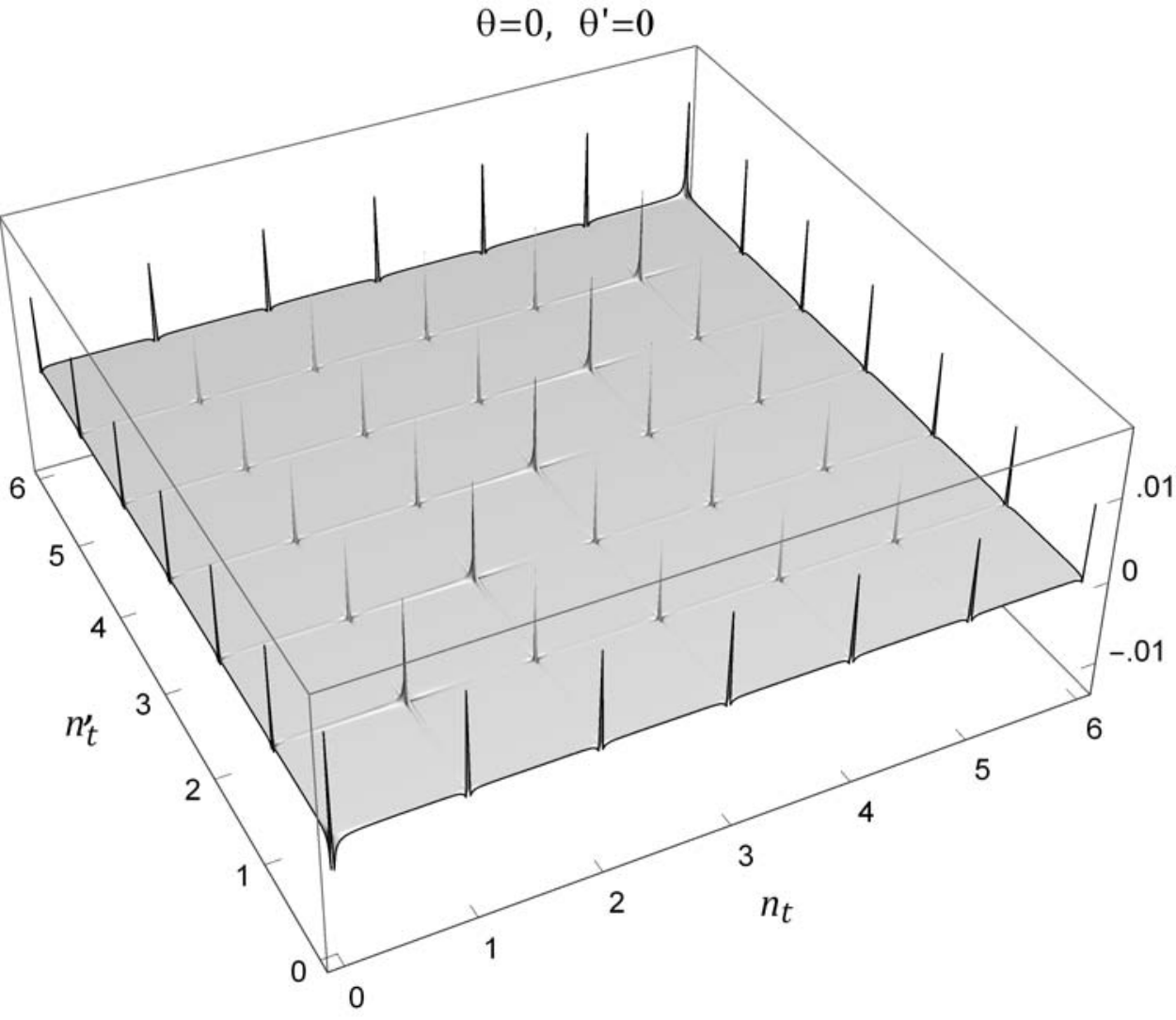}\hspace{.01cm} 
\includegraphics[width=7.2cm]{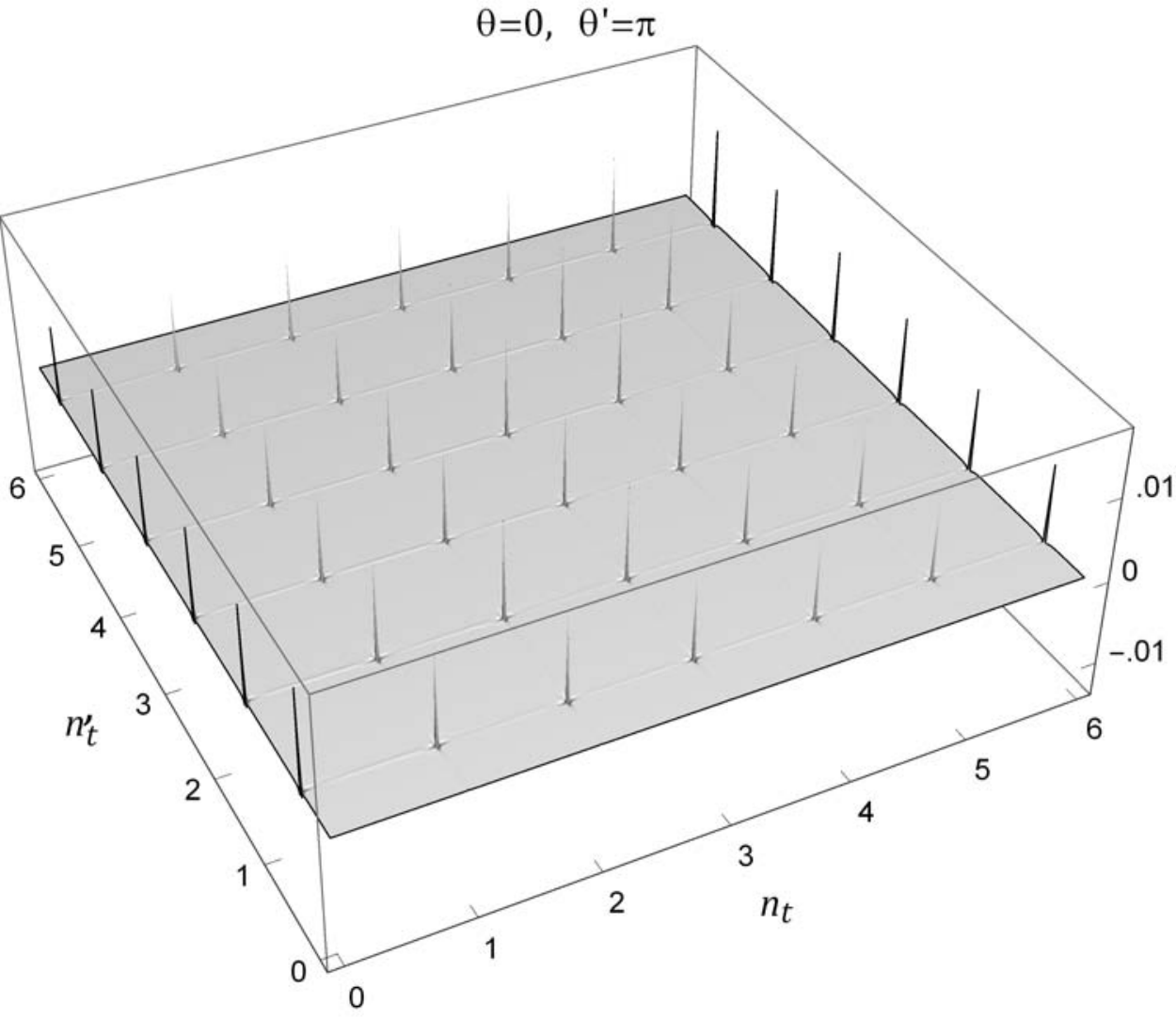}\\ 
\includegraphics[width=8.2cm]{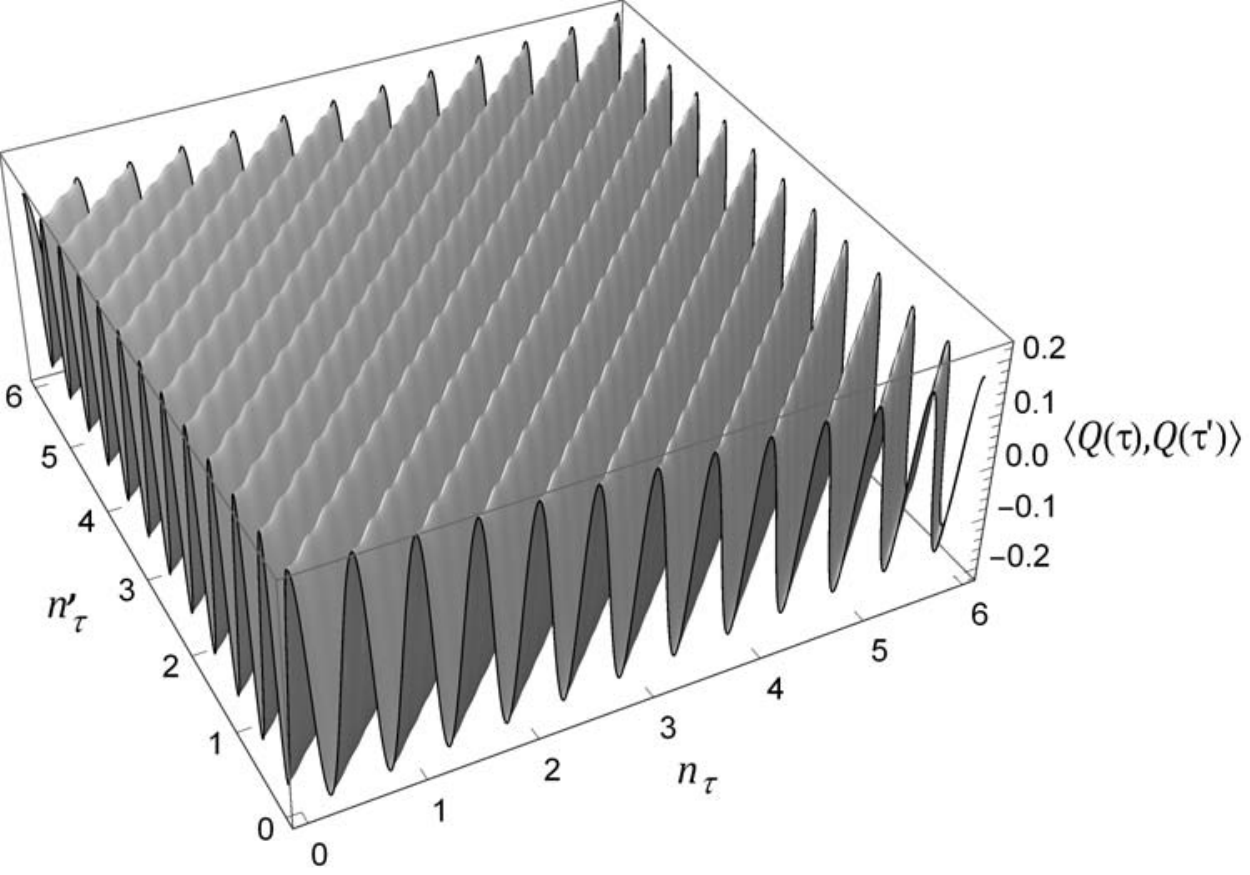}
\includegraphics[width=3.1cm]{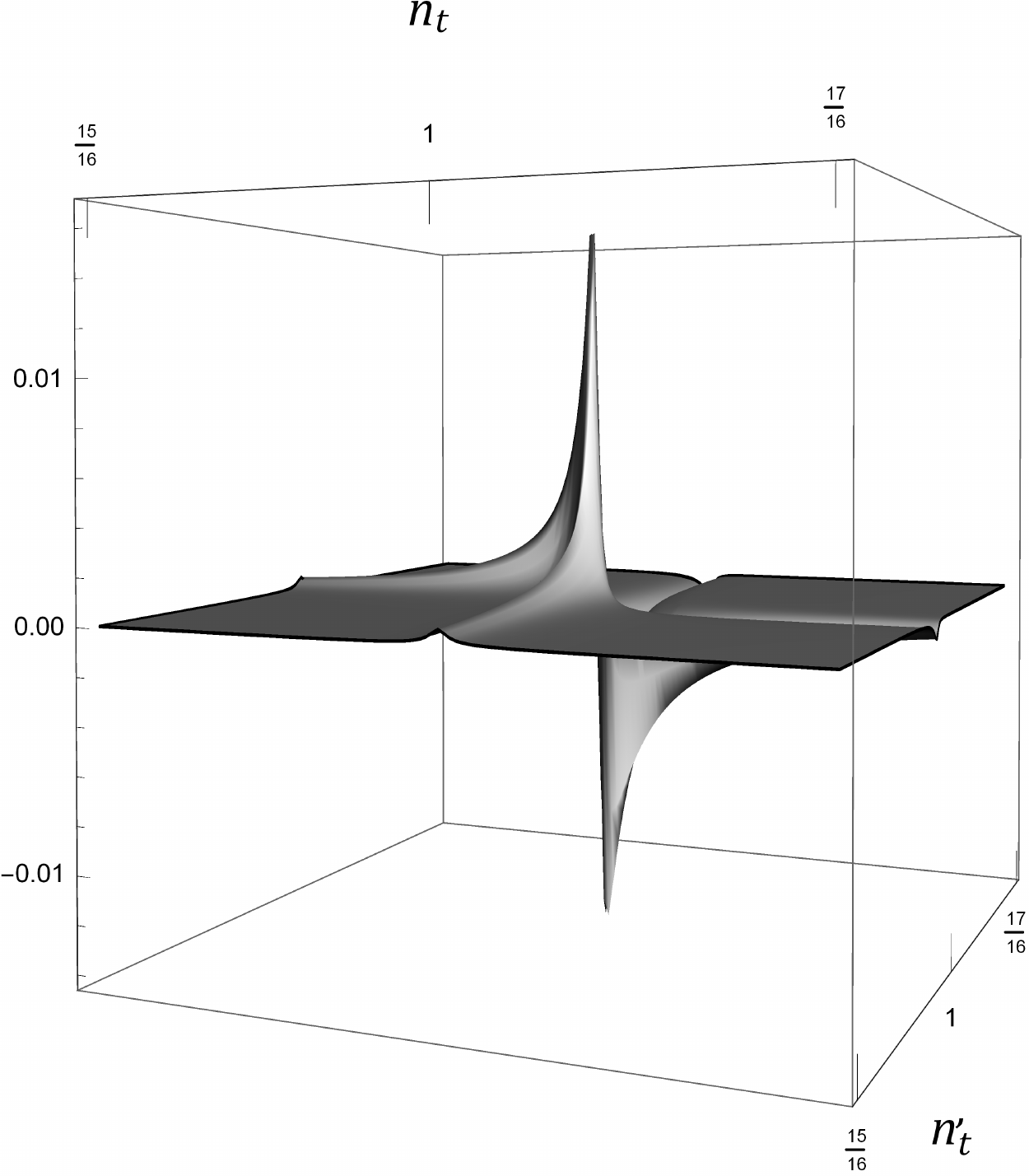}
\includegraphics[width=3.1cm]{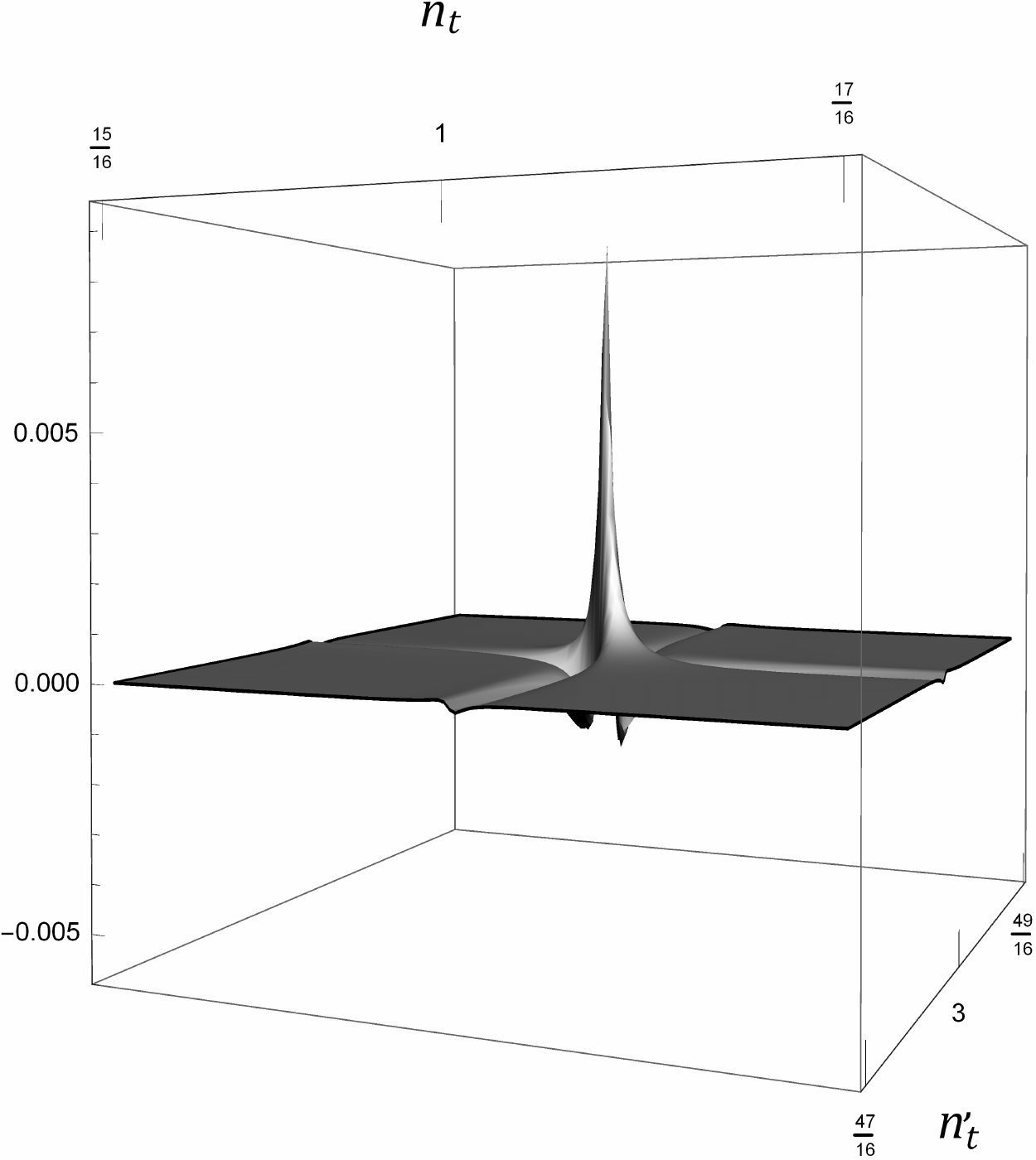}
\caption{(Upper row) The scaled, symmetrized, late-time renormalized field correlators $r^2\langle\hat{\Phi}_{\bf x}(t),
\hat{\Phi}_{\bf x'}(t')
\rangle_{\rm ren}$ evaluated at ${\bf x}$ with $\theta=0$ and ${\bf x'}$ with $\theta'=0$ (left) and $\theta'=\pi$ (right), both in the 
radiation zone. Here $\gamma=0.01$, $\Omega=2.3$, $a_0=4$, the period of a cycle of oscillatory motion in the proper time of the detector is 
exactly two times of the natural period of the internal HO ($\tau_p = 2 \times (2\pi/\Omega)$), 
$n^{}_t \equiv (t_0/t_p)-(1/4)-N_0$,  $n'_t \equiv (t'_0/t_p)-(1/4)-N_0$ with some large integer $N_0$.
(Lower row) Close-ups of the peaks in the upper-left plot around $(n^{}_t, n'_t) = (1,1)$ (middle) and $(n^{}_t, n'_t) = (1,3)$ (right) 
for $\theta=\theta'=0$. All the peaks in the upper-right plot for $(\theta,\theta') =(0, \pi)$ have similar shape to the one in the 
lower-right plot. The lower-left plot is the symmetrized two-point correlator $\langle\hat{Q}(\tau),\hat{Q}(\tau') \rangle$ of the detector, 
which dominates the signs of the above field correlators.}
\label{PulsCor}
\end{figure}

Interesting enough, a wide range of the right slope of the right hump in Figure \ref{HHG2} (right) behaves like 
$\omega^3/(e^{\hbar\omega/(k^{}_BT'_B)}-1)$, which is the spectrum of the black-body radiation. 
The parameter $T'_{B} \equiv T_{B} \sqrt{(c+v_{\rm max})/(c-v_{\rm max})}$ is blue-shifted from an effective temperature $T^{}_{B}$. 
We find $T^{}_B \approx 1.08 \bar{a}$ ($\hbar=k^{}_B=c=1$) in the interval $500\le n \le 1500$ in Figure \ref{HHG2} (right). 
As we increase the value of $a^{}_0$ from $4$ to $40$ and so $v_{\rm max}$ goes deeper into the relativistic regime, we find 
that $T^{}_B$ goes from $1.08\bar{a}$ to $1.005\bar{a}$, which suggests $T^{}_B \to \bar{a}$ as $a^{}_0\to\infty$.  
Further analysis on this observation is ongoing.
Note that the value of $T^{}_B$ here is not equal to the effective temperature $T_{\rm eff}$ experienced by the UD detector, 
which is a little less than the naive Unruh temperature $\bar{a}/2\pi$ at large averaged proper acceleration $\bar{a}$. Note also that 
in our fitting for $T^{}_B$ we only count the harmonics ($\omega = n\omega_0$, $n\in {\bf Z}$) in Figure \ref{HHG2} (right). The absolute 
value of the full, continuous spectrum of the radiated power in our case is not as smooth or isotropic as the spectrum of the black-body 
radiation: it depends on $\theta$, and at each $\theta$ it looks like a comb with the peak values located at the harmonics in the frequency 
domain (cf. Figure \ref{PulsCor} (upper left)).

\begin{figure}
\includegraphics[width=7.2cm]{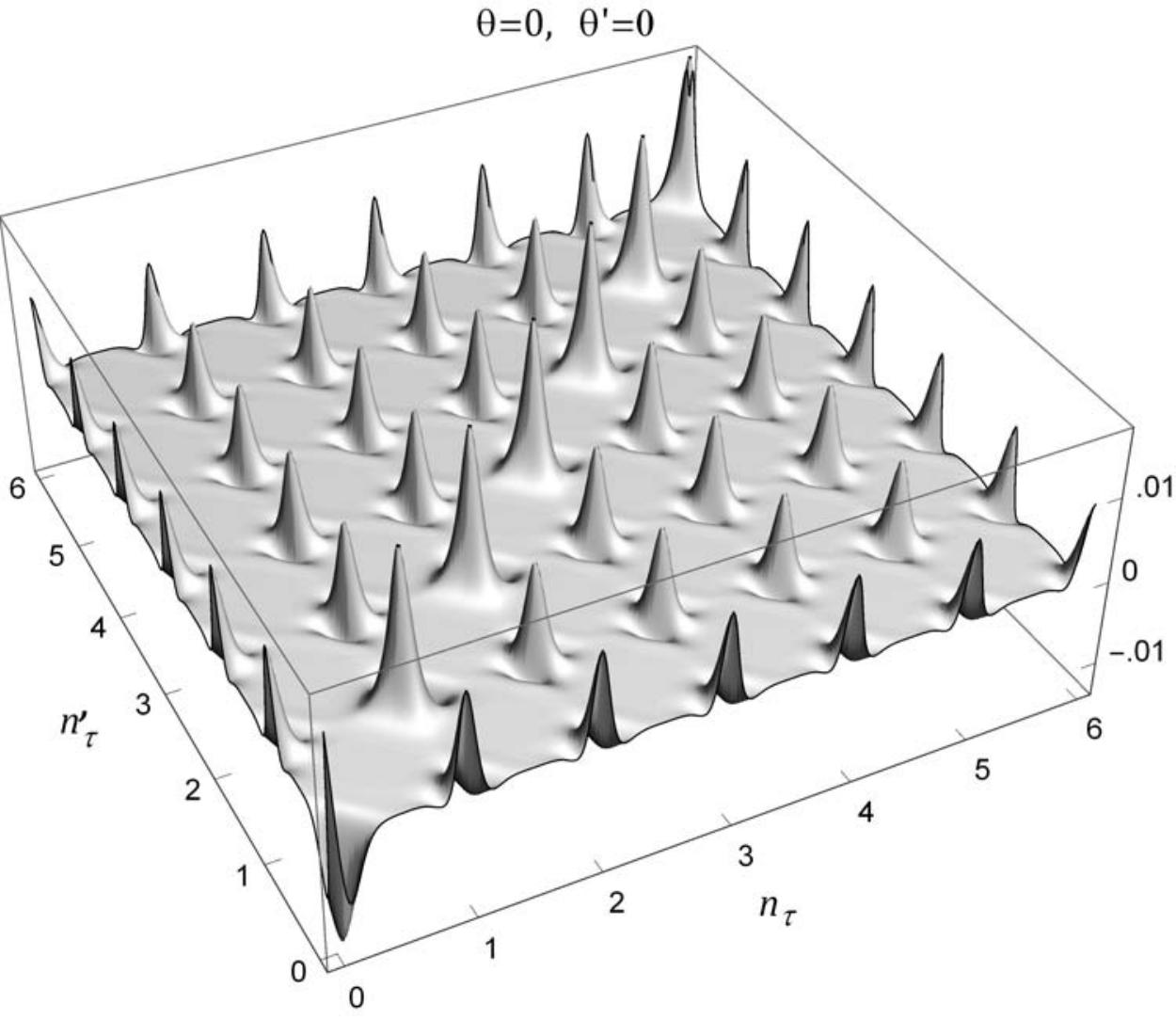}
\includegraphics[width=7.2cm]{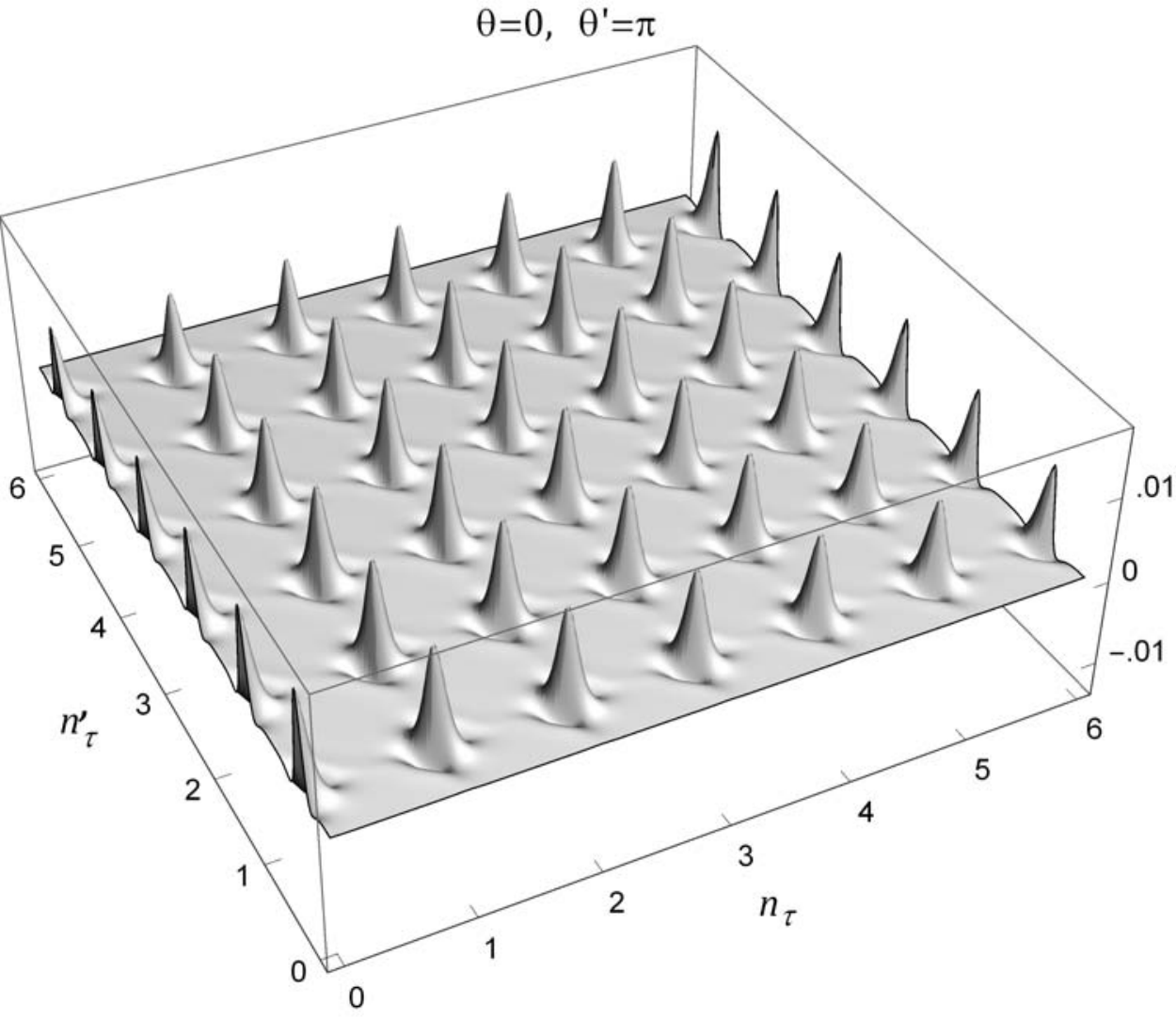}
\caption{The same scaled, symmetrized, late-time renormalized field correlators as those in Figure \ref{PulsCor} (upper row) are plotted as 
functions of $\tau=\tau^{}_-(x)$ and $\tau' =\tau^{}_-(x')$. Here $n^{}_\tau\equiv (\tau/\tau_p)-(1/4)-N_0$ and $n'_\tau\equiv (\tau'/\tau_p)
-(1/4)-N_0$. Note that some peak values are lost in the upper plots of Figure \ref{PulsCor} due to the limited resolution of the plots.}
\label{PulsCorTau}
\end{figure}

\subsection{Correlations of pulses}

In Figure \ref{PulsCor} (upper row) we show the scaled, symmetrized, late-time renormalized correlators $r^2\langle \hat{\Phi}_{\bf x}(t), 
\hat{\Phi}_{\bf x'}(t')\rangle_{\rm ren}|_{r\to \infty}$, where $\langle A, B\rangle \equiv \langle(AB+BA)\rangle/2$,
of the field amplitudes observed at different times $t$ and $t'$ on the same side ($\theta=\theta'=0$, upper-left plot) and opposite 
sides ($\theta=0$ and $\theta'=\pi$, upper-right plot) of the axis of the detector motion. 
Denoting $n^{}_t \equiv (t_0/t_p)-(1/4)-N_0$ and $n'_t \equiv (t'_0/t_p)-(1/4)-N_0$ for $t_0\equiv t-r$ and $t'_0\equiv t'-r$ with some 
very large $r \gg 2\pi/\omega_0$ (in the radiation zone) and some large integer $N_0 \gg (\gamma t_p)^{-1}$ (at late times) 
\footnote{The actual value of $N_0$ is totally unimportant in the on-resonance case, anyway.},
one can see that the correlations of the field amplitudes are amplified and peaked around $n^{}_t \in {\bf Z}$ and $n'_t \in {\bf Z}$ 
for $(\theta,\theta')=(0,0)$, and around $n^{}_t\in {\bf Z}$ and $n'_t-1/2 \in {\bf Z}$  
for $(\theta,\theta')=(0,\pi)$, respectively. The corresponding $t$ and $t'$ are the moments that the pulses reach the observers at 
$\theta=0$ ($t-r= [n+N_0 +(1/4)]t_p$, $n=1,2,3,\cdots$) and $\theta'=0$ ($t'-r= [n'+N_0 +(1/4)]t_p$) or $\pi$ ($t'-r= [n'+N_0 +(3/4)]t_p$). 
Thus the pulses of the radiated power are correlated.
The same scaled correlators as functions of the retarded proper times $\tau^{}_-$ are shown in Figure \ref{PulsCorTau}. It is obvious that 
the peaks in the upper plots of Figure \ref{PulsCor} are compressed from the ones in Figure \ref{PulsCorTau} through the nonlinear relation 
between the observer's time $t$ and the retarded proper time $\tau^{}_-$ of the detector at fixed $r$ and $\theta$. 

Looking more closely to the peaks, one can see that the shape of the peaks on the diagonal axes ($t=t'$) in Figure \ref{PulsCor} 
(upper-left) is different from the shape of the off-diagonal peaks. 
Around the diagonal peaks, the value of $r^2\langle\hat{\Phi}_{\bf x}(t),\hat{\Phi}_{\bf x'}(t')\rangle_{\rm ren}$ at $\theta=\theta'=0$
is positive for both $(t$ mod $t_p)$ and $(t'$ mod $t_p)$ are a little less than $\tau_p/4$, and is negative
for both $(t$ mod $n)$ and $(t'$ mod $n)$ are a little greater than $\tau_p/4$ (Figure \ref{PulsCor} (lower-middle)).
Around the off-diagonal peaks, if the peak value of the renormalized field correlator is positive (negative), the value will be always 
positive (negative) in $t-t'$ direction, and will become negative (positive) in $t+t'$ direction when getting close to but not
immediately neighboring to the peak (Figure \ref{PulsCor} (lower-right)).
These features are even clearer in Figure \ref{PulsCorTau}. 

In the example we consider here, the sign of the off-diagonal peaks of the field correlation is dominated by the sign of $\langle \hat{Q}(\tau_{-}(t)),\hat{Q}(\tau_{-}(t'))\rangle$ in $G^{[11]}$ in Eq. (\ref{G11}). This is evident by comparing Figure \ref{PulsCor} (upper-left) 
and Figure \ref{PulsCorTau} (left) with Figure \ref{PulsCor} (lower-left). As $|n-n'|$ increases, the absolute peak value 
of the correlation roughly decreases like $e^{-\gamma |n-n'|\tau_p}$.

\subsection{Correlation of harmonics, and down conversion}

\begin{figure}
\includegraphics[width=7.2cm]{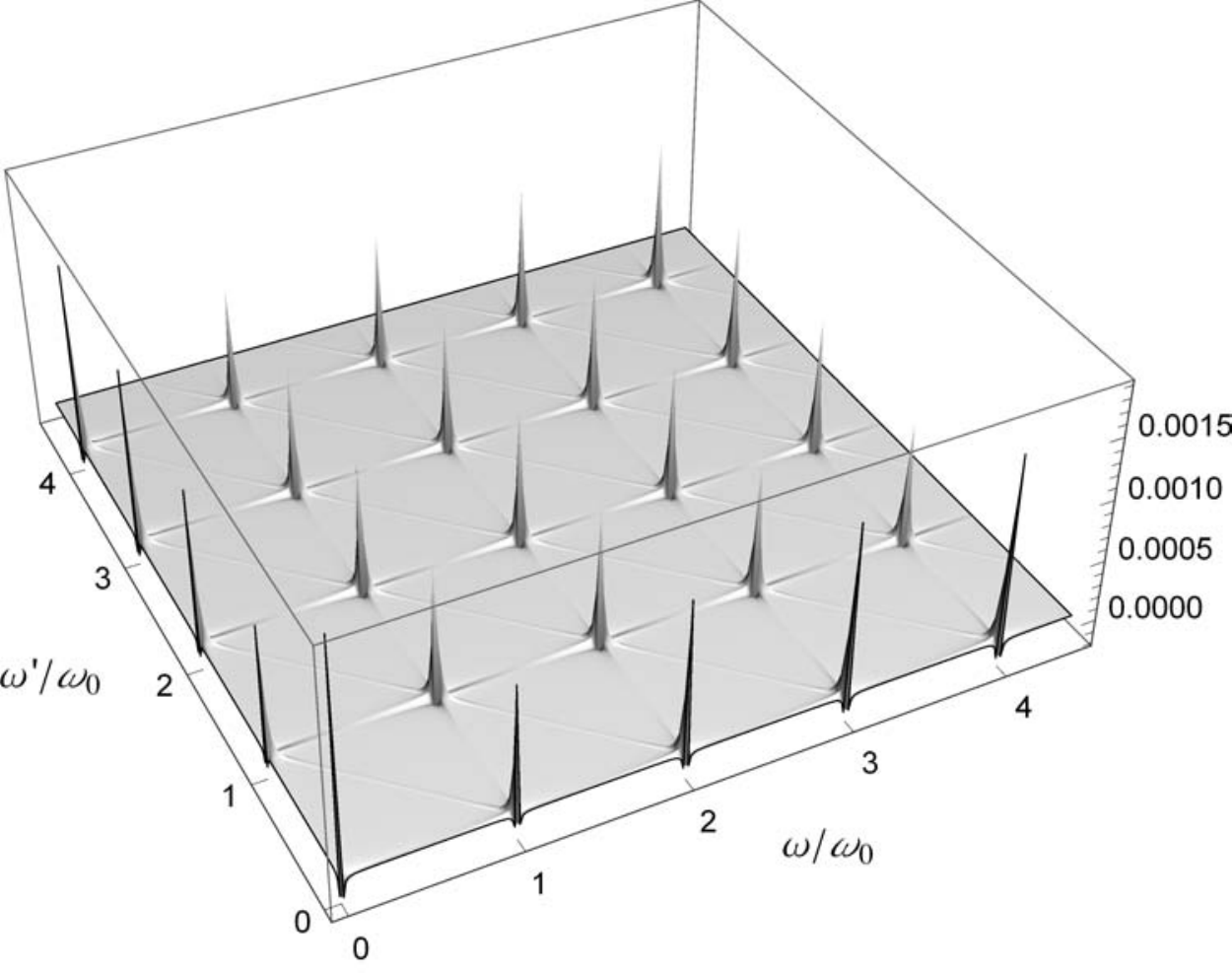}
\includegraphics[width=7.2cm]{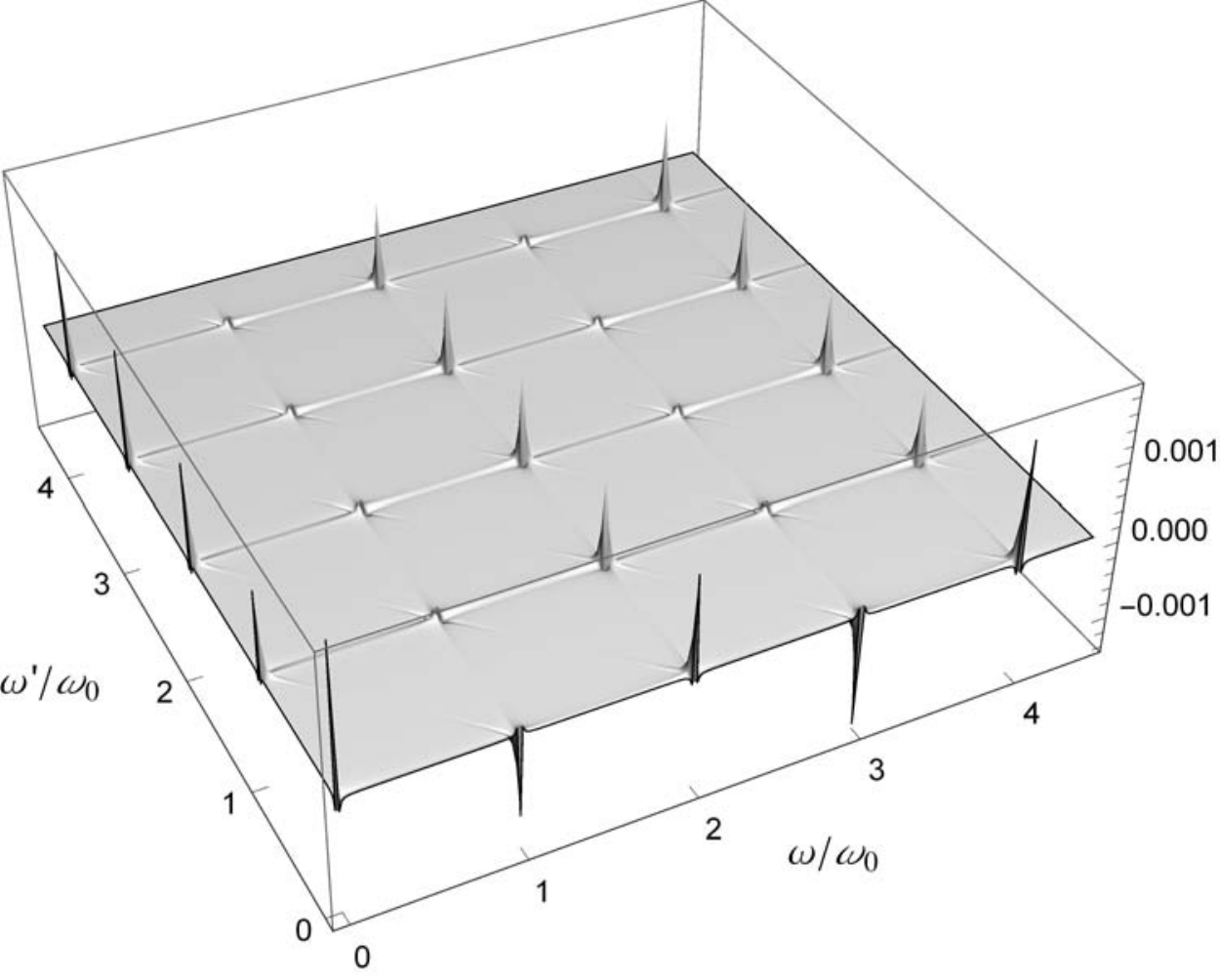}\\ 
\includegraphics[width=7.2cm]{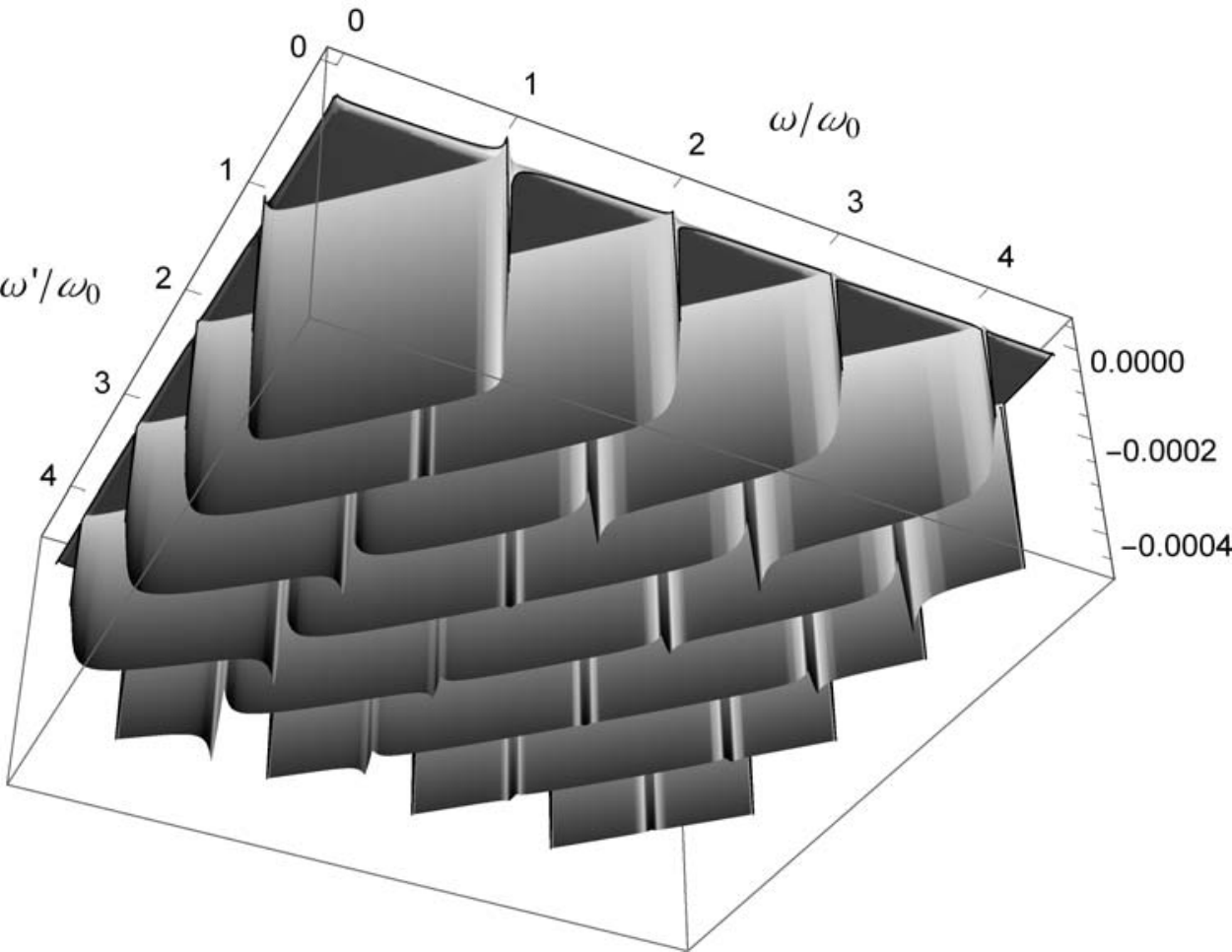} 
\includegraphics[width=7.2cm]{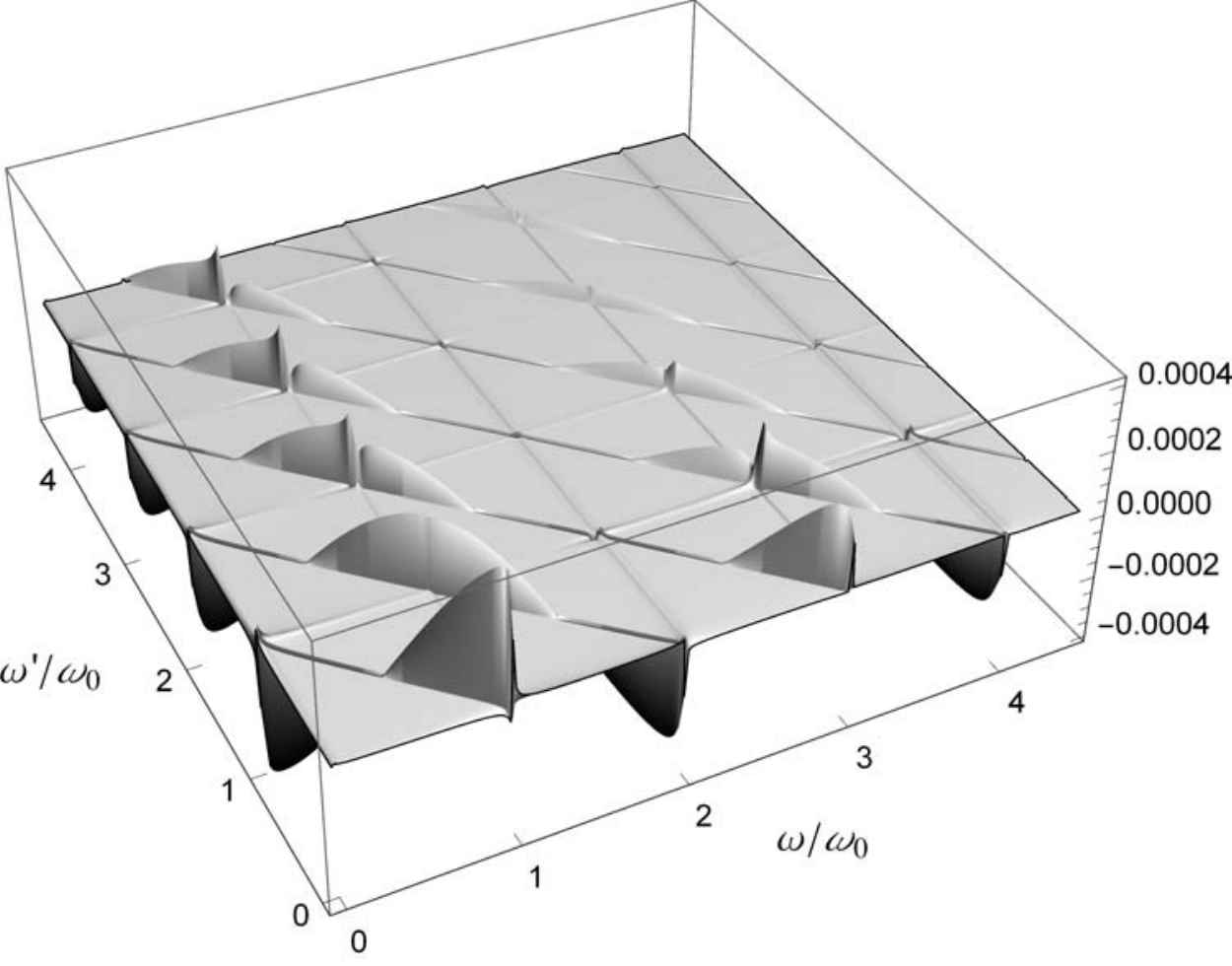}\\ 
\caption{The scaled correlators of the field in the frequency domain Fourier-transformed from the ones in Figure \ref{PulsCor} for 
$(\theta,\theta')=(0,0)$ (left) and for $(\theta, \theta')=(0,\pi)$ (right). 
The real parts and imaginary parts of the correlators are shown in the upper and the lower rows, respectively. 
Here 
we have $t$ and $t'$ running across a time interval of $64 t_p$ to get sub-harmonic structures. }
\label{HHG1}
\end{figure}

\begin{figure}
\includegraphics[width=7.2cm]{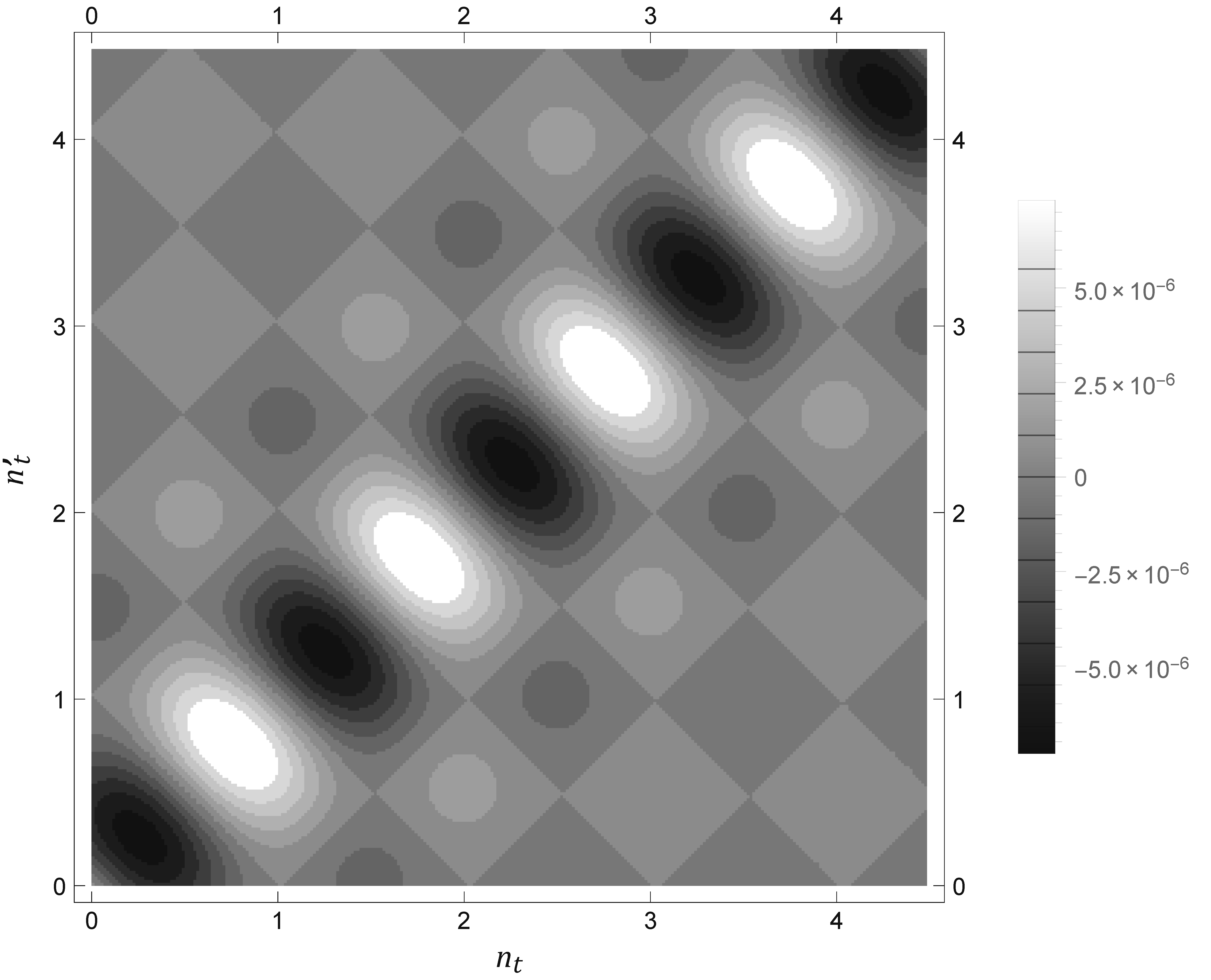}
\includegraphics[width=7.2cm]{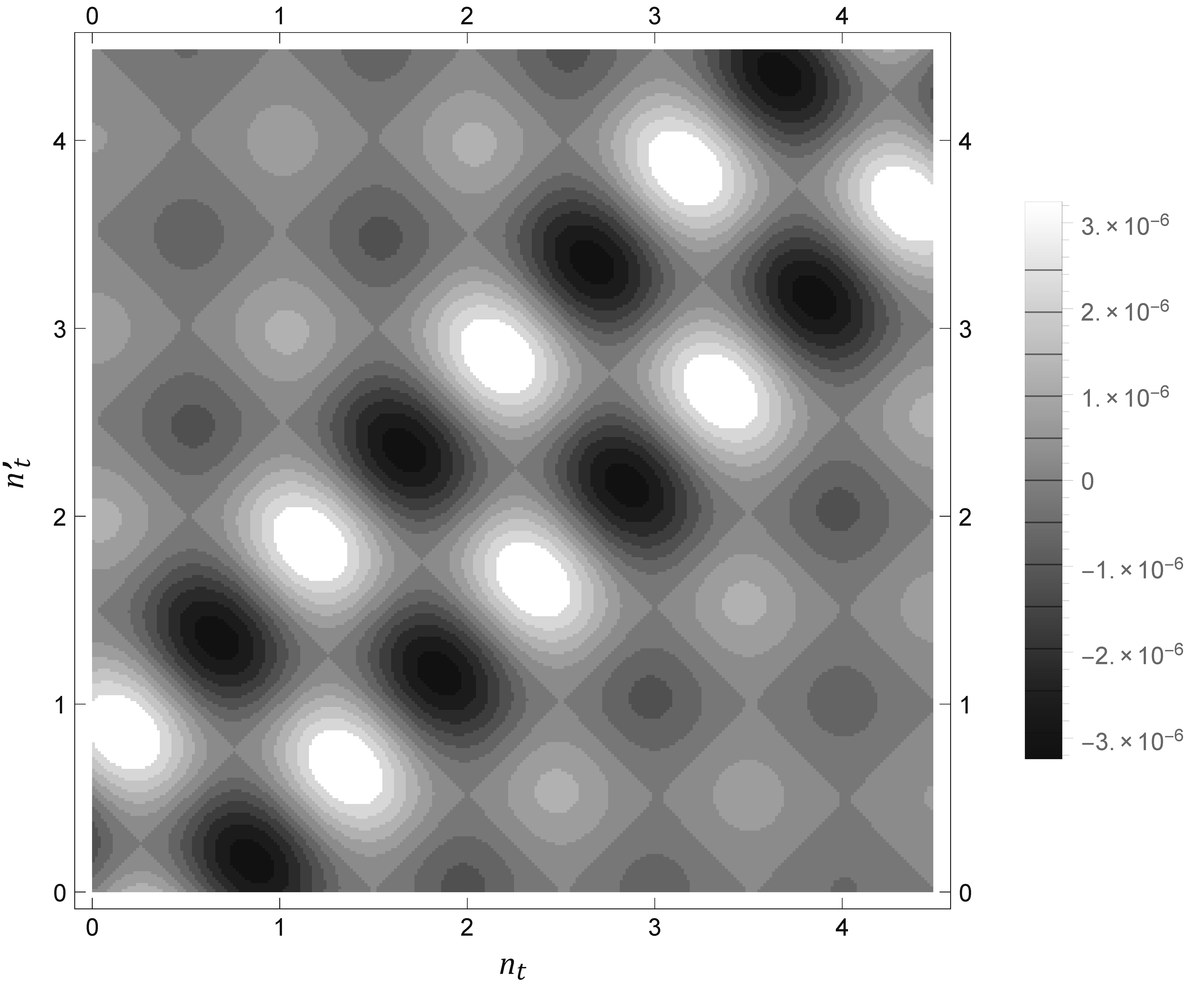}
\includegraphics[width=7.2cm]{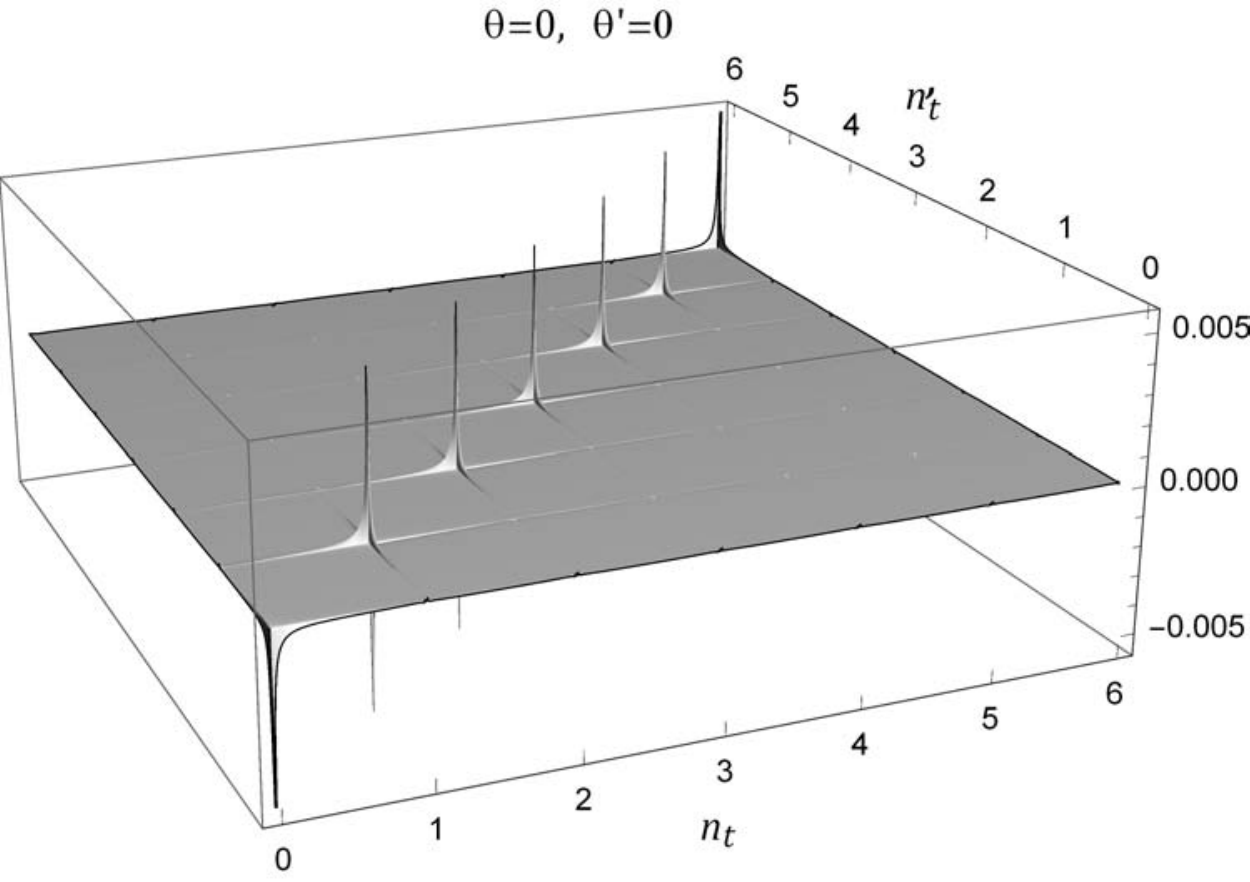}
\includegraphics[width=7.2cm]{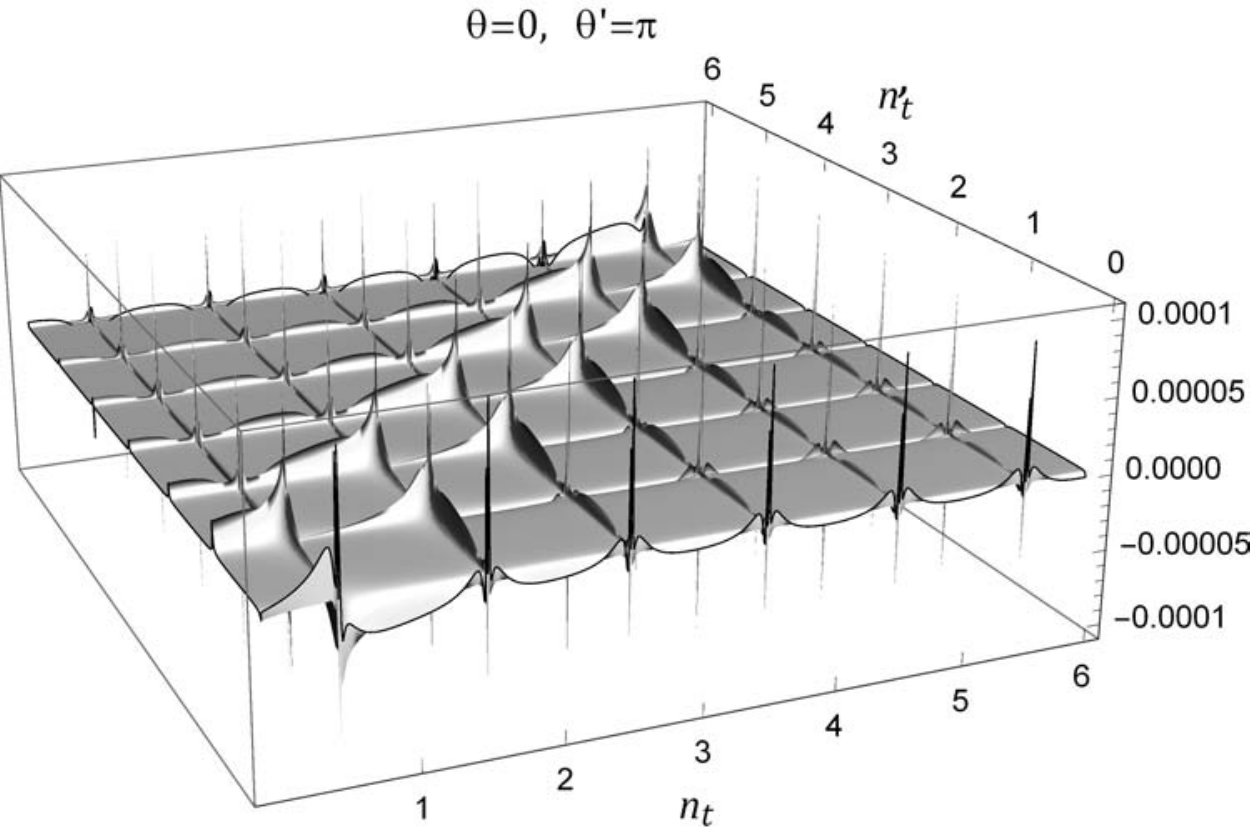}
\caption{(Upper row) The inverse-Fourier-transformed results only including the thin walls with $\omega+\omega'=\pm \omega_0$ of the 
imaginary part of the Fourier-transformed field correlators in the lower plots of Figure \ref{HHG1} for $(\theta,\theta')=(0,0)$ 
(left) and $(\theta, \theta')=(0,\pi)$ (right). (Lower row) The inverse-Fourier-transformed results of the whole imaginary 
part of the Fourier-transformed scaled correlator of the field.}
\label{DC2}
\end{figure}

Figure \ref{HHG1} shows the correlators of the radiation field transformed from those in Figure \ref{PulsCor} to the frequency domain. 
The real parts of the Fourier transformed correlators of the field at $(\theta,\theta')=(0,0)$ and 
$(\theta,\theta')=(0,\pi)$ are presented in the upper-left and the upper-right plots, respectively. 
One can immediately see that in both cases there are strong correlations at the lattice points $(\omega, \omega')=
(m\omega_0, m'\omega_0)$ with $m,m' = 0,1,2,3,\cdots$, indicating that different harmonics are coherent, as we claimed earlier. 
 
The plots in the lower row of Figure \ref{HHG1} show the imaginary parts of the Fourier transformed correlators.
In the lower-left plot for $\theta=\theta'=0$, one can clearly see the anti-diagonal, negative valued ``thin walls" in $45$ degrees in the 
unit cells of the lattice of the harmonics. They satisfy the conditions $\omega+\omega' = n \omega_0$, $n=1,2,3,\cdots$.
The one with $\omega+\omega'=\omega_0$ indicates that in quantum radiation there exist correlated pairs of field quanta with both 
$\omega$ and $\omega'$ even lower then the frequency $\omega_0$ of the oscillatory motion driven by some external agent. 
This may be interpreted as a down conversion process. The similar anti-diagonal thin walls with $\omega+\omega' = n\omega_0$, 
$n=2, 3, 4,\cdots$, correspond to the down conversions from the higher harmonics.

As pointed out in Ref.~\cite{UW84}, in the leading-order perturbative result for the expectation value of the stress energy tensor, 
there exist interference terms corresponding to the creation of a pair of the Minkowski particles.  
In Refs. ~\cite{SSH06, SSH08}, it is found that electrons driven by a strong, periodic electromagnetic field can also convert vacuum 
fluctuations into entangled photon pairs in the first order perturbation, which can be understood in terms of the Unruh effect.
The whole process can be viewed as a down conversion from the photons at the frequency $\omega_0$ of the intense laser to the lower 
frequencies $\omega_1$, $\omega_2$ of the entangled photon pair with $\omega_1+\omega_2 = \omega_0$.
Since the perturbative results in the time-dependent perturbation theory is valid in transient rather than in equilibrium conditions
\cite{HLL12}, it is natural to see a similar down conversion in our full result for a detector in linear oscillatory motion, which is 
non-equilibrium.

In (\ref{G11}) and (\ref{DDG11}), the factors such as ${\cal R}_{,t}{\cal R'}_{,r}/({\cal RR'})^2$ and $\eta_{-,\mu}\eta_{-,\nu'}/
({\cal RR'})$ are products of periodic functions $\sim f(t)g(t')$.
If a correlation function of $t$ and $t'$ can be arranged in the form $C(t,t')=\sum_{p=1,2,\cdots}f^{}_p(t)\times g^{}_p(t')$, where all
the functions $f^{}_p(t)$ and $g^{}_p(t')$ are periodic at the frequency $\omega_0=2\pi/t_p$, then 
$C(t,t')= \sum_p \left(\sum_{n\in {\bf Z}} \tilde{f}^{}_{pn}e^{in\omega_0 t}\right)\times \left(\sum_{m\in {\bf Z}} 
\tilde{g}^{}_{pm}e^{im\omega_0 t}\right) = \sum_{p,n,m}\tilde{f}^{}_{pn}\tilde{g}^{}_{pm} e^{i\omega_0(n t+m t')}$ will not 
contribute any nonzero frequency lower than $\omega_0$ with respect to $t$ or $t'$. 
Since the field correlators have the property $G(t,t')\not= G(t+t_p, t')$ or $G(t, t'+t_p)$ in general,
they are not in the form of $\sum_{p=1,2,\cdots}f^{}_p(t)g^{}_p(t')$, and in (\ref{G11}) and (\ref{DDG11}) the correlators of the 
detector such as $\langle \hat{Q}(\eta_-(t)), \hat{Q}(\eta_-(t'))\rangle$ could produce down conversion.
Similarly in (\ref{DDG10}), the Wightman function $D^+$ could do. Thus the cause of down conversion in this model must be quantum.

Performing inverse Fourier transforms on the imaginary parts of the renormalized correlator of the field with $\omega+\omega'=\pm\omega_0$
only, we obtain the results in the time domain in the upper row of Figure \ref{DC2}. One can clearly see that the anti-diagonal structures
in the lowest unit cells in the lower plots of Figure \ref{HHG1}, namely, the down conversions of the fundamental frequency $\omega_0$, 
correspond to the periodicity in $t+t'$ direction ($+45^\circ$ on the $n^{}_t n'_t$-plane) with period $2t_p$ in $t+t'$ for the 
renormalized correlators of the field in Figure \ref{PulsCor}. 
The symmetric and antisymmetric behavior of the anti-diagonal structures in the lowest unit cells about the line $\omega=\omega'$ 
in the frequency domain also propagate to the symmetric and antisymmetric structures about the line $t=t'$ in the time domain
in the upper plots of Figure \ref{DC2}.

The thin walls of the field correlators in the frequency domain in the lower-left plot of Figure \ref{HHG1} for 
$\theta=\theta'=0$ spread widely in $\omega-\omega'$ direction, 
thus when the inverse Fourier transform includes all the imaginary part of the renormalized field correlators in the frequency domain,
the result will be highly concentrated around the diagonal lattice points in the time domain as shown in Figure \ref{DC2} (lower-left). 
The negative correlation around $(n^{}_t, n'_t)=(n+\epsilon, n+\epsilon)$ with $n\in {\bf Z}$ and $\epsilon>0$ in the 
vicinity of the diagonal peak centered around $(n^{}_t, n'_t) =(n,n)$ (see the lower-middle plot of Figure \ref{PulsCor}) 
are mainly contributed by these imaginary parts.

The radiation field amplitudes at $\theta=0$ and $\theta'=0$ can satisfy the phase-matching conditions $\omega+\omega'=n\omega_0$ and 
${\bf k}+{\bf k}'=n\omega_0 \hat{x}^3$ simultaneously. This may explain the significance of the correlation along the anti-diagonal thin 
walls. In contrast, for the correlators of the field amplitudes at $\theta=0$ and $\theta'=\pi$, the phase matching conditions cannot be 
fulfilled, and the anti-diagonal thin walls are significant only in the unit cells of the lowest few harmonics, as shown in Figure 
\ref{HHG1} (lower right).
The contribution by the whole imaginary parts of the Fourier-transformed field correlator at $(\theta, \theta')=(0,\pi)$ (Figure \ref{DC2} 
(lower-right)) to the original correlator in the time domain is small and widespread compared with the case of $(\theta, \theta')=(0,0)$ 
(Figure \ref{DC2} (lower-left)).

In the off-resonance cases, the pattern of the late-time field correlators has no exact periodicity in $t+t'$ direction. While the
details of different pulses would not be exactly the same at late times, each pulse would still be concentrated around $t_0$ mod $t_p = 
t_p/4$ at $\theta=0$ or $3t_p/4$ at $\theta=\pi$. As down conversions involve low frequency quanta which are not sensitive to the detail 
of each pulse, we expect down conversions would still appear in the off-resonance cases. 

\section{Squeezing in the asymptotic reduced state}
\label{SecARS}

\begin{figure}
\includegraphics[width=7.2cm]{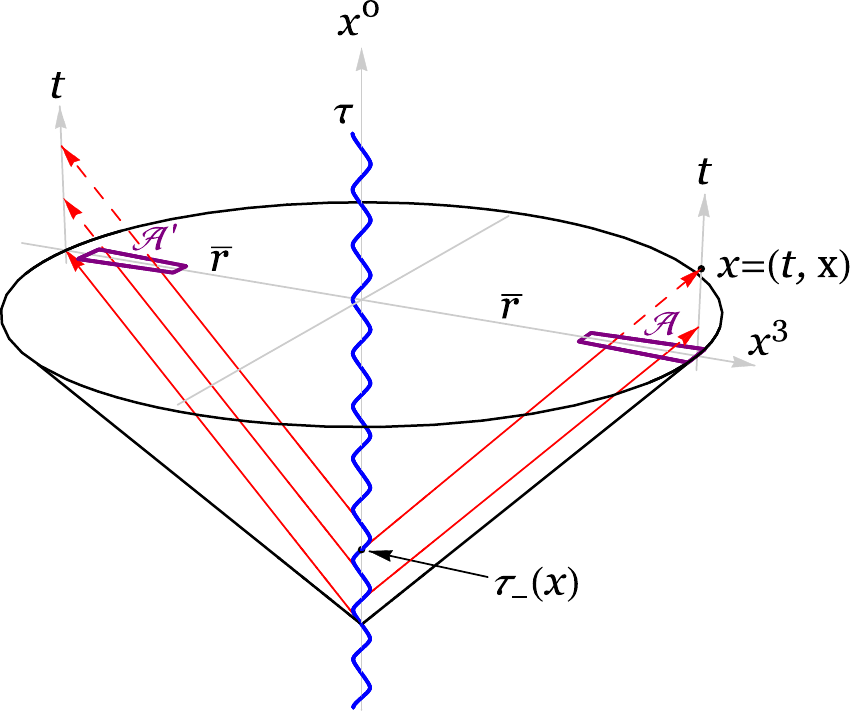}
\caption{The sampling regions ${\cal A}$ and ${\cal A}'$ for the asymptotic state, with boundaries colored in purple, in the radiation
zone ($r\gg 2\pi/\omega_0$).
The blue curve represents the worldline of the detector and the red arrows represent the emitted pulses in the oscillatory plane.}
\label{ARS}
\end{figure}

It has been argued that the presence of the negative radiated power in Section \ref{SecNegaE} indicates that the Unruh radiation corresponds 
to a multi-mode squeezed state of the field~\cite{SSH06, SSH08}. 
However, negative energy density can also arise in many field states with even particle numbers, the two-particle Fock state of some 
field mode could be the simplest \cite{KF93}. It is thus interesting to examine more details of the field state for the Unruh radiation 
observed in the radiation zone. 

\subsection{Asymptotic reduced states of the field}

A quantum state of the combined detector-field system (\ref{Stot1}) can be described by the density matrix
$\bar{\rho}[(Q,\Phi_{\bf x}), (Q',\Phi'_{\bf x});x^0]$,  
or equivalently in the $(K,\Delta)$-representation \cite{UZ89}   
\begin{equation}
  \rho[{\bf K}, {\bf \Delta}; x^0] = \int {\cal D}{\bf \Sigma}\, e^{{i\over\hbar} {\bf K}\cdot {\bf \Sigma}}
    \bar{\rho}\left[ {\bf \Sigma} - {{\bf \Delta}\over 2}, {\bf \Sigma} + {{\bf \Delta}\over 2} ; x^0 \right],
\end{equation}
where we write $(Q,\Phi_{\bf x})={\bf \Sigma}-({\bf \Delta}/2)$ and $(Q',\Phi'_{\bf x})= {\bf \Sigma}+ ({\bf \Delta}/2)$ with the 
boldface letters ${\bf \Sigma}$ and ${\bf \Delta}$ denoting the combined detector and field vectors in the configuration space. 
Our combined system is linear and started with a Gaussian state (\ref{initstat}) with $\langle\hat{\Phi}^{}_\mu\rangle = 
\langle\hat{\Pi}^{}_\mu\rangle = 0$, so the reduced state of the field (with the detector's degree of freedom traced out) simply reads
\begin{equation}
  \rho^R[{\cal K}_j; x^0] = \exp -{1\over 2\hbar^2} {\cal K}_i {\cal V}^i{}_j(x^0) {\cal K}^j,
\label{Fstate}
\end{equation}
where the indices $i,j$ run over a time-slice in the position or wave-vector space depending on the representation of the field we chose, 
${\cal K}_j = (K_j, \Delta_j)$ and ${\cal K}^j = ({\cal K}_j)^T$, and ${\cal V}$ is the covariance matrix with each element 
\begin{equation}
   {\cal V}^i{}_j(x^0) = \left( \begin{array}{cc}
        \langle \hat{\Phi}_i(x^0), \hat{\Phi}_j(x^0)\rangle & \langle \hat{\Phi}_i(x^0), \hat{\Pi}_j(x^0)\rangle \\
        \langle \hat{\Pi}_i(x^0), \hat{\Phi}_j(x^0)\rangle & \langle \hat{\Pi}_i(x^0), \hat{\Pi}_j(x^0)\rangle \end{array} \right)
\label{covmtx}				
\end{equation}
a $2\times 2$ matrix of the symmetrized two-point correlators of the field \cite{HLL12}. Then one can extract quantum information such as 
squeezing and entanglement from the covariance matrix. 

Nevertheless, no apparatus localized in the radiation zone is possible to get the full information of the field in the whole space, nor do
our numerical calculations for the Unruh radiation. What can be measured or numerically calculated in our setting are the {\it asymptotic} 
correlators of the field defined in a finite spatial or temporal region in the radiation zone (Figure \ref{ARS}),
with the bound field (counterpart of the velocity field in electrodynamics \cite{Hi02, Ja98}) totally ignored.
Inserting these asymptotic correlators of the field into (\ref{covmtx}) and (\ref{Fstate}), what we get is the asymptotic reduced state 
of the field, which is certainly not the true reduced state of the field.

\subsection{Mode decomposition}

The Minkowski vacuum state of a massless scalar field is the simplest in terms of the field modes: it is a direct product of the ground 
states for the (complex) HOs $\Phi_{\bf k} = \int d^3 x e^{-i{\bf k\cdot x}} \Phi_{\bf x}$ with the natural frequencies 
$\omega = |{\bf k}|$ \cite{Hat92, LCH10}, thus separable in terms of field modes $\Phi_{\bf k}$. In quantum optics, conventional discussions
on the squeezed state are also based on the field modes. To compare with the vacuum state and look into the squeezing in the radiation, 
therefore, we would represent the asymptotic reduced state of the Unruh radiation in terms of the field modes.

Since the motion of our source (the UD detector) is quite localized in space, if we are looking at the complete field state, we may have
to expand the retarded field in terms of the solutions of the massless Klein-Gordon equation in spherical coordinates,
which is not easy to deal with in calculations.
Fortunately, we are only interested in the asymptotic states here, and so the partial wave analysis applies:
The spherical waves with the spherical Hankel function $h^{(1)}_l(kr) \approx e^{ikr}/kr$ locally look like the plane waves in the radiation 
zone ($kr\to \infty$), namely, $e^{ik(\bar{r}+\Delta r)}/[k(\bar{r}+\Delta r)]\approx f(\bar{r}) e^{ik\Delta r}$ for a large constant 
$\bar{r}$. 
Below we are considering the asymptotic reduced state of the field in the conventional plane-wave field modes, which has the translational
symmetry that a true state of the field scattered by a localized source is lacking. 
As long as the decomposition is linear, the asymptotic reduced state of the field (\ref{Fstate}) will always be Gaussian.

\subsection{Squeezed thermal harmonic modes}

To get a clearer picture and make the numerical calculation more economic, we will further restrict ourselves to the most significant 
subset of the field degrees of freedom.
We have learned in the previous sections that the quantum radiation emitted by the detector is concentrated around $\theta=0$ and $\pi$,
or $+x^3$ and $-x^3$ directions, respectively.
In both directions the field modes of the harmonics with frequency $n \omega_0$, $n=1,2,3,\cdots$ are the most interesting. We can construct 
the covariant matrix of these field modes using the late-time correlators $\langle \hat{\cal R}^{}_{\pm n{\bf k}_0}, 
\hat{\cal R}'_{\pm n'{\bf k}_0}\rangle$, ${\cal R}$, ${\cal R}'= \Phi, \Pi$, and ${\bf k}_0 = (0, 0, \omega_0)$ in the radiation zone. 
Note that one should not subtract out the correlators of the vacuum state of the free field 
\begin{eqnarray} 
  \langle \hat{\Phi}^{^{[0]}}_{n{\bf k}_0}, \hat{\Phi}^{^{[0]}}_{-n'{\bf k}_0} \rangle &=& \frac{\hbar}{2 \omega_n} V \delta_{n,n'} 
    \label{vaccorF}\\
  \langle \hat{\Pi}^{^{[0]}}_{n{\bf k}_0}, \hat{\Pi}^{^{[0]}}_{-n'{\bf k}_0} \rangle &=& \frac{\hbar \omega_n}{2} V \delta_{n,n'} 
	  \label{vaccorP}
\end{eqnarray}
with $\omega_n \equiv \sqrt{(n \omega_0)^2 + k_\perp^2}$ from the elements of the covariant matrix \cite{LCH10}, 
though they are singular as the volume $V$ of the sampling region goes to infinity ($V\delta_{nn'}\to (2\pi)^3 \delta^3({\bf k}-{\bf k}')$
as $V\to\infty$). 

In our example, we first calculate the sum of the scaled renormalized field correlators at a late 
time $x^0 =\bar{t}\equiv [N_0+(3/4)]t_p+\bar{r}$ with some integer $N_0 \gg (\gamma t_p)^{-1}$ and some $\bar{r}\gg 2\pi/\omega_0$ in the 
two finite sampling regions ${\cal A}$ around $\theta=0$ and ${\cal A}'$ around $\theta=\pi$ as
\begin{eqnarray} 
		r r' \langle \hat{\cal R}^{}_{r}, \hat{\cal R}'_{r'}\rangle_{\rm ren} &\equiv& 
		r r'\left[ \langle \hat{\cal R}^{}_{r \hat{x}^3}, \hat{\cal R}'_{r' \hat{x}^3}\rangle_{\rm ren} +
    \langle \hat{\cal R}^{}_{-\tilde{r} \hat{x}^3}, \hat{\cal R}'_{r' \hat{x}^3}\rangle_{\rm ren} +\right.\nonumber\\ & &\left.
    \langle \hat{\cal R}^{}_{r \hat{x}^3}, \hat{\cal R}'_{-\tilde{r}' \hat{x}^3}\rangle_{\rm ren} +
    \langle \hat{\cal R}^{}_{-\tilde{r} \hat{x}^3}, \hat{\cal R}'_{-\tilde{r}' \hat{x}^3}\rangle_{\rm ren}\right]_{x^0=\bar{t}} 
\end{eqnarray}
where ${\cal R},{\cal R}'=\Phi$ or $\Pi$, $\tilde{r}=r-(t_p/2)$, $\tilde{r}'=r'-(t_p/2)$, $r, r' \gg 2\pi/\omega_0$, $r \hat{x}^3, 
r' \hat{x}^3 \in {\cal A}$ and $-\tilde{r}\hat{x}^3,-\tilde{r}\hat{x}^3 \in {\cal A}'$, to suppress the imaginary part produced by the 
Fourier transform to be done below. 
Here the $rr'$ factor is to balance the $1/r$ and $1/r'$ dependence of $\hat{\cal R}^{}_{\bf x}$ and $\hat{\cal R}^{}_{\bf x'}$ in the 
radiation zone. We write $r=\bar{r}+\Delta r$ and $r' = \bar{r}+\Delta r'$ with $\Delta r, \Delta r' \ll \bar{r}$. 
Since ${\cal A}$ and ${\cal A}'$ are finite regions, the wave-vector space is discrete with the minimal line element $dk = 2\pi/L$, 
where $L$ is the radial length scale of the sampling regions. 
We further assume that $d\Omega$ is so small and $\bar{r}$ is so large that $\bar{r}^2 d\Omega \sim L^2$ while 
$\langle \hat{\cal R}^{}_{\bf x}, \hat{\cal R}'_{\bf x'}\rangle_{\rm ren} =\langle \hat{\cal R}^{}_{\bf x}, \hat{\cal R}'_{\bf x'}\rangle -
\langle \hat{\cal R}^{^{[0]}}_{\bf x}, \hat{\cal R}'^{^{[0]}}_{\bf x'}\rangle $ is constant in $\theta$ and $\varphi$ directions.
Although the Fourier transformed correlator should be 
\begin{equation}
  \langle \hat{\cal R}^{}_{\bf k}(\bar{t}), \hat{\cal R}'_{\bf k'}(\bar{t})\rangle_{\rm ren} = \int d^3 {\bf x} d^3 {\bf x'} 
	e^{-i{\bf k\cdot x}} e^{-i{\bf k'\cdot x'}}\langle \hat{\cal R}^{}_{\bf x}(\bar{t}), \hat{\cal R}'_{\bf x'}(\bar{t})\rangle_{\rm ren} ,
\end{equation}
integrated over the whole space, we replace the integral by an approximated one to obtain
\begin{eqnarray}
    && \langle \hat{\cal R}^{}_{k\hat{x}^3}(\bar{t}), \hat{\cal R}'_{k'\hat{x}^3}(\bar{t})\rangle_{\rm ren} \nonumber\\ &\equiv& 
		\int_{-L}^0 d\Delta r d\Delta r' \left[\int r^2 d\Omega e^{-i {\bf k}_\perp\cdot {\bf r}_\perp} \right]
		\left[ \int r'^2 d\Omega' e^{-i {\bf k}'_\perp\cdot{\bf r}'_\perp} \right] \frac{e^{-i k \Delta r-i k' \Delta r'}}{r r'}
	  r r'\langle \hat{\cal R}^{}_{\bar{r}+\Delta r}, \hat{\cal R}'_{\bar{r}+\Delta r'}\rangle_{\rm ren}\nonumber\\
    &\approx& \frac{O(L^4)}{\bar{r}^2}\int_{-L}^0 d\Delta r d\Delta r' e^{-i k \Delta r-i k' \Delta r'} 
	  \left[ r r'\langle \hat{\cal R}^{}_{\bar{r}+\Delta r}, \hat{\cal R}'_{\bar{r}+\Delta r'}\rangle_{\rm ren} \right],
\label{RkRkren}
\end{eqnarray}
where we set a positive IR cutoff $\epsilon = k_{\perp}^2 \ll (dk)^2$ to keep $1/\omega^{}_{n=0}$ in (\ref{vaccorF}) regular. 
One can see that $\langle \hat{\cal R}^{}_{k\hat{x}^3}, \hat{\cal R}'_{k'\hat{x}^3}\rangle_{\rm ren}$ would be further suppressed by a 
factor of $\bar{r}^{-2}$ from $\langle \hat{\cal R}^{^{[0]}}_{k\hat{x}^3}, \hat{\cal R}'^{^{[0]}}_{k'\hat{x}^3}\rangle$ if 
$r r'\langle \hat{\cal R}^{}_{\bar{r}+\Delta r}, \hat{\cal R}'_{\bar{r}+\Delta r'}\rangle_{\rm ren}$ and 
$\langle \hat{\cal R}^{^{[0]}}_{\bar{r}+\Delta r}, \hat{\cal R}'^{^{[0]}}_{\bar{r}+\Delta r'}\rangle$ could be of the same order.

We then arrange the covariance matrix ${\cal V}$ as 
\begin{equation}
{\cal V} = \left( \begin{array}{ccccccc}
  \ddots &            &             &  \vdots 	&           &          & {\scriptstyle .}^{{\scriptstyle .}^{\scriptstyle .}​}\\
         & {\cal V}^{-2}{}_{2}  &  {\cal V}^{-1}{}_{2}  & {\cal V}^{0}{}_{2}  & {\cal V}^{1}{}_{2}  & {\cal V}^{2}{}_{2}  & \\
         & {\cal V}^{-2}{}_{1}  &  {\cal V}^{-1}{}_{1}  & {\cal V}^{0}{}_{1}  & {\cal V}^{1}{}_{1}  & {\cal V}^{2}{}_{1}  & \\				
  \cdots & {\cal V}^{-2}{}_{0}  &  {\cal V}^{-1}{}_{0}  & {\cal V}^{0}{}_{0}  & {\cal V}^{1}{}_{0}  & {\cal V}^{2}{}_{0}  & \cdots\\ 
  			 & {\cal V}^{-2}{}_{-1} &  {\cal V}^{-1}{}_{-1} & {\cal V}^{0}{}_{-1} & {\cal V}^{1}{}_{-1} & {\cal V}^{2}{}_{-1} & \\
		  	 & {\cal V}^{-2}{}_{-2} &  {\cal V}^{-1}{}_{-2} & {\cal V}^{0}{}_{-2} & {\cal V}^{1}{}_{-2} & {\cal V}^{2}{}_{-2} & \\
	{\scriptstyle .}^{{\scriptstyle .}^{\scriptstyle .}​}&            &             &  \vdots   &           &          & \ddots
\end{array} \right) \label{CMreduce}
\end{equation}
with each element ${\cal V}^m{}_n$ ($m,n\in {\bf Z}$) a $2\times 2$ matrix 
\begin{equation}
    {\cal V}^{m}{}_{n} = \left( \begin{array}{cc}
        \langle \hat{\Phi}_{m{\bf k}_0}(\bar{t}),\hat{\Phi}_{n{\bf k}_0}(\bar{t})\rangle & 
				\langle \hat{\Phi}_{m{\bf k}_0}(\bar{t}),\hat{\Pi}_{n{\bf k}_0}(\bar{t})\rangle \\
        \langle \hat{\Pi}_{m{\bf k}_0}(\bar{t}),\hat{\Phi}_{n{\bf k}_0}(\bar{t})\rangle & 
				\langle \hat{\Pi}_{m{\bf k}_0}(\bar{t}),\hat{\Pi}_{n{\bf k}_0}(\bar{t})\rangle \end{array} \right).\label{CMreducemn}
\end{equation}
Inserting $\langle \hat{\cal R}^{}_{\bf k}, \hat{\cal R}'_{\bf k'}\rangle_{\rm ren}$ in (\ref{RkRkren}) and the vacuum correlators 
(\ref{vaccorF}) and (\ref{vaccorP}) into (\ref{CMreduce}) and then diagonalizing it, we find that each eigen-vector is dominated by a pair 
of $\Phi^{}_{\pm{\bf k}}$ or $\Pi^{}_{\pm {\bf k}}$ with the same $|{\bf k}|=|n|\omega_0$ because the correlators 
(\ref{vaccorF}) and (\ref{vaccorP}) for the vacuum state of the free field are much greater than the corrections by the radiation field 
(\ref{RkRkren}). Denoting the approximated eigenvectors $\Phi^{\pm}_{\bf k} = (\Phi^{}_{\bf k}\pm \Phi_{-\bf k})/\sqrt{2}$ 
and $\Pi^{\pm}_{\bf k} = (\Pi^{}_{\bf k}\pm \Pi^{}_{-\bf k})/\sqrt{2}$, the corresponding eigenvalues of ${\cal V}$ are approximately 
\begin{eqnarray}
  \lambda(\Phi^\pm_{n{\bf k}_0}) &\approx& \langle \hat{\Phi}^{^{[0]}}_{n{\bf k}_0},\hat{\Phi}^{^{[0]}}_{-n{\bf k}_0}\rangle + 
	  \bar{r}^{-2} \ell^\pm_n(\Phi) , \label{EVFFapx}\\
  \lambda(\Pi^\pm_{n{\bf k}_0}) &\approx & \langle \hat{\Pi}^{^{[0]}}_{n{\bf k}_0},\hat{\Pi}^{^{[0]}}_{-n{\bf k}_0}\rangle + 
	  \bar{r}^{-2} \ell^\pm_n(\Pi) ,
	  \label{EVPPapx}
\end{eqnarray}
for $n = 1,2,3,\cdots$, respectively, where 
\begin{equation}
  \ell^{\pm}_n({\cal R}) \equiv \bar{r}^2\langle \hat{\cal R}^{}_{n{\bf k}_0},\hat{\cal R}^{}_{-n{\bf k}_0}\rangle_{\rm ren} 
    \pm s^{}_{\cal R} \sqrt{\bar{r}^2\langle \hat{\cal R}^{}_{n{\bf k}_0},\hat{\cal R}^{}_{n{\bf k}_0}\rangle_{\rm ren}\,
		    \bar{r}^2\langle\hat{\cal R}^{}_{-n{\bf k}_0},\hat{\cal R}^{}_{-n{\bf k}_0}\rangle_{\rm ren}},         \label{rFapx}
\end{equation}
with $s^{}_{\cal R} \equiv {\rm sign}\{$Re $\langle \hat{\cal R}^{}_{n{\bf k}_0},\hat{\cal R}^{}_{n{\bf k}_0}\rangle_{\rm ren}\}$
taking the value $+1$ or $-1$.

In fact, the amplitude of a real scalar field in the $k$-space is complex subject to the relation $\Phi^{}_{-{\bf k}}=\Phi^*_{\bf k}$. 
To fit the standard form of a Gaussian state (\ref{Fstate}) for a collection of quantum mechanical HOs, with different 
indexes $i$ and $j$ referring to different degrees of freedom, it would be convenient to split $\Phi^{}_{\bf k}$ into the real and 
imaginary parts, namely, ${\rm Re}\,\Phi{}_{\bf k} = (\Phi^{}_{\bf k}+ \Phi^{}_{-\bf k})/\sqrt{2}$ and ${\rm Im}\,\Phi^{}_{\bf k} =
(\Phi^{}_{\bf k}-\Phi^{}_{-\bf k})/(i\sqrt{2})$, which turn out to be the approximated eigenvectors $\Phi^+_{\bf k}$ and 
$\Phi^-_{\bf k}$ for the above covariance matrix ${\cal V}$ up to a factor $i$. They can be thought of as two HOs with index ${\bf k}$ in 
a half of the $k$-space while $\Phi^{\pm}_{-{\bf k}}$ $(=\pm \Phi^{\pm}_{\bf k})$ are not independent degrees of freedom.

According to (\ref{EVFFapx}) and (\ref{EVPPapx}), the asymptotic reduced state of the field harmonics in $\theta=0$ and $\pi$ directions is 
approximately a product state of each $(n{\bf k}_0, -n{\bf k}_0)$ mode-pair in terms of $\Phi^+_{n{\bf k}_0}$ and $\Phi^-_{n{\bf k}_0}$.
We will see that each asymptotic reduced state of a mode-pair $\Phi^{}_{\pm n{\bf k}_0}$ looks like a two-mode squeezed thermal state 
(so $\Phi^{}_{n{\bf k}_0}$ and $\Phi^{}_{-n{\bf k}_0}$ are entangled while $\Phi^+_{n{\bf k}_0}$ and $\Phi^-_{n{\bf k}_0}$ are 
approximately separable).

\begin{figure}
\includegraphics[width=7cm]{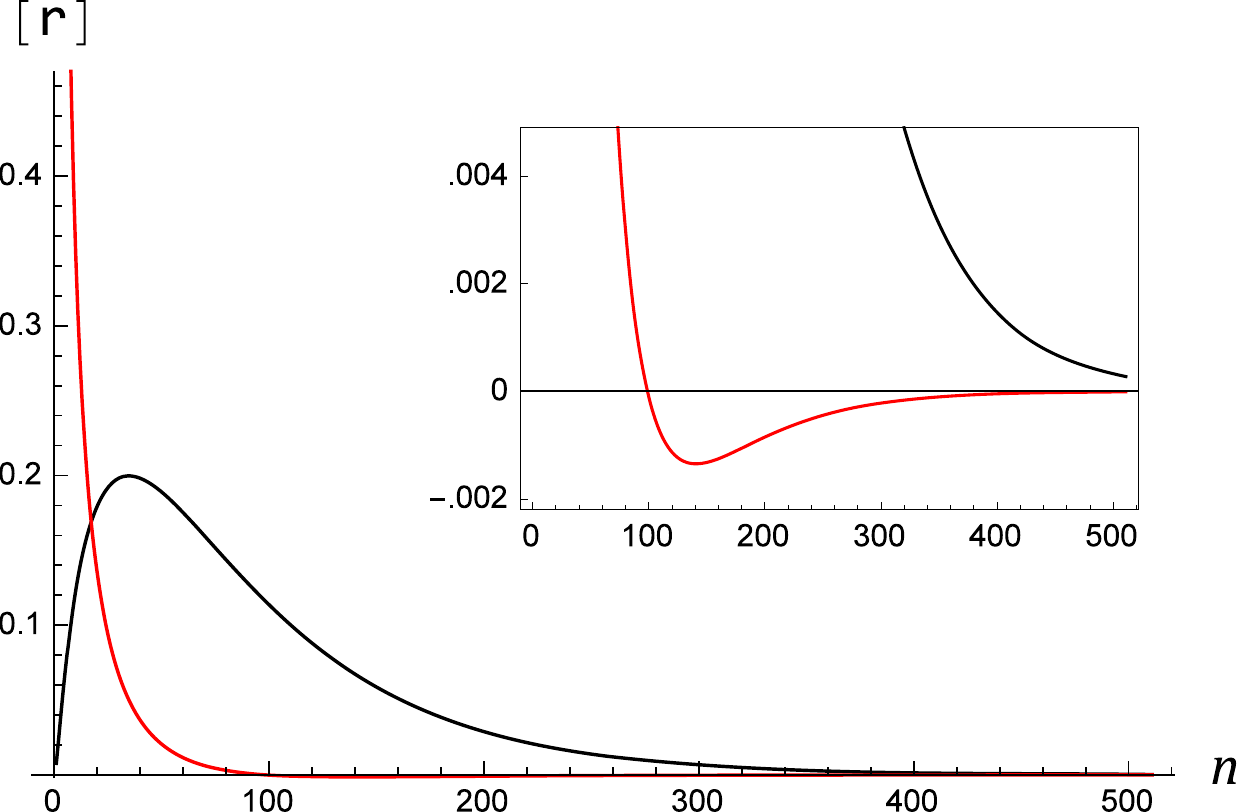}
\includegraphics[width=7.2cm]{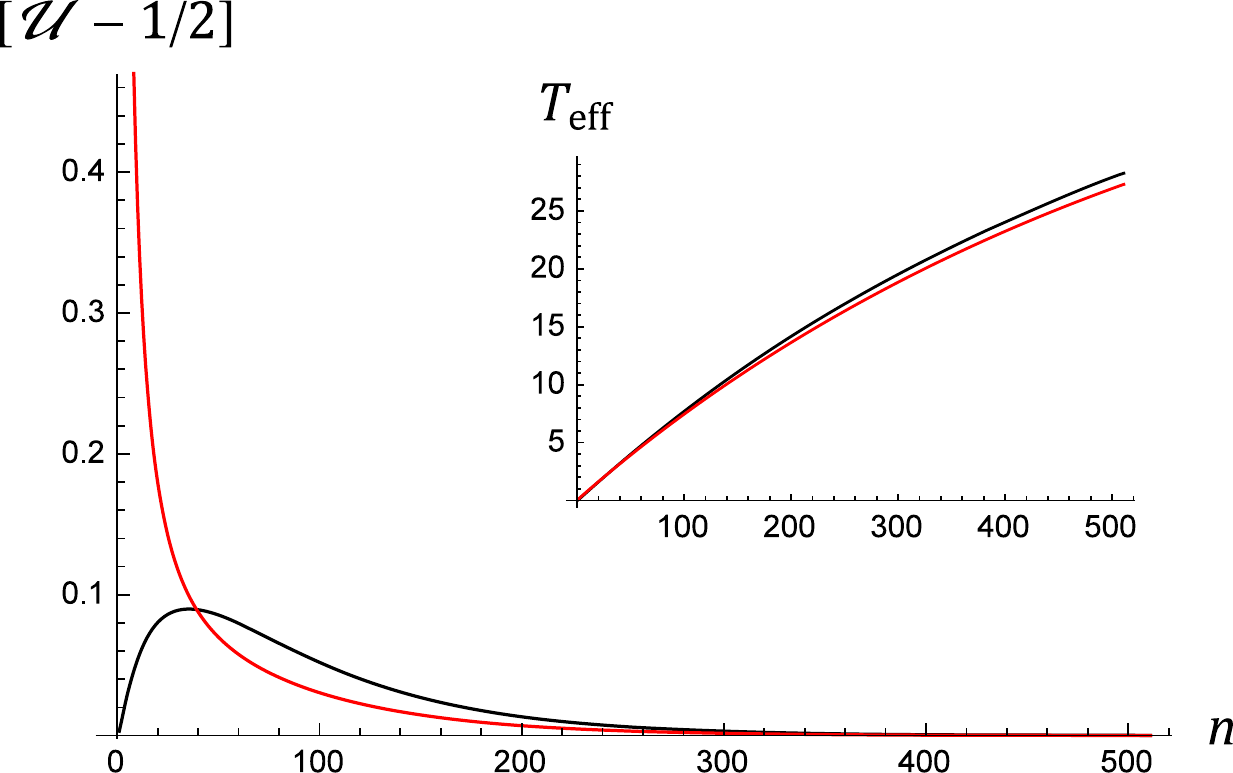}
\caption{(Left) The normalized squeeze parameter $[{\sf r}^+]\equiv V\bar{r}^2 {\sf r}^+$ (red) and $[{\sf r}^-]\equiv V\bar{r}^2 {\sf r}^-$ 
(black). The inset represents the same curves with a different scale. 
(Right) The normalized uncertainty function $[{\cal U}^+ -(\hbar/2)] \equiv V\bar{r}^2 ({\cal U}^+ - (\hbar/2))$ (red) and $[{\cal U}^- 
-(\hbar/2)]\equiv V\bar{r}^2  [{\cal U}^+ - (\hbar/2)]$ (black). The inset represents the corresponding effective temperature 
$T_{\rm eff}^\pm(n)$ of each mode, which is un-normalized and so depends on the value of $V$ and $\bar{r}$. 
Here $\gamma=0.01$, $\Omega=2.3$, $a_0=4$, $\bar{a} =7.6005$, $\omega_0 \approx 1.1826$, 
$dk = \omega_0/4$, and the IR cutoff $k_{\perp}=10^{-6} dk$. 
For $n=1$, $[{\sf r}^+] \approx 
[{\cal U}^+ -(\hbar/2)] \approx 4.01$, while 
$[{\sf r}^-] > 2\times [{\cal U}^- - (\hbar/2)] \approx 0.006$.}
\label{Sqeez}
\end{figure}

For a HO with the natural frequency $\Omega$ in a steady squeezed thermal state with the squeeze parameter ${\sf r}$ and temperature $T$, 
one has $\langle Q^2 \rangle = e^{-2{\sf r}}\hbar/(2\Omega)\coth(\Omega/(2T))$ and $\langle P^2 \rangle = e^{2{\sf r}} (\hbar\Omega/2)  
\coth(\Omega/(2T))$ 
provided that $\langle Q, P\rangle=0$ \cite{BV87}. 
Thus we extract the squeeze parameter and the effective temperature by computing
\begin{eqnarray}
  {\sf r}^{\pm}(n) &\equiv& \frac{1}{4}\ln\left( \frac{\lambda(\Pi^\pm_{n{\bf k}_0})}{\omega_n^2 \lambda(\Phi^\pm_{n{\bf k}_0})} \right),\\
  T^\pm_{\rm eff}(n) &\equiv& \frac{\omega_n}{2\coth^{-1}\left( 2 {\cal U}^\pm(n)/\hbar\right)},  \label{Teffdef}
\end{eqnarray}
respectively, where 
\begin{equation}
  {\cal U}^\pm(n) \equiv \sqrt{\lambda(\Phi^\pm_{n{\bf k}_0})\lambda(\Pi^\pm_{n{\bf k}_0})/V^2}
\end{equation}
is the uncertainty function satisfying the uncertainty relation ${\cal U}^\pm \ge \hbar/2$.
Note that the above definition of $T^{\pm}_{\rm eff}(n)$ is identical to the effective temperature defined in Eq.(33) of Ref. \cite{LH07},
though the latter is for the UD detector rather than the field modes.
Since $V$ and $\bar{r}$ are large parameters, one has the approximated values  
\begin{eqnarray}
  {\sf r}^{\pm} &\approx& \frac{1}{2\hbar V\bar{r}^2}\left( \omega_n^{-1} \ell^{\pm}_n(\Pi)-\omega_n \ell^{\pm}_n(\Phi)\right) + 
	    O(V^{-2}\bar{r}^{-4}),\\
	{\cal U}^\pm - \frac{\hbar}{2} &\approx& \frac{1}{2V\bar{r}^2}\left( \omega_n^{-1} \ell^{\pm}_n(\Pi)+\omega_n \ell^{\pm}_n(\Phi)\right) + 
	    O(V^{-2}\bar{r}^{-4}),
\end{eqnarray}
in the radiation zone.
To eliminate the dependence on the observer's parameters $V$ and $\bar{r}$, we define the normalized squeeze parameter and the normalized
uncertainty relation in the radiation zone as
\begin{eqnarray}
   \left[ {\sf r}^\pm \right] &\equiv& V\bar{r}^2 {\sf r}^\pm, \\
   \left[ {\cal U}^\pm - \frac{\hbar}{2} \right] &\equiv& V\bar{r}^2 \left( {\cal U}^\pm - \frac{\hbar}{2} \right).
\end{eqnarray}
Their values for the same case in Figure \ref{HHG2} are shown in Figure \ref{Sqeez}.

In the left plot of Figure \ref{Sqeez}, 
we find that as $n$ increases, $[{\sf r}^+]$ drops from the maximum value at $n=1$, 
which is positive, then becomes negative for $n\ge 99$, and reaches the minimum at $n=141$. 
$[{\sf r}^-]$ increases from a very small value at $n=1$, 
reaches the maximum value at $n=35$. It is positive for all $n$. 
$[{\cal U}^+-(\hbar/2)]$ starts from the maximum value at $n=1$, 
then monotonically decays to zero as $n$ increases,
while $[{\cal U}^--(\hbar/2)]$ increases from a small value at $n=1$, 
reaches the maximum value at $n=36$. 
Both $[{\cal U}^\pm-(\hbar/2)]$ are always positive, as they should not violate the uncertainty principle.

The values of the effective temperature $T_{\rm eff}^\pm(n)$ depend on the parameters of the combined system as well as 
those of the observer (such as $V$ and $\bar{r}$). When these parameters are fixed, in the inset of Figure \ref{Sqeez} (left), one can see 
that $T_{\rm eff}^\pm(n)$ behaves differently from $[{\sf r}^\pm]$ and $[{\cal U}^\pm-(\hbar/2)]$  
(monotonically increasing in the same domain of $n$).
From (\ref{Teffdef}) the effective temperature $T_{\rm eff}^{\pm}(n)$ for ${\cal U}^+$ is higher than the one for ${\cal U}^-$ if and 
only if ${\cal U}^+ > {\cal U}^-$.

Compare Figure \ref{Sqeez} with Figure \ref{HHG2}, 
we find that the most (or least) squeezed or uncertain harmonics ($n>1$) are quite different from those harmonics contributing the 
maximal or minimal radiated power.
So far we do not have any simple equality connecting the effective temperature of the harmonics $T^\pm_{\rm eff}(n)$ and
the effective temperature read off from the frequency spectrum of the radiated power ($T^{}_B\approx \bar{a}$) in Section \ref{cHHG},
or the effective temperature experienced by the UD HO detector ($T_{\rm eff} \approx \bar{a}/(2\pi)$) in \cite{DLMH13}.

\section{Summary}
\label{SecConcl}

We calculated the radiated power of an Unruh-DeWitt harmonic-oscillator detector in a linear oscillatory motion in the Minkowski 
vacuum of a massless scalar field in (3+1)D. The radiation field is determined in the radiation zone for a laboratory observer and is 
independent of the cutoff corresponding to the time resolution of the detector. 
We find that there are two pulses in a cycle of the oscillatory motion, one is concentrated around $\theta=0$, the other around $\theta=\pi$, 
and is highly compressed in the time domain when the detector motion is in the relativistic regime.
There are periods of negative radiated power in each cycle at each fixed observing angle in the laboratory frame.
The negative radiated power is due to the interference between vacuum fluctuations driving the HO and the retarded field emitted by the 
driven HO, which tends to suppress the signal of the Unruh effect in the radiation.
Anyway, the averaged radiated power over a cycle is always positive as guaranteed by the quantum inequalities.

While the detector in the oscillatory motion is emitting the radiation, the effective temperature inside the detector is close to the
Unruh temperature with the averaged proper acceleration. If we replace the correlators at the effective temperature by those at zero 
temperature, the result of the radiated power will be lower. The difference becomes significant when the detector is moving at a higher 
averaged acceleration and a shorter oscillating cycle, even with the interference terms included. This suggests that one may be able to distinguish 
the Unruh effect in the radiation by an Unruh-DeWitt detector in highly non-equilibrium conditions.

The compression of the pulse shape in the time domain corresponds to the coherent high harmonic generation in the frequency domain. 
Strong correlations can be found between the pulses in the time domain, and between the harmonics in the frequency domain. 
The correlation between two radiation fields emitted at two different retarded times $\tau$ and $\tau'$ decreases as the interval of the 
retarded times $|\tau-\tau'|$ increases, while the correlation between two radiation field harmonics at two different frequencies
$n \omega_0$ and $n'\omega_0$ decreases as $|n-n'|$ increases. At late times the correlations shows quasi-periodicity in $t+t'$ direction 
in the time domain, which can be related to the down conversion of the field quanta in the frequency domain. 

We further construct an asymptotic reduced state of the harmonics of the field around $\theta=0$ and $\pi$ in the radiation zone. 
We find that the harmonic modes of different wavelengths are approximately separable, due to the overwhelming largeness of the vacuum 
correlators of the free field. For each pair of field modes of the same harmonic frequency but in opposite directions, 
the reduced state looks like a two-mode squeezed thermal state. 

\begin{acknowledgments}
I thank Ming-Chang Chen, Pisin Chen, Larry Ford, Jen-Tsung Hsiang, Bei-Lok Hu, Tim Ralph, Ralf Sch\"utzhold, and Daiqin Su for 
illuminating discussions. I also thank the hospitality of the Department of Physics and Astronomy, University of Waterloo and 
Perimeter Institute for Theoretical Physics where part of this work was done during my visit from August 2014 to July 2015.
This work is supported by the Ministry of Science and Technology of Taiwan under Grants Nos. 102-2112-M-018-005-MY3, 103-2918-I-018-004,
105-2112-M-018-003 and 106-2112-M-018-002-MY3, and in part by the National Center for Theoretical Sciences, Taiwan.
\end{acknowledgments}

\appendix

\section{Two-point correlators with detectors in uniform motion and uniform acceleration}
\label{Sec2pt}

At late times the initial information in the UD detector has gone away with the retarded fields, and so the two point correlators of the detector as well as those of the field at finite distances from the detector are dominated by the v-parts \cite{LH06}, with the initial moment $\tau^{}_I \to -\infty$ and the interaction duration $\eta = \tau-\tau^{}_I \to \infty$.

For a UD detector in the model (\ref{Stot1}) uniformly accelerated along the worldline $z^\mu_{UA}(\tau)=(a^{-1}\sinh a\tau, 0,0,
a^{-1}\cosh a\tau)$ with the proper acceleration $a$, 
the v-parts of the symmetrized correlators of the detector (with ``mass" $m_0=1$ in \cite{LH06}) in the Minkowski vacuum are given by 
\cite{OLMH12} 
\begin{eqnarray}
  & &\langle \hat{Q}(\eta),\hat{Q}(\eta')\rangle_{\rm v}^{UA} \nonumber\\
	&=& \frac{\lambda_0^2}{\Omega^2} \rm{Re}\int_{\tau^{}_I}^\tau d\tilde{\tau}
   \int_{\tau'_I \to \tau^{}_I}^{\tau'} d\tilde{\tau}' K(\tau-\tilde{\tau})K(\tau'-\tilde{\tau}')D^+(z(\tau),z(\tau')) \nonumber\\ 
	&=& \frac{1}{2\pi \Omega} {\rm Re} \left\{ 
	 a e^{-a\tau'-\gamma\tau} \left[-\frac{i e^{-i\Omega\tau}}{a +\gamma +i \Omega} F_{\gamma+i\Omega}(e^{-a \tau'}) \right.\right.\nonumber\\
	& &\hspace{1cm} +\left.\frac{1}{a-(\gamma+i\Omega)}\left( \frac{\gamma}{\Omega} e^{-i\Omega\tau} 
	  -\left(\frac{\gamma}{\Omega}+i\right) e^{i\Omega\tau} \right) F_{-(\gamma+i\Omega)}(e^{-a \tau'}) \right] \nonumber\\
  & &+ a e^{-a\tau-\gamma\tau'}\left[-\frac{i e^{-i\Omega \tau'}}{a +\gamma +i \Omega}F_{\gamma+i\Omega}(e^{-a \tau}) \right.\nonumber\\
	& &\hspace{1cm} +\left.\frac{1}{a-(\gamma+i\Omega)} \left(\frac{\gamma}{\Omega} e^{-i\Omega\tau'} - 
  	\left(\frac{\gamma}{\Omega}+i\right) e^{i\Omega\tau'} \right)  F_{-(\gamma+i\Omega)}(e^{-a \tau})\right]\nonumber\\
  & &+\frac{4\gamma}{\Omega} \left(\Lambda_0 - \ln\frac{a}{\Omega}\right) e^{-\gamma(\tau+\tau')}\sin\Omega\tau\,\sin\Omega\tau' \nonumber\\
	& &+\frac{a}{\Omega} \left(e^{-(\gamma+i\Omega)(\tau+\tau')} \frac{\gamma}{\gamma+i\Omega} - 
    	 e^{-(\gamma-i\Omega)\tau -(\gamma+i\Omega)\tau'}\right) \nonumber\\
  & &+\frac{\gamma}{\Omega} \psi\left(1-\frac{\gamma+i\Omega}{a}\right) e^{-(\gamma+i\Omega)(\tau+\tau')}
		    \left[ 2 - \left(1+ i\frac{\Omega}{\gamma}\right)\left( e^{2i\Omega\tau}+e^{2i\Omega\tau'}\right) \right]\nonumber\\
	& &+i\pi e^{-(\gamma+i\Omega)|\tau-\tau'|} \cot \frac{\pi(\gamma+i\Omega)}{a}\nonumber\\
	& &+\left.a e^{-a|\tau-\tau'|}\left[ \frac{iF_{-(\gamma+i\Omega)} (e^{-a|\tau-\tau'|}) }{a-(\gamma+i\Omega)} + 
        \frac{i F_{\gamma+i\Omega} (e^{-a |\tau-\tau'|})}{a+\gamma+i\Omega}\right] \right\}, \label{QT1QT2}
\end{eqnarray}
while
$\langle \hat{P}(\eta),\hat{Q}(\eta')\rangle_{\rm v}^{UA}=\partial_\tau \langle \hat{Q}(\eta),\hat{Q}(\eta')\rangle_{\rm v}^{UA}$,
$\langle \hat{Q}(\eta),\hat{P}(\eta')\rangle_{\rm v}^{UA}=\partial_{\tau'} \langle \hat{Q}(\eta),\hat{Q}(\eta')\rangle_{\rm v}^{UA}$, and
$\langle \hat{P}(\eta),\hat{P}(\eta')\rangle_{\rm v}^{UA}=\partial_\tau\partial_{\tau'} \langle \hat{Q}(\eta),\hat{Q}(\eta')
\rangle_{\rm v}^{UA}$. 
Here $F_s(X)\equiv {}_2F_1(1+(s/a),1,2+(s/a),X)$, $K(X)\equiv e^{-\gamma X}\sin\Omega X$, $D^+(x,x')$ is the positive-frequency Wightman 
function (\ref{WightmanD}) for the massless scalar field in Minkowski vacuum, and we have used the identity 
\begin{equation}
  \psi(1+W) - \psi(1-W) = \frac{1}{W} - \pi \cot \pi W .
\end{equation}
The cutoff $\Lambda_0 = -\gamma_E -\ln \Omega|\tau'_I-\tau^{}_I|$ 
is present at early times since
\begin{equation}
  \lim_{T\to 0} \frac{e^{-T}}{1+W}\,{}_2F_1(1+W, 1, 2+W,e^{-T}) = -\psi(1+W)-\gamma_E - \ln T, \label{2F1div}
\end{equation}
and we have the initial moment $\tau'_I \to \tau^{}_I$ \cite{LH06, LH07}. The same logarithmic divergence arises
in the last line of (\ref{QT1QT2}) in the coincidence limit $\tau'\to \tau$, and 
we introduce another cutoff $\Lambda_1 = -\gamma_E -\ln \Omega|\tau'-\tau|$ 
to control it. In the coincidence limit $\tau'\to\tau$ for $\langle \hat{Q}(\eta),\hat{Q}(\eta')\rangle_{\rm v}^{UA}$ in (\ref{QT1QT2}), however, the $\Lambda_1$ term is purely imaginary 
and so should be dropped. For $\langle \hat{P}(\eta),\hat{Q}(\eta')\rangle_{\rm v}^{UA}$ and 
$\langle \hat{Q}(\eta),\hat{P}(\eta')\rangle_{\rm v}^{UA}$, the situation is similar, and
$\Lambda_1$ is present only in $\lim_{\tau'\to \tau}\langle \hat{P}(\eta),\hat{P}(\eta')\rangle_{\rm v}^{UA}$.
Then the above correlators in the coincidence limit reduce to the results given in \cite{LH06, LH07}.

The two-point correlators of a detector in uniform motion can be obtained by letting $a\to 0$ and applying
\begin{eqnarray}
   && \lim_{a\to 0} \frac{e^{-a\eta}}{1+(w/a)}\,{}_2F_1\left(1+(w/a), 1, 2+(w/a), e^{-a\eta}\right) = e^{w\eta}\Gamma(0,w \eta),\\
	 && \lim_{a\to 0} \frac{e^{-a\eta}}{1+(w/a)}\,{}_2F_1\left(1+(w/a), 1, 2+(w/a),-e^{-a\eta}\right) = 0,
\end{eqnarray}
with finite $\eta>0$ to (\ref{QT1QT2}). 

The two-point correlators of the field in the presence of the above uniformly accelerated detector going along $z^\mu_{UA}(\tau)$
can also be written in closed form. For the naive terms $\langle \hat{\cal R}^{{}^{[1]}}(x),\hat{\cal R}^{{}^{[1]}}(x')\rangle_{\rm v}^{UA}$ 
and their derivatives (${\cal R}=\Phi$, $\Pi$), one can insert the above two-point correlators of the detector to Eqs. (\ref{G11}) and 
(\ref{DDG11}). For the interference terms, the simplest one is
\begin{eqnarray}
  \langle \hat{\Phi}^{{}^{[1]}}(x), \hat{\Phi}^{{}^{[0]}}(x')\rangle_{\rm v}^{UA} &=& 
	\frac{\hbar\gamma}{4\pi^2 \Omega {\cal RR}'} {\rm Re}
	\left\{ e^{(-\gamma+i\Omega)\eta^{}_-(x)} \left[ h^{}_0\left(\eta^{}_-(x')\right) -h^{}_{\cal F}\left(\eta^{}_+(x')\right)\right] -\right. 
	\nonumber\\& & \left. \left[ h^{}_0\left(\eta^{}_-(x')-\eta^{}_-(x)\right) -h^{}_{\cal F}\left(\eta^{}_+(x')-\eta^{}_-(x)\right)\right] \right\}
	\label{F1F0}
\end{eqnarray}
where $\eta^{}_\pm(x) \equiv \tau^{}_\pm(x) - \tau^{}_I$ with the advanced time $\tau^{}_+(x)$ and the retarded time $\tau^{}_-(x)$ of the 
worldline $z^\mu_{UA}(\tau)$ for the observer at $x$ \cite{LH06}, ${\cal F} = 1$ if $x'$ is in the Rindler wedge F
for the worldline $z^\mu_{UA}(\tau)$, 
and ${\cal F} = 0$ if $x'$ is in the L- or R-wedge. The function $h^{}_f(X)$ is given by
\begin{eqnarray}
  h^{}_f(X) = \left\{ \begin{array}{lcc}
    \frac{-ia}{a+\gamma-i\Omega} e^{-a X-i\pi f} F_{\gamma-i\Omega}\left(e^{-a X-i\pi f}\right) +\frac{a}{i\gamma+\Omega} 
		    & \;{\rm for} & X>0;\\
    \frac{-ia}{a-\gamma+i\Omega} e^{a X+i\pi f} F_{-\gamma+i\Omega}\left(e^{a X+i\pi f}\right) + & & \\
		\hspace{2cm}\pi e^{(\gamma-i\Omega)X} 
		    \frac{\cosh \left[\frac{\pi}{a}(\Omega+i\gamma)(f-1)\right]}{\sinh\left[\frac{\pi}{a}(\Omega+i\gamma)\right]} & \;{\rm for} & X<0.
    \end{array}\right.
\end{eqnarray}
According to (\ref{2F1div}), $h^{}_0(X)$ 
has a logarithmic divergence as $X\to 0$, while 
$h^{}_1(0)$ is regular.
Other two-point correlators of the field can be obtained straightforwardly from (\ref{F1F0}) by 
$\langle \hat{\Pi}^{{}^{[1]}}(x), \hat{\Phi}^{{}^{[0]}}(x')\rangle_{\rm v}^{UA} = 
\partial^{}_{x^0}\langle \hat{\Phi}^{{}^{[1]}}(x), \hat{\Phi}^{{}^{[0]}}(x')\rangle_{\rm v}^{UA}$, 
$\langle \hat{\Phi}^{{}^{[1]}}(x), \hat{\Pi}^{{}^{[0]}}(x')\rangle_{\rm v}^{UA} = 
\partial^{}_{x'^0}\langle \hat{\Phi}^{{}^{[1]}}(x), \hat{\Phi}^{{}^{[0]}}(x')\rangle_{\rm v}^{UA}$, and 
$\langle \hat{\Pi}^{{}^{[1]}}(x), \hat{\Pi}^{{}^{[0]}}(x')\rangle_{\rm v}^{UA} = 
\partial^{}_{x^0}\partial^{}_{x'^0}\langle \hat{\Phi}^{{}^{[1]}}(x), \hat{\Phi}^{{}^{[0]}}(x')\rangle_{\rm v}^{UA}$, with the 
identity
\begin{eqnarray}
  \partial_X \left[ \pm e^{-X} {}_2F_1(1+W,1,2+W,\pm e^{-X})\right] &=& \frac{1+W}{1\mp e^{+X}} \pm\nonumber\\ 
	& &  W e^{-X}\, {}_2F_1(1+W,1,2+W,\pm e^{-X}), \label{dFdX}
\end{eqnarray}
which can be easily seen from the relation
\begin{equation}
   \frac{y}{1+W}\, {}_2F_1\left(1+W,1,2+W, y\right) =\sum_{n=1}^\infty \frac{y^n}{n+W} 
\end{equation}  
with $y = \pm e^{-X}$. In the coincidence limit $X\to 0$ the first term in the right hand side of (\ref{dFdX}) diverges like $1/X$ 
and so the derivative of (\ref{dFdX}) diverges like $1/X^2$. Fortunately these divergences are either purely imaginary thus absent in the 
symmetrized two-point correlators of the field, or canceled in the sum  $\langle \hat{\Phi}^{{}^{[1]}}(x), \hat{\Phi}^{{}^{[0]}}
(x')\rangle_{\rm v}^{UA} + \langle \hat{\Phi}^{{}^{[0]}}(x), \hat{\Phi}^{{}^{[1]}}(x')\rangle_{\rm v}^{UA}$. 

\end{document}